\documentclass[amstex,epsf,amssymb,usenatbib]{mn2e}
\usepackage{epsf,graphics}
\usepackage{amsmath,amssymb,natbib}
\usepackage{rotating,times,pictex,graphicx,latexsym,color,longtable}
\def\aj{AJ}                   % Astronomical Journal
\def\araa{ARA\&A}             % Annual Review of Astron and Astrophys
\def\apj{ApJ}                 % Astrophysical Journal
\def\aap{A\&A}                % Astronomy and Astrophysics
\def\mnras{MNRAS}             % Monthly Notices of the RAS
\def\pasj{PASJ}               % Publications of the ASJ
\title{The Structure of Molecular Clouds: I - All Sky Near Infrared Extinction
Maps}

\author[J. H. Rowles, D. Froebrich] {Jonathan Rowles $^{1}$\thanks{E-mail:
jr262@kent.ac.uk, } {Dirk Froebrich $^{1}$\thanks{E-mail: df@star.kent.ac.uk}}\\
$^1$Centre for Astrophysics \& Planetary Science, The University of Kent,
Canterbury, Kent CT2 7NH, U.K.} 

\date{Accepted .....
      Received ..... ;
      in original form .....}

\pagerange{\pageref{firstpage}--xxx}
\pubyear{2008}

\begin{document}
\maketitle
\label{firstpage}

\begin{abstract} 

We are studying the column density distribution of all nearby giant molecular
clouds. As part of this project we generated several all sky extinction maps.
They are calculated using the {\it median} near infrared colour excess technique
applied to data from the Two Micron All-Sky Survey (2MASS). Our large scale
approach allows us to fit spline functions to extinction free regions in order
to accurately  determine  the colour excess values. Two types of maps are
presented: i) Maps with a constant noise and variable spatial resolution; ii)
Maps with a constant spatial resolution and variable noise. Our {\it standard}
$A_V$ map uses the nearest 49 stars to the centre of each pixel for the
determination of the extinction. The one sigma variance is constant at 0.28\,mag
$A_V$ in the entire map. The distance to the 49$^{th}$ nearest star varies from
below 1\arcmin\ near the Galactic Plane to about 10\arcmin\ at the poles, but is
below 5\arcmin\ for all giant molecular clouds ($|b|<$\,30\degr). A comparison
with existing large scale maps shows that our extinction values are
systematically larger by 20\,\% compared to Dobashi et al. and 40\,\% smaller
compared to Schlegel et al.. This is most likely caused by the applied star
counting technique in Dobashi et al. and systematic uncertainties in the dust
temperature and emissivity in Schlegel et al.. Our superior resolution allows us
to detect more small scale high extinction cores compared to the other two
maps. 

\end{abstract}

\section{Introduction}
\label{intro}

Stars form within giant molecular clouds (GMCs). A large fraction of these
stars originate in embedded clusters (Lada \& Lada \cite{2003ARA&A..41...57L}).
The properties of these forming clusters, such as the mass function and the
star formation efficiency, are hence inextricably linked to the properties of
their natal molecular cloud. Of particular importance is the turbulence within
the clouds. Thus, the study of the dynamics, structure and chemistry of GMCs
plays a pivotal role in our understanding of the star formation process.

In order to study the turbulence within GMCs we can investigate their column
density distribution and compare it to theoretical predictions for the volume
density distribution which is found to be log-normal for isothermal flows (e.g.
Passot \& V\'{a}zquez-Semadeni \cite{1998PhRvE..58.4501P} and references
therein). For a large number of clouds along the line of sight with a log-normal
volume density distribution the column density distribution becomes normal.
Otherwise the column density distribution is log-normal (V\'{a}zquez-Semadeni \&
Garc\'{i}a \cite{2001ApJ...557..727V}). 

There are in principle a variety of ways in which the column density
distribution of GMCs can be investigated. These include e.g. thermal continuum
imaging of the cold dust, spectral line mapping of molecules such as CO, or the
determination of extinction maps. Each of these methods has its own
shortcomings, considering the accuracy and selection effects with which the
real column density distribution in clouds can be mapped. A detailed discussion
of this topic can be found in Goodman et al. \cite{2008arXiv0806.3441G}. The
authors conclude that near infrared extinction mapping might provide the 'best'
or least biased way to determine the column density distribution of clouds,
given a constant gas to dust ratio.

The mapping of extinction at optical and near infrared (NIR) wavelengths can be
performed using star counting (e.g. Wolf \cite{1923AN....219..109W}), colour
excess techniques (e.g. NICE - Lada et al. \cite{1994ApJ...429..694L}; NICER -
Lombardi \& Alves \cite{2001A&A...377.1023L}), or a combination thereof (e.g.
Lombardi \cite{2005A&A...438..169L}). With the availability of large scale
homogeneous surveys such as the Digitized Sky Survey (DSS; Lasker
\cite{1994AAS...184.3501L}) or the Two Micron All Sky Survey (2MASS; Skrutskie
et al. \cite{2006AJ....131.1163S}) the possibility to map the extinction of
entire cloud complexes in the Milky Way has arisen. In recent years a number of
large scale maps have been created by various authors: Dobashi et al.
\cite{2005PASJ...57S...1D} (hereafter D05) used optical star counts to map the
extinction along the entire Galactic Plane with $|b|$\,$<$\,40$\degr$;
Froebrich et al. \cite{2005A&A...432L..67F} used star counts in 2MASS to
determine the relative extinction in the entire Galactic Plane with
$|b|$\,$<$\,20$\degr$; Large scale maps (e.g. Pipe Nebula, Ophiuchus, Lupus)
based on the NICER technique are presented e.g. by Lombardi et al.
\cite{2006A&A...454..781L, 2008A&A...489..143L}; A
127$\degr$\,$\times$\,63$\degr$ region towards the Galactic Anticentre region
has been mapped by Froebrich et al. \cite{2007MNRAS.378.1447F} using the NICE
technique. Based on far infrared emission maps an all sky extinction map has
been created by Schlegel et al. \cite{1998ApJ...500..525S} (hereafter S98).

We plan to study the column density distribution of all clouds in the Milky Way
accessible with the NICE technique. As a first step we present here our all sky
near infrared extinction maps based on 2MASS. We have created a variety of maps
using a fixed grid (constant spatial resolution) method, as well as a nearest
neighbour (constant noise) approach. The resulting column density distributions
of the clouds will be analysed in a forthcoming publication.

Our paper is organised as follows. In Sect.\,\ref{method} we describe in detail
the method used to determine the extinction maps, including the calibration and
the uncertainties. The results are presented in Sect.\,\ref{results} where we
also compare our maps with the existing large scale extinction maps from D05
and S98. A summary is presented in Sect.\,\ref{summ}.

\begin{table}
\caption{\label{avail} The different maps created for the project. See text for
details.}
\centering
\begin{tabular}{lll}
 Number & Description & Type \\ \hline
 1 & $A_V$ - 25 stars & Con-noise-map \\
 2 & $A_V$ - 49 stars & Con-noise-map \\
 3 & $A_V$ - 100 stars & Con-noise-map \\
 4 & $\sigma_{A_V}$ - 25 stars & Con-noise-map \\
 5 & $\sigma_{A_V}$ - 49 stars & Con-noise-map \\
 6 & $\sigma_{A_V}$ - 100 stars & Con-noise-map \\
 7 & Distance to 25$^{th}$ neighbour & Con-noise-map \\
 8 & Distance to 49$^{th}$ neighbour & Con-noise-map \\
 9 & Distance to 100$^{th}$ neighbour & Con-noise-map \\ \hline
 10 & $A_V$ & Con-resolution-map \\
 11 & $\sigma_{A_V}$ & Con-resolution-map \\
\end{tabular}
\end{table}

\section{Extinction map determination method}
\label{method}

\subsection{The extinction maps}

For the determination of our extinction maps we used the 2MASS Point Source
catalogue (Skrutskie et al. \cite{2006AJ....131.1163S}). To ensure that only
high photometric quality sources are used, we extracted all objects with the
highest quality flag in each of the three filters (Qflag=`AAA') from the
catalogue. All other sources were disregarded. We extracted positions of all
sources from the catalogue, as well as the J, H, and K magnitudes and their
associated errors. The coordinates were converted into the Galactic coordinate
system using the IRAF\footnote{IRAF is distributed by the National Optical
Astronomy Observatories, which are operated by the Association of Universities
for Research in Astronomy, Inc., under cooperative agreement with the National
Science Foundation.} task \textsc{convert}. 

We have created different maps for the project to facilitate different needs. A
summary of all available maps can be found in Table\,\ref{avail}. There are two
main types of maps: i) Maps that use a fixed number of nearest stars at each
position. ii) Maps that use all stars within a fixed area around the position.
Effectively the first type of map can be considered as a constant noise -
variable spatial resolution map, and the second map as a constant spatial
resolution - variable noise map.

i) {\it Con-noise-maps}: These maps use either the 25, 49, or 100 nearest stars
to the central pixel position to determine the {\it median} J-H, H-K, and J-K
colour of stars. The choice of the number of stars basically creates three maps
with different spatial resolutions and noise. Thus, the spatial resolution
varies by about a factor of 2 between the map using 25 stars and the map using
100 stars. Additionally, we determine maps of the distance to the 25$^{th}$,
49$^{th}$, and 100$^{th}$ nearest neighbour (i.e. the spatial resolution at
each point). Furthermore a map of the uncertainty in the median colour at each
point is created.

ii) {\it Con-resolution-maps}: For these maps we fixed a radius around the
center of each pixel, and all stars closer than that radius were used in the
{\it median} J-H, H-K, and J-K colour determination. As radii we choose 1.7
times the pixel size for an oversampling of about three. We further determined
a map of the uncertainties in the median colours.

The pixel size of our maps varies with galactic latitude, according to stellar
density. For maps with variable spatial resolution we ensure that the
oversampling is on average between 2.5 and 3.5, except in areas of high
extinction where larger values are obtained. The pixel size in the maps was
varied from 0.5\arcmin\ at $b$\,=\,0$^\circ$ to 2\arcmin\ at
$|b|$\,=\,90$^\circ$ with increments of 0.5\arcmin. For the maps with constant
resolution the same pixel size as for the {\it Con-noise-maps} are chosen. With
the fixed oversampling this results in a fixed spatial resolution of these
maps. The details of how the pixel sizes and spatial resolution depends on
$|b|$ are given in Table\,\ref{varies}.

\begin{table}
\caption{\label{varies} Pixel sizes and spatial resolution of our maps
depending on Galactic Latitude $|b|$. The 'N-A' in the spatial resolution
column for the {\it Con-noise-maps} indicates that the spatial resolution
depends on the number of stars (25, 49, or 100) used and also varies with
position. A map of the spatial resolution for the map using the nearest 49
stars is shown in the right panel of Fig.\,\ref{mainmap}.}
\centering
\begin{tabular}{l|lcc}
Map type       & $|b|$ range & Pixel size [\arcmin] & Spatial resolution [\arcmin] \\ \hline
Con-noise      & 90$\degr$ -- 50$\degr$      & 2.0  & N-A \\
               & 50$\degr$ -- 40$\degr$      & 1.5  & N-A \\
               & 40$\degr$ -- 20$\degr$      & 1.0  & N-A \\
               & 20$\degr$ -- \,\,\,0$\degr$ & 0.5  & N-A \\ \hline
Con-resolution & 90$\degr$ -- 50$\degr$      & 2.0  & 6.8 \\
               & 50$\degr$ -- 40$\degr$      & 1.5  & 5.1 \\
               & 40$\degr$ -- 20$\degr$      & 1.0  & 3.4 \\
               & 20$\degr$ -- \,\,\,0$\degr$ & 0.5  & 1.7 \\
\end{tabular}
\end{table}

\subsection{Extinction determination and calibration}

\subsubsection*{Median colours}

We first determine the median colour of all stars at the position of each pixel
(depending on the type of map a different number of stars is included in this
calculation). For example we prepare a map where each pixel contains the value:

\begin{equation}
[J-H] = Median(m^i_{\rm J} - m^i_{\rm H})
\end{equation}

where $i$ runs from 1 to $N$, with $N$ being the total number of stars included
in the calculation at each pixel. $[J-H]$ symbolises the median colour of stars
in a pixel (in this case the J-H colour). $m^i_{\rm J}$ and $m^i_{\rm H}$ are
the individual 2MASS magnitudes of the stars used for the colour determination. 

We use the median colour instead of the mean because of the following: i) It
eliminates intrinsically red, young stars, which would systematically influence
the mean. This is the case as long as there are more foreground and background
stars than young stars. There are only two obvious cases in our map where the
determined colour is dominated by intrinsically red objects and hence not
representative of the extinction. These are the areas of the Orion Nebula
Cluster and the core of the L\,1688 cloud in Ophiuchus (see below). ii)
Similarly, clusters of stars with colours different from background and
foreground stars are eliminated if they do not dominate the numbers. iii) Most
importantly the median ensures that we either measure the correct colour
excess/extinction of a cloud or we measure no extinction (i.e. do not detect the
cloud at all). As long as there are more background stars to the cloud the
median ensures we determine the colour of the background stars, and hence the
correct colour excess. If there are more foreground stars to the cloud the
median corresponds to the colour of the foreground stars and we do not detect
the cloud. For details and a simulation of the recovery of extinction values
using this method see Froebrich \& del Burgo \cite{2006MNRAS.369.1901F}. This
last point is the most important for our work, since we will analyse the column
density distribution within the clouds. The use of the mean colour would distort
the column density distribution in a non-linear and position/extinction
dependent way, which would not allow any of our planned analysis.

\subsubsection*{Colour excess}

Each of our median colour maps has to be converted into the respective colour
excess map in order to determine the extinction/column density of material. The
median colour of the stars in regions without dust clouds varies with position
in the Galaxy due to different stellar populations. We hence need to determine
(fit) a function describing the median colour of stars in extinction free
regions in the maps. This was done in an iterative way described in the
following: 

\begin{enumerate}

\item[i)] The median colour maps were rebinned to a 10\arcmin\ per pixel map,
and median filtered with a filter size of one degree. 

\item[ii)] We masked out regions which are influenced by dust clouds.
Identification of clouds was initially achieved using the S98 maps and
excluding data with E($B-V$)\,$>$\,0.125\,mag. Once a preliminary $A_V$ map had
been produced (see steps iii to v) we repeated this stage but using our
extinction map created from $[H-K]$. In this final step we masked out all
pixels with $A_V$\,$>$0.5\,mag and extended the mask by 40\arcmin\ towards
unmasked regions to ensure all clouds with significant $A_V$ values are masked
out. 

\item[iii)] We then fit a spline function to the unmasked extinction free
regions to establish the colour excess zero point at all positions. At first
1D-spline functions were fit along 10\arcmin\ wide columns of constant galactic
longitude with knots every degree except in the masked regions. This enabled us
to fill in the gaps around the Galactic Plane where no extinction free data is
available. 

\item[iv)] We then fit 1D-spline functions along 10\arcmin\ wide rows of
constant galactic latitude with knots every degree to the results of step iii).
This final background colour array was expanded to the original size and
smoothed with a 5\arcmin\ radius. 

\item[v)] We then subtract the result of step iv) from the median colour maps to
determine the colour excess, i.e.:

\begin{equation}
\left< J-H \right>_{(l,b)} = [J-H]_{(l,b)} - fit[J-H]_{(l,b)}
\end{equation}

where $(l,b)$ denotes the position in the map. 

\end{enumerate}

The colour excess maps are then converted into H-band extinction, assuming a
power law for the extinction in the NIR $A_\lambda \propto \lambda^{-\beta}$,
with $\beta$ the near infrared extinction power-law index. The value for $\beta$
given in the literature varies between 1.6 and 2.0 (see e.g. Martin \& Whittet
\cite{1990ApJ...357..113M}, Drain \cite{2003ARA&A..41..241D}). Here we use a
value of $\beta$\,=\,1.7. Hence the extinction in the H-band can be determined
as:

\begin{equation}
A_{H, \left< J-H \right>} = \frac{\left< J-H
\right>}{\left(\frac{\lambda_H}{\lambda_J}\right)^{\beta} - 1} 
\end{equation}

or

\begin{equation}
A_{H, \left< H-K \right>} = \frac{\left< H-K \right>}{1 - \left(
\frac{\lambda_K}{\lambda_H}\right)^{-\beta}}    
\end{equation}

Both extinction values can be averaged and converted into optical extinction
($A_V$) via:

\begin{equation}
A_V = \frac{5.689}{2} \cdot \left( A_{H, \left< J-H \right>} + A_{H, \left< H-K
\right>} \right)
\end{equation}

where the factor 5.689 is the conversion of H-band into optical extinction
following Mathis \cite{1990ARA&A..28...37M}. The (in principle) available third
colour excess $\left< J-K \right>$ is not used, since it is not independent of
the two other colours. 

\subsubsection*{Calibration offsets}

The extinction maps determined from different colour excess (e.g. $\left< J-H
\right>$ and $\left< H-K \right>$) should, if we apply the correct extinction
law, result in the same extinction values. This also implies that our fit of the
median background colour is correct. Near the Galactic Plane and the Galactic
Center, however, there are not enough data points/regions that are free of
extinction, and hence the fit might not be 100\,\% accurate. This is in
particular a problem close to the Galactic Center, where the population of stars
is dominated by giants, compared to away from the Galactic Plane, where normal
dwarf stars dominate the stellar population in 2MASS. Thus, the majority of the
regions used for the fit of the background colour is dominated by dwarfs. Hence,
our fitted function describes the median colours of dwarf stars. In case of the
$[H-K]$ colours, there is no severe systematic problem with this, since dwarfs
and giants of the same spectral type have basically the same $[H-K]$ colours.
Hence, we can assume that our fit in $[H-K]$ is correct everywhere. The
situation is very different for $[J-H]$ and $[J-K]$, where giants and dwarfs
have different colours. Since we fit these colours using mostly regions
dominated by dwarfs, the fit in regions close to the Galactic Center will have
systematic offsets. Since giants have redder $[J-H]$ and $[J-K]$ colours than
dwarfs of the same spectral type, this will generate a systematic positive
offset in the extinction maps created from these colours, compared to the maps
generated from $[H-K]$. 

We have hence used the extinction maps created from $[H-K]$ as a control and
subtracted it from the other maps to check for differences. When determining the
difference $D_{A_V}$ between extinction maps obtained from $[J-H]$ and $[H-K]$,
we clearly find systematic residuals (see Fig.\ref{sysres}). Due to the redder
colour of the giants in $[J-H]$ an offset of almost 2\,mag in $A_V$ is found in
regions close to the Galactic Center. The size scale of these offsets is very
large, up to 120$^\circ$ in $l$. This effect of different colours of giants and
dwarfs can also be seen in the Large and Small Magellanic Clouds, which are
clearly apparent in the difference map. There are also negative offsets near
the Galactic Plane. These offsets can be caused by two effects: 

\begin{figure}
\centering
\includegraphics[width=8cm]{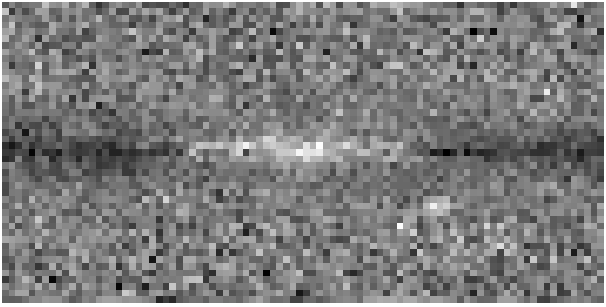}

\caption{\label{sysres} Grey scale representation of the difference map
calculated as $D_{A_V} = 5.689 \cdot \left( A_{H, \left< J-H \right>} - A_{H,
\left< H-K \right>} \right)$ (see text for details). $A_V$ values are scaled
linearly from -1.5\,mag (black) to +1.5\,mag (white). The image size is
360$\degr$\,$\times$\,180$\degr$ with the Galactic Centre in the middle. }

\end{figure}

\begin{enumerate}

\item[i)] The region is dominated by young stellar objects with an intrinsic
$H-K$ excess due to a disk. While this is the case in the Orion Nebula Cluster 
and the core of L\,1688 in Ophiuchus, it cannot be the cause for the large
scale offsets along the entire Galactic Plane. 

\item[ii)] The region contains a large fraction of higher mass stars (spectral
types O-F). Their $[J-H]$ colour is measurably smaller than in the population
of low mass dwarfs away from the Galactic Plane used for the fit. Hence, a
smaller extinction value is measured. The $[H-K]$ colours are also smaller, but
the effect seems to be not as pronounced (i.e. our extinction map calculated
from $[H-K]$ does not show negative $A_V$ `holes' near the Galactic Plane as
does the map determined from $[J-H]$). 

\end{enumerate}

\begin{figure*}
\centering
\includegraphics[width=8.5cm]{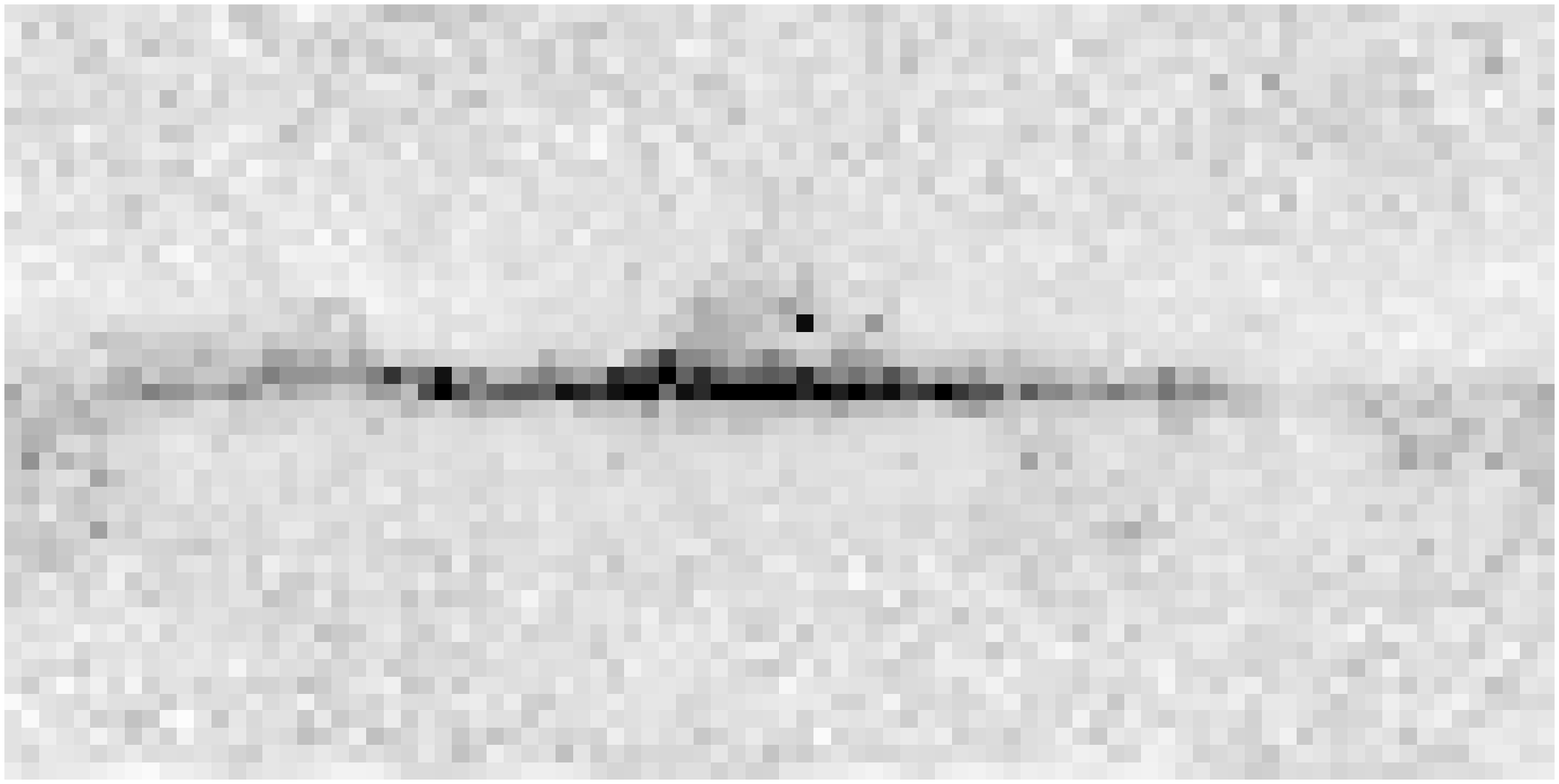} \hfill
\includegraphics[width=8.5cm]{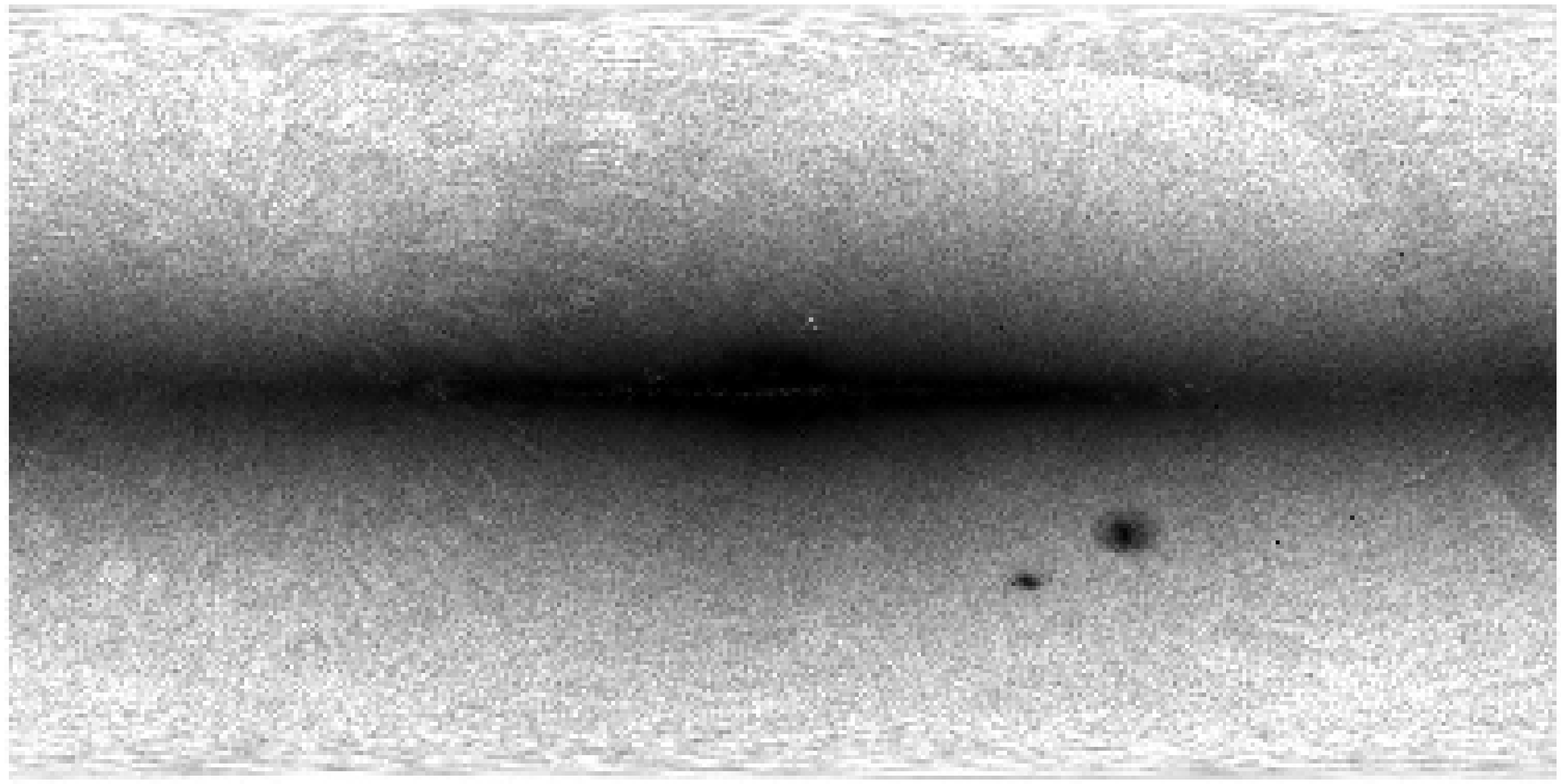}

\caption{\label{mainmap} {\bf Left Panel:} Our all sky extinction map using the
nearest 49 stars. The image size is 360$\degr$\,$\times$\,180$\degr$ with the
Galactic Centre in the middle. Optical extinction values are scaled linearly
from -1\,mag (white) to +7\,mag (black). {\bf Right Panel:} Grey scale map of
the distance to the 49$^{th}$ nearest neighbour, i.e. a spatial resolution map
of our data. Grey values are scaled from 1\arcmin\ (black) to 11\arcmin\
(white). The image scale is the same as in the left panel.}

\end{figure*}

We thus assume that point ii) is the reason for the large scale negative
offsets. There are further small scale variations caused by the effects of
variable observing conditions in 2MASS (see below). To correct for all large
scale offsets we filtered the difference image $D_{A_V}$ with a filter radius
of 4.5$^\circ$ in $l$ and 0.3$^\circ$ in $b$ and subtracted this filtered image
from the extinction map created from $[J-H]$. This will calibrate the
extinction maps compared to the maps determined from $[H-K]$, and this is also
the final map that is used in the determination of the $A_V$ maps presented in
this paper (see left panel of Fig.\,\ref{mainmap} for one example). Note that
this assumes that our extinction maps obtained from $[H-K]$ do not show any
systematic offsets. Our comparison with the known extinction maps from D05 and
S98 in Sect.\,\ref{comp} will show that this assumption is valid within the one
sigma uncertainties of our map.

\subsection{Uncertainties of the extinction maps}
\label{stat_uncertain}

\subsubsection*{Statistical uncertainties}

The uncertainties/variances $\sigma_{A_V}$ of the final extinction maps are
calculated via error propagation from 2MASS photometry. The uncertainties of the
median colours of stars in each pixel are e.g.:

\begin{equation}
\sigma_{[J-H]} = \frac{1.253}{N} \cdot \sqrt{ \sum\limits_{i=1}^{N}
(\sigma_J^i)^2 + \sum\limits_{i=1}^{N} (\sigma_H^i)^2 }
\end{equation}

where N denotes the total number of stars used for this pixel. $\sigma^i_J$ and
$\sigma^i_H$ are the individual J and H-band uncertainties in the photometry of
the star $i$. The factor 1.253 is caused by the use of the median instead of the
mean (Kenney \& Keeping \cite{1962..............K}). The uncertainty of the
median is larger by this factor than the uncertainty of the mean for large
samples and a normal distribution (which we assume for our individual
uncertainties).

To determine the uncertainties in the $A_V$ map we assume that the statistical
errors of the colour excess values are identical to the variance in the median
colour maps. That said, we assume that we only introduce systematic offsets but
no additional statistical uncertainties when converting the median colour maps
into colour excess maps (see below for details on these systematic offsets).
Hence:

\begin{equation}
\sigma_{\left< J-H \right>} = \sigma_{[J-H]}
\end{equation}

The corresponding errors in the H-band extinction determined from the colour
excess are then:

\begin{equation}
\sigma_{A_{H, \left< J-H \right>}} = \frac{\sigma_{\left< J-H
\right>}}{\left(\frac{\lambda_H}{\lambda_J}\right)^{\beta} - 1}  
\end{equation}

and

\begin{equation}
\sigma_{A_{H, \left< H-K \right>}} = \frac{\sigma_{\left< H-K \right>}}{1 -
\left( \frac{\lambda_K}{\lambda_H}\right)^{-\beta}} 
\end{equation}

When determining the variance of the averaged $A_V$ map we have to take into
account that the variances of the colours are not independent of each other
(see e.g. Froebrich \& del Burgo \cite{2006MNRAS.369.1901F}). We hence need to
consider the covariance between the extinction maps determined from $\left< J-H
\right>$ and $\left< H-K \right>$ colour excess. Hence the variance of the
$A_V$ map can be determined by:

\begin{equation}\label{sigma_av}
\sigma^2_{A_V} = \left( \frac{5.689}{2} \right)^2 \cdot \left( \sigma^2_{A_{H,
\left< J-H \right>}} + \sigma^2_{A_{H, \left< H-K \right>}} + 2 \cdot
\sigma_{cov} \right)
\end{equation}

where $\sigma_{cov}$ is the covariance of the two extinction maps. This
covariance can in principle be determined in extinction free regions (i.e. no
colour excess) in our map. We can calculated it as the average pixel value in
the product image of the two H-band extinction maps in areas with no
extinction, i.e:

\begin{equation}\label{cov_eq}
\sigma_{cov} = \frac{1}{N} \cdot \sum\limits^N_{j=0} \left( \left(A_{H, \left<
J-H \right>}\right)_j \cdot \left(A_{H, \left< H-K \right>}\right)_j \right)
\end{equation}

where $j$ runs over all pixels $N$ in the map with no colour excess. The
covariance values found by this way are about
$\sigma_{cov}$\,=\,0.0015\,mag$^2$. However, the H-band extinction values in
our maps are not completely independent from neighbouring pixels, since our
spatial resolution is larger than the pixel size. We hence cannot use
Eq.\,\ref{cov_eq} to determine $\sigma_{cov}$. Instead we determine the
covariance in the following way: i) Determine a histogram of extinction values
in cloud free regions in our final extinction map. ii) Measure the FWHM or
variance $\sigma_{A_V}$ of the distribution. iii) Using the knowledge of the
contribution from $\sigma_{\left< J-H \right>}$ and $\sigma_{\left< H-K
\right>}$ in Eq.\,\ref{sigma_av} we can determine the covariance.

Using this method we measure typical covariance values from 0.0025\,mag$^2$ to 
0.0055\,mag$^2$ in the map for the nearest 49 stars. We use
$\sigma_{cov}$\,=\,0.004\,mag$^2$ in the final maps. Table\,\ref{covreg} lists
some individual covariance values for selected areas. Note that this high value
for the $\sigma_{cov}$ means that the covariance dominates the total statistical
uncertainties in our maps. We empirically find that the average covariance
values for our other constant noise maps can approximately be described by a
power-law of the form $\sigma_{cov}(N) \propto N^{-0.8}$, where $N$ denotes the
number of stars used to determine the extinction at each pixel. We use this
equation to determine the covariance values for the constant resolution map,
because this map uses a different number of stars at each pixel and our
$\sigma_{cov}$ determination method described above cannot be applied.

Figure\,\ref{errAC} shows the distribution of the individual statistical
uncertainties for all pixels in a 25 square degree region near the Galactic
Plane (2.5$^\circ$$<$$b$$<$7.5$^\circ$; 221.67$^\circ$$<$$l$$<$226.67$^\circ$)
for all of our maps. In Table\,\ref{sigmatab} we list the peak position of the
$\sigma_{A_V}$ distribution for a sample of areas in our map using the nearest
49 stars.

\begin{table}
\centering
\caption{\label{covreg} Examples of the covariance values $\sigma_{cov}$
determined in a number of extinction free regions.}
\begin{tabular}{ccc}
 $l$ range & $b$ range & $\sigma_{cov}$\,[mag$^2$] \\  \hline
163.3$\degr\ $ -- 158.3$\degr\ $                   & $+$32.3$\degr\ $ -- \,\,$+$37.3$\degr\ $ & 0.0055 \\
\,\,\,31.3$\degr\ $ -- \,\,\,26.3$\degr\ $         & $+$31.3$\degr\ $ -- \,\,$+$36.3$\degr\ $ & 0.0031 \\
313.3$\degr\ $ -- 308.3$\degr\ $                   & $+$31.9$\degr\ $ -- \,\,$+$36.9$\degr\ $ & 0.0026 \\
272.0$\degr\ $ -- 267.0$\degr\ $                   & $+$31.0$\degr\ $ -- \,\,$+$36.0$\degr\ $ & 0.0030 \\
143.0$\degr\ $ -- 138.0$\degr\ $                   & $-$35.7$\degr\ $ -- \,\,$-$30.7$\degr\ $ & 0.0039 \\
\,\,\,85.1$\degr\ $ -- \,\,\,80.1$\degr\ $         & $-$36.0$\degr\ $ -- \,\,$-$31.0$\degr\ $ & 0.0037 \\
225.5$\degr\ $ -- 220.5$\degr\ $                   & $-$36.1$\degr\ $ -- \,\,$-$31.1$\degr\ $ & 0.0047 \\
244.2$\degr\ $ -- 239.2$\degr\ $                   & $-$37.5$\degr\ $ -- \,\,$-$32.5$\degr\ $ & 0.0052 \\
\end{tabular}
\end{table}

\subsubsection*{Systematic uncertainties}

So far we have only considered the statistical uncertainties in our maps. The
fitting of the background median colours to obtain the colour excess values, as
well as potential changes in the extinction law, will add further uncertainties
to our $\sigma_{A_V}$ estimate. The variances due to these points should in
principle be added into Eq.\,\ref{sigma_av}. Since all these uncertainties are
systematic in nature and on different scales, we do not add them into the
uncertainty maps presented here, but will discuss the scale and magnitude of
the terms in the following.

i) The fit of the background colours will introduce systematic offsets into
the extinction maps if it is not perfect. The angular scale of these offsets
will, however, be large, due to the nature of the fitted function. Hence the
scale of any systematic offsets will be much larger than 4.5$\degr$ in $l$ and
0.3$\degr$ in $b$, the smallest filter size used in the process of calibration.
These offsets will hence not influence any analysis of the column density
distributions of molecular clouds in the maps, in particular away (more than
1$\degr$) from the Galactic Plane. They will only lead to small offsets in the
absolute $A_V$ values. The magnitude of these offsets can be estimated from
the systematic comparison of our map with the maps of D05 and S98. We find in
Sect.\,\ref{comp}, that the offsets are generally not larger than the one
sigma variance of the extinction. 

ii) There are small scale (typically $\sim$14\arcmin\ by $\sim$6\arcmin) 
systematic offsets caused by the 2MASS observing procedure. Different observing
conditions have lead to varying completeness limits in the 2MASS data. The
different completeness limits cause the median colours of the stars to change
locally in the catalogue. For example at better conditions more fainter stars
are detected and these are usually redder. This is particularly obvious in
$[H-K]$, since deeper observations generally pick up more nearby dwarf stars,
which tend to be redder in $[H-K]$, but will not influence the $[J-H]$ colour.
We do not cap the stars included in the determination of our map at a global
completeness limit, however, since this would reject too many objects and
seriously degrade the achievable spatial resolution in many parts of the map.
The magnitudes of the extinction offsets are up to 1\,mag $A_V$ in the maps
determined from $[H-K]$, and only up to 0.5\,mag $A_V$ in the maps determined
from $[J-H]$. Hence, the maps determined from median $[J-H]$ colours suffer much
less and should be used if a locally absolutely calibrated $A_V$ map is
required. Note that in our final averaged $A_V$ maps these small scale
variations can be as large as two times the statistical variance. Hence, care
needs to be taken when analysing the column density distribution in these maps.
In particular, any cloud structures aligned with the 2MASS observing pattern
should be considered carefully.

iii) A spatially variable extinction law could introduce offsets that are
non-systematic and small scale. A change of the extinction law from the
interstellar value might be expected in higher column density regions, where
coagulation of dust grains and/or the formation of ice mantels might change the
optical properties (e.g. Ossenkopf \cite{1993A&A...280..617O}, Preibisch et al.
\cite{1993A&A...279..577P}, Larson \& Whittet \cite{2005ApJ...623..897L}). If
the  $\beta$ value of the extinction law changes by $\pm$\,0.2 compared to our 
applied value of 1.7, a change in optical extinction of about 10\,\% will be
the result. Given our average statistical uncertainties of about 0.28\,mag for
the map using the nearest 49 stars (see Table\,\ref{sigmatab}), possible
variations in the extinction law will only lead to measurable effects at $A_V$
values larger than 8\,mag (three sigma detections). The applied constant value
of $\beta = 1.7$ is hence justified. Furthermore, there is no residual cloud
structure visible in the $D_{A_V}$ map (Fig.\,\ref{sysres}), emphasising that
in general the assumption of a constant $\beta$ value is valid. We will
investigate possible small scale variations of $\beta$ in a forthcoming paper.

\begin{table*}
\caption{\label{sigmatab} Summary table of the properties of selected regions
in our map. We list the region name, the area, the peak in the distribution of
uncertainties in the map using the nearest 49 stars, the slope and offsets of
the extinction when compared between our map, D05 and S98. The ($^*$) indicates
slopes and offsets that have been fit by eye, since no satisfactory linear
regression fit could be achieved.}
\begin{tabular}{clllccccccc}
Region & Name & $l$ range & $b$ range & $\sigma_{A_V}$ & \multicolumn{2}{l}{D05 vs Our map} & \multicolumn{2}{l}{S98 vs Our map} & \multicolumn{2}{l}{S98 vs D05} \\
 &  & & & [mag] & offset & slope & offset & slope & offset & slope \\  \hline
1 & Camelopardalis        & 157.2$\degr\ $ - 152.2$\degr\ $            & 1.4$\degr$ - 6.4$\degr\ $ & 0.286 &  0.22 & 1.21 & -0.52\,\,\,& 0.52\,\,\,& -0.61\,\,\,& 0.36\,\,\,\\
2 & Serpens               & \,\,\,31.3$\degr\ $ - \,\,\,26.3$\degr\ $  & 1.3$\degr$ - 6.3$\degr\ $ & 0.282 &  0.17 & 1.28 & -1.70$^*$  & 1.00$^*$  & -1.40$^*$  & 0.80$^*$  \\
3 & Lupus                 & 341.8$\degr\ $ - 336.8$\degr\ $            & 1.9$\degr$ - 6.9$\degr\ $ & 0.282 &  0.82 & 1.06 & -0.34\,\,\,& 0.73\,\,\,& -0.80$^*$  & 0.60$^*$  \\
4 & Vela                  & 272.0$\degr\ $ - 267.0$\degr\ $            & 1.0$\degr$ - 6.0$\degr\ $ & 0.281 &  0.22 & 1.41 & -0.86\,\,\,& 0.70\,\,\,& -0.48\,\,\,& 0.37\,\,\,\\
5 & Taurus                & 166.0$\degr\ $ - 161.0$\degr\ $            & 4.9$\degr$ - 9.9$\degr\ $ & 0.282 &  0.19 & 1.13 & -0.31\,\,\,& 0.62\,\,\,& -0.54\,\,\,& 0.56\,\,\,\\
6 & North America Nebula  & \,\,\,85.0$\degr\ $ - \,\,\,80.0$\degr\ $  & 1.0$\degr$ - 6.0$\degr\ $ & 0.278 &  0.19 & 1.34 & -0.74\,\,\,& 0.73\,\,\,& -0.42\,\,\,& 0.46\,\,\,\\
7 & Monoceros             & 225.5$\degr\ $ - 223.5$\degr\ $            & 1.0$\degr$ - 6.0$\degr\ $ & 0.281 & -0.01 & 1.23 & -0.48\,\,\,& 0.58\,\,\,& -0.70$^*$  & 0.55$^*$  \\
8 & Auriga                & 185.0$\degr\ $ - 180.0$\degr\ $            & 3.0$\degr$ - 8.0$\degr\ $ & 0.281 &  0.11 & 1.23 & -0.58\,\,\,& 0.65\,\,\,& -0.63\,\,\,& 0.46\,\,\,\\
\end{tabular}
\end{table*}

\subsection{Other Methods}
 
There are, of course, other methods available to determine extinction maps based
on NIR photometry. These are in particular the NICER method presented by
Lombardi \& Alves \cite{2001A&A...377.1023L}, or the work presented by Lombardi
\cite{2005A&A...438..169L}. These methods are based on the minimisation of the
variance of the determined $A_V$ values by optimally combining the available
colour and position information. Naturally, these methods will have a smaller
variance than our method. The reduction in noise when using the NICER method
with respect to the simpler averaging of colour excess values applied by us is
typically of the order of about 20\,\% (see e.g. Froebrich et al.
\cite{2007MNRAS.378.1447F}). This would e.g. lower the 3\,$\sigma$ detection
limit in our standard map from 0.84\,mag to 0.70\,mag. However, possible
variations/uncertainties in the applied extinction law, as well as small scale
changes in the completeness limit (see Sect.\,\ref{stat_uncertain}) result in
similar size variations of the $A_V$ values. Furthermore, this small gain in
signal-to-noise has to be weighted against the immense additional computational
costs of the more sophisticated methods. These computational costs, as well as 
the simplicity and robustness of the colour excess technique are hence the
essential arguments for our choice of technique.

\section{Results}
\label{results}

\subsection{$A_V$ and spatial resolution maps}

We have created a number of all sky near infrared extinction maps for our
project. The available maps are listed in Table\,\ref{avail}. They will all be
made available on the CDS and can also be downloaded at {\tt
http://astro.kent.ac.uk/extinction}. 

In the left panel of Fig.\,\ref{mainmap} we show a grey scale representation of
our all sky $A_V$ map using the nearest 49 stars. Extinction values are scaled
linearly between -1\,mag (white) and +7\,mag (black) of optical extinction. The
general distribution of dust along the plane of the Galaxy is immediately
visible, as well as the more nearby higher latitude clouds such as Orion,
Ophiuchus, Chameleon, Taurus, etc.. The lack of any significant dust away from
the Galactic plane is also clear. Features such as the Large and Small
Magellanic Clouds can also be seen, and are caused by their red giants (see the
discussion above about the calibration of the maps). Some low $A_V$ regions are
noticeable in the southern half of the image, at latitudes of $b
\approx$\,-70\degr. The $E(B-V)$ map of S98 shows material in these positions as
well, indicating that the clouds are real. In Appendix\,\ref{highres} we show
detailed high resolution cut-outs of the entire 49$^{th}$ nearest neighbour map
for clarity. There we present 32\degr\,$\times$\,41\degr\ sized images. For
better contrast the extinction values are scaled by square root from 0\,mag to
15\,mag optical extinction.

The map of the distance to the 49$^{th}$ nearest neighbour star is shown in the
right panel of Fig.\,\ref{mainmap}. It varies from about 1\arcmin\ at $b
\approx$\,0\degr\ to 10\arcmin\ at $|b| \approx$\,90\degr. This clearly shows
that the spatial resolution varies by about a factor of 10 from near the
Galactic Plane to the poles in the image. The map further shows that the spatial
resolution is better than 3\arcmin\ to 5\arcmin\ in all regions of interest,
i.e. regions that contain giant molecular clouds, and generally better than
10\arcmin\ for $|b| <$\,30\degr. 

\begin{figure}
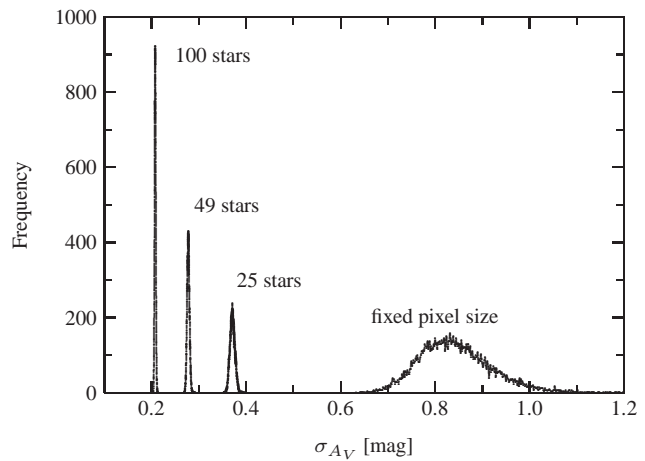

\beginpicture
\setcoordinatesystem units <62.825mm,0.05mm> point at  0 0
\setplotarea x from 0.1 to 1.2 , y from 0 to 1000
\axis left label {\begin{sideways}Frequency\end{sideways}}
ticks in long numbered from 0 to 1000 by 200
      short unlabeled from 0 to 1000 by 100 /
\axis right label {}
ticks in long unlabeled from 0 to 1000 by 200
      short unlabeled from 0 to 1000 by 100 /
\axis bottom label {$\sigma_{A_V}$\,[mag]}
ticks in long numbered from 0.2 to 1.2 by 0.2
      short unlabeled from 0.2 to 1.2 by 0.1 /
\axis top label {}
ticks in long unlabeled from 0.2 to 1.2 by 0.2
      short unlabeled from 0.2 to 1.2 by 0.1 /

\input hist_data_25.txt
\input hist_data_49.txt
\input hist_data_100.txt
\input hist_data_fixed.txt

\put {100 stars} at 0.33 900
\put {49 stars} at 0.36 500
\put {25 stars} at 0.45 300
\put {fixed pixel size} at 0.8 200

\endpicture  

\caption{\label{errAC} Distribution of the individual statistical uncertainties
of our $A_V$ values in a 25 square degree region near the Galactic Plane
(2.5\degr\,$<$\,$b$\,$<$\,7.5\degr; 221.67\degr\,$<$\,$l$\,$<$\,226.67\degr). We
show the one sigma variance distribution $\sigma_{A_V}$ for the maps using the
nearest 25 (peak at about 0.21\,mag), 49 (peak at about 0.28\,mag), 100 stars
(peak at about 0.37\,mag), and the constant pixel size map (peak at about
0.83\,mag). Note that the peaks of the distributions will not change with
position for the {\it con-noise} maps (see e.g. Table\,\ref{sigmatab}) since the
same number of stars is included in the $A_V$ calculation. The spatial
resolution of the 49$^{th}$ nearest neighbour map in the area for which the
histograms are plotted is about a factor of three higher compared to the fixed
pixel size map, hence the larger noise in the latter.}

\end{figure}

\subsection{Distribution of uncertainties}

Most of our maps use a fixed number of stars around the pixel position to
determine the extinction. Since most of the selected stars have very similar
photometric uncertainties, we named those maps {\it constant noise maps}. On
the other hand, the map using a fixed pixel size will use a different number of
stars and hence result in a more variable noise. 

In Fig.\,\ref{errAC} we plot the distribution of the one sigma variance in our
map for a 25 square degree region near the Galactic Plane
(2.5\degr\,$<$\,$b$\,$<$\,7.5\degr; 221.67\degr\,$<$\,$l$\,$<$\,226.67\degr).
For the constant noise maps the peaks in the distributions are very narrow
(justifying the name), while for the constant resolution map the uncertainties
vary by a much larger amount even in this rather small field of view. The main
cause is that the covariance dominates the variance in the maps, as seen in
Sect.\,\ref{stat_uncertain}. There are some pixels in the constant resolution
map where the noise and the extinction cannot be determined since no stars are
found in these pixels. Note that the spatial resolution of the 49$^{th}$ nearest
neighbour map in the area for which the histograms are plotted is about a factor
of three larger compared to the fixed pixel size map, which explains the larger
values of the uncertainties in the latter.

For the constant noise maps the peaks in the variance distribution are at 
0.21\,mag, 0.28\,mag, and 0.37\,mag for the 25$^{th}$, 49$^{th}$, and 100$^{th}$
nearest neighbour map respectively. This means, that e.g. in the 49$^{th}$
nearest neighbour map, the important threshold for self-shielding of molecular
clouds of 1\,mag $A_V$ (Hartmann et al. \cite{2001ApJ...562..852H}) is detected
always with a signal-to-noise ratio of larger than three. 

\begin{figure*}
\centering
\includegraphics[width=5.75cm]{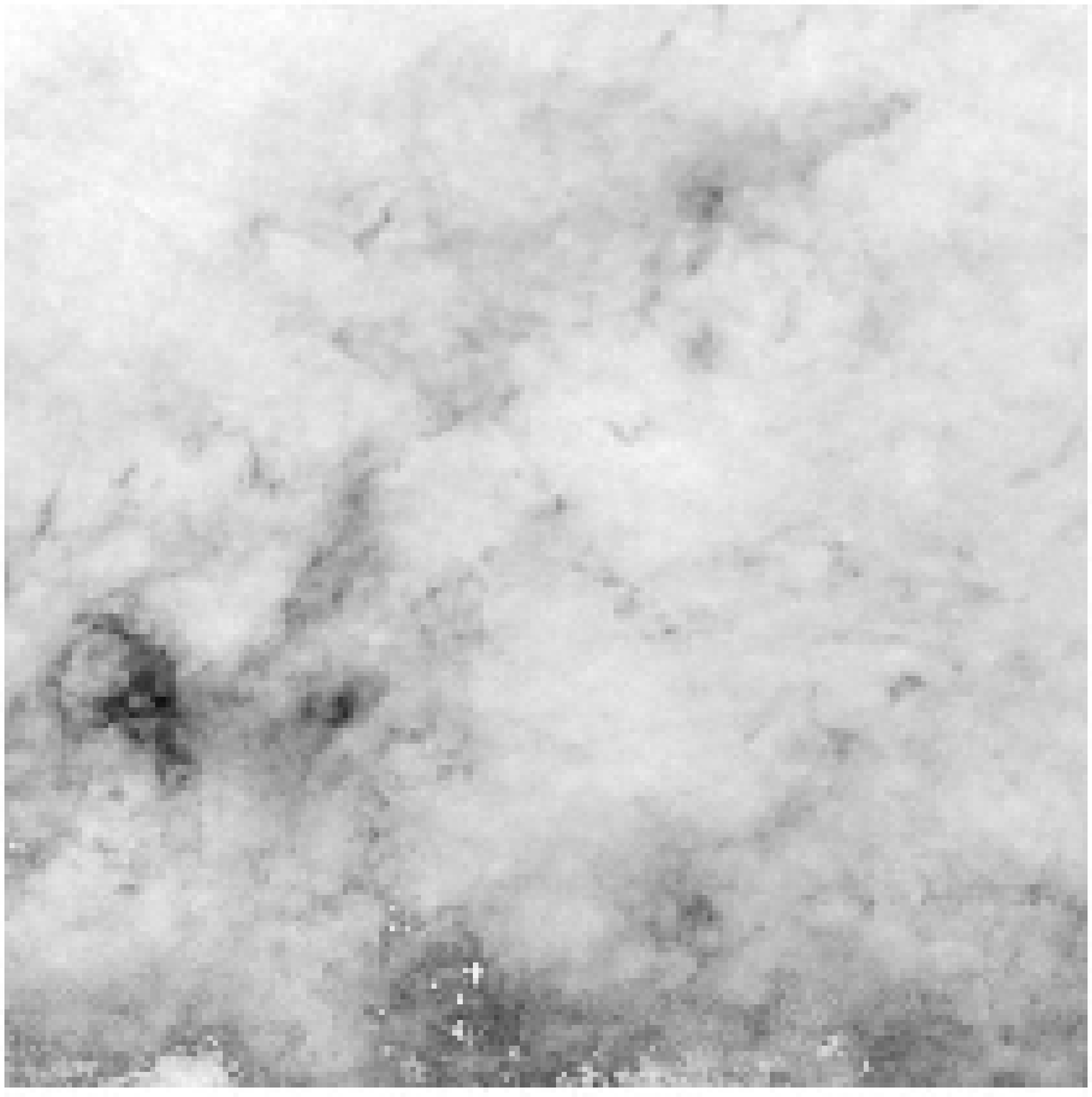} \hfill
\includegraphics[width=5.75cm]{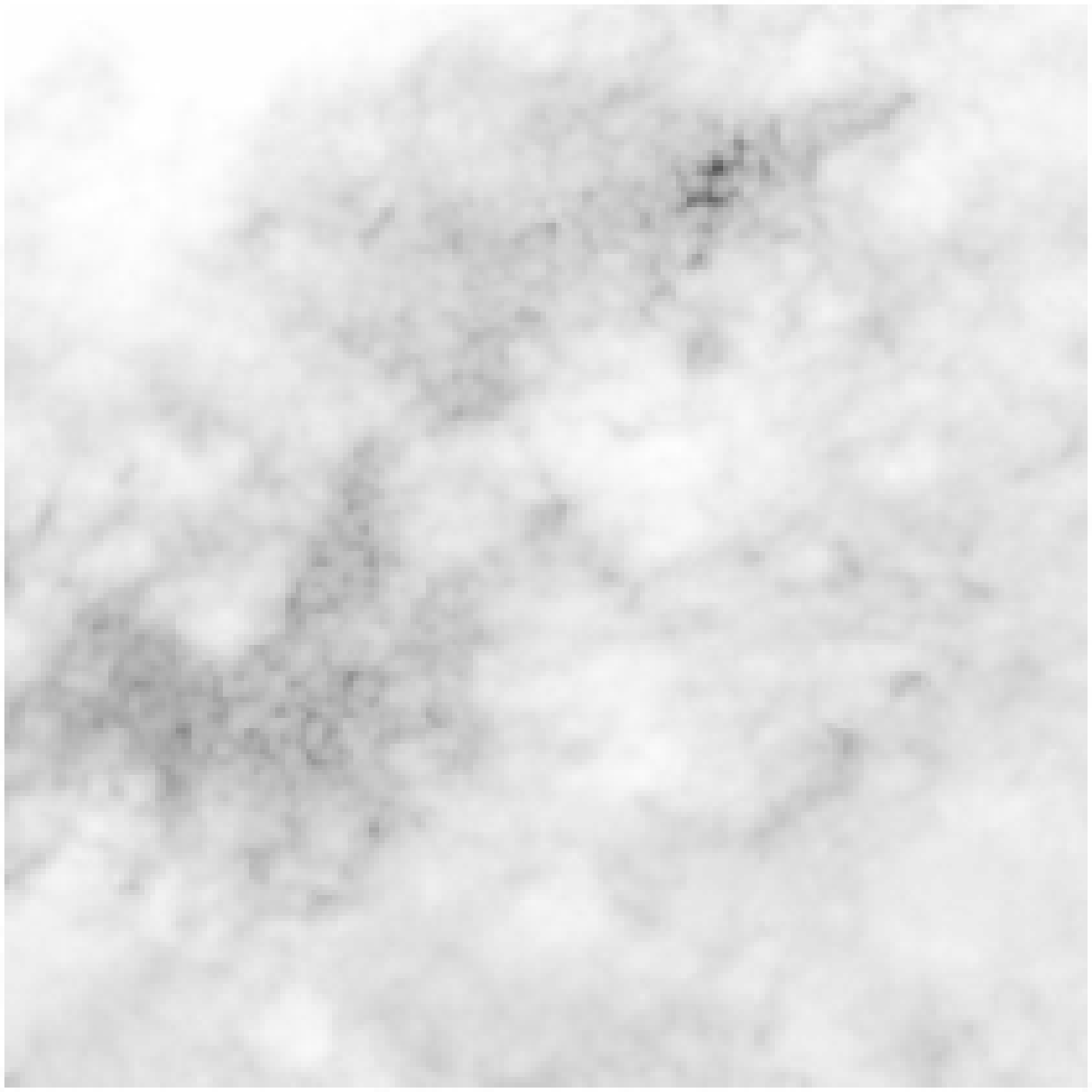} \hfill
\includegraphics[width=5.75cm]{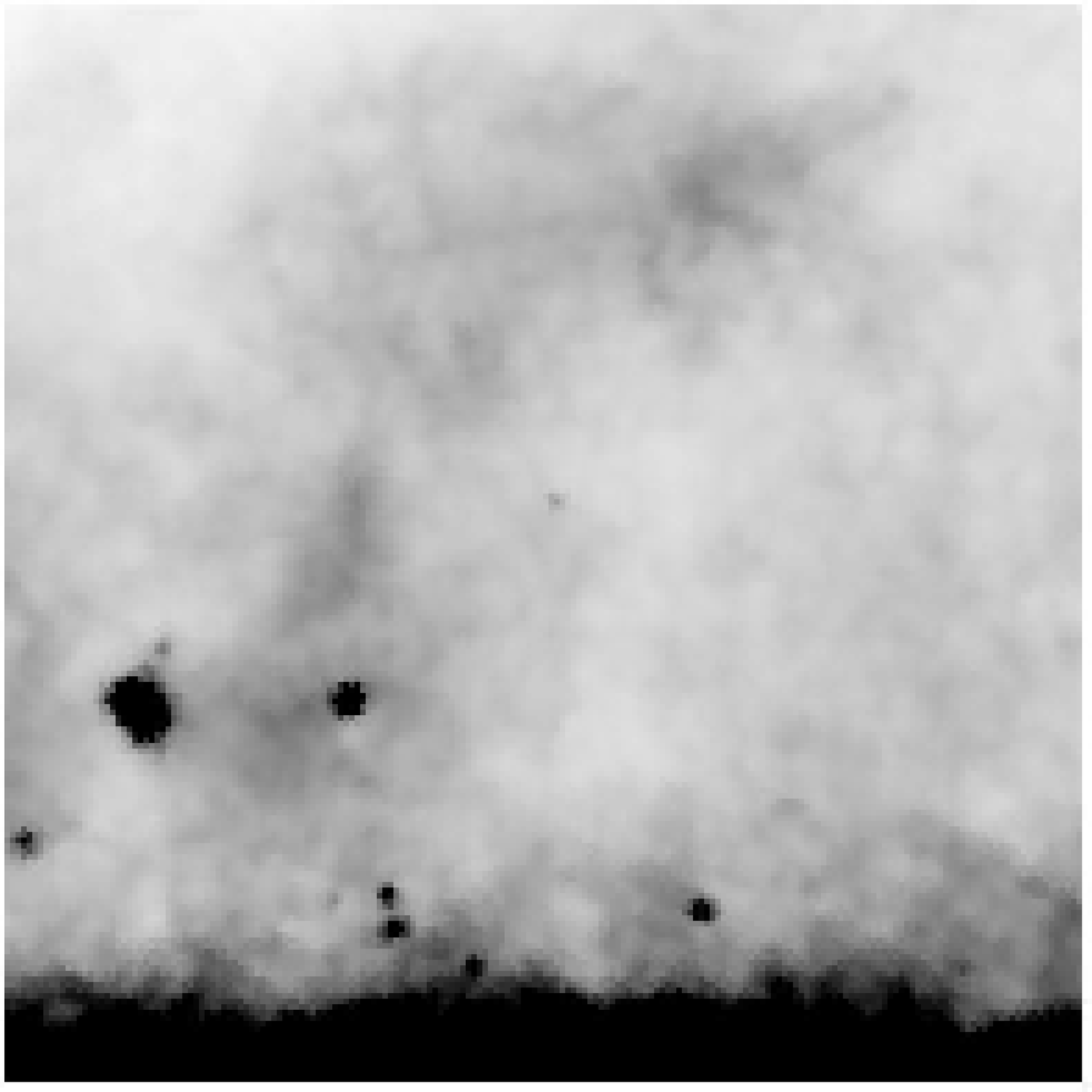}

\caption{\label{serpens_comp} Grey scale representation of region 2 (Serpens)
of our $A_V$ map (left panel), the map of D05 (middle panel), and the map of
S98 (right panel). The images are 10\degr\,$\times$\,10\degr\ in size and
encompass a coordinate range of 20\degr\,$< l <$\,30\degr\ and 0\degr\,$< b
<$\,10\degr. The extinction values in all maps are scaled linearly. White
represents 0.5\,mag optical extinction, while black represents 20\,mag $A_V$. }

\end{figure*}

\subsection{Comparison with other large scale extinction maps}
\label{comp}

To validate our calibration and to compare our new map with existing works we
facilitate the widely used all sky extinction map from S98 and the $|b| \leq
$40\degr\ map from D05. 

D05 produced $A_V$ maps at angular resolutions of 6\arcmin\ and 18\arcmin,
covering the sky in the galactic latitude range $|b| \leq $\,40\degr. They
performed star counts in the optical data from the Digitized Sky Survey (Lasker
\cite{1994AAS...184.3501L}). The wavelengths used lead to a better tracing of
low extinction regions in their maps, but underestimate $A_V$ in areas of high
extinction.   

S98 produced an all sky map of $E(B-V)$, intended for use as a new estimator of
Galactic extinction. They used data from COBE/DIRBE and IRAS/ISSA maps. The
diffuse emission from dust in the infrared can be used as a measure of column
density once the temperature and emissivity is known. The resolution of the
final $E(B-V)$ map of S98 is 6.1\arcmin. 

\subsubsection*{Visual Comparison}

In Fig.\,\ref{serpens_comp} we compare a section of our $A_V$ map (left panel)
with the maps of D05 (middle) and S98 (right). Extinction values are scaled
linearly between 0.5 and 20\,mag. The maps show a region (20\degr\,$< l
<$\,30\degr\ and 0\degr\,$< b <$\,10\degr) containing the Serpens molecular
cloud complex, that nicely illustrates the differences between the maps.

The most obvious difference is the higher spatial resolution of our map
(between 2\arcmin\ and 4\arcmin) compared to the other two maps (about
6\arcmin). Hence, much finer details, such as small, high column density cores
and small scale structures are visible only in our new map. 

When comparing our map with D05 there are two further main differences: i)
There is a number of very high $A_V$ regions not detected in the D05 map (e.g.
the main Serpens core). The reason is that optical star counts are not able to
trace such high column densities. ii) Near the Galactic Plane (positioned along
the bottom of the image) all but a few clouds are undetected in D05. Again the
reason is the rather low penetration depth of the optical star counting
technique. The near infrared colour excess method applied by us is able to
trace much higher column density regions. However, high $A_V$ regions very
close to the plane are also not picked up by our technique. These are regions
where foreground stars dominate the stellar population and our median colour
excess approach breaks down. Such regions are apparent as `white' low
extinction regions near the bottom in the left panel of
Fig.\,\ref{serpens_comp}.

A comparison of our map with S98 also shows two main differences: i) On average
the $A_V$ values from S98 are about a factor of 1.5 to 2 higher. This has
already been commented on by D05, and is probably caused by systematic
uncertainties in the assumed dust temperature and emissivity in S98. ii) There
are very high extinction regions in the densest cores and within 1\degr\ of the
Galactic Plane visible only in S98. The isolated high extinction cores are
probably superimposed point sources and/or regions with significantly different
dust temperatures. The high $A_V$ region near the plane is partly caused by the
same effect. Additionally the far infrared map basically traces the entire
emission from the galactic disk, while our technique has a limit of about
5-8\,kpc (e.g. Marshall et al. \cite{2006A&A...453..635M}, Froebrich \& del
Burgo \cite{2006MNRAS.369.1901F}).

\subsubsection*{Detailed comparison with Dobashi et al. (2005)}

\begin{figure}
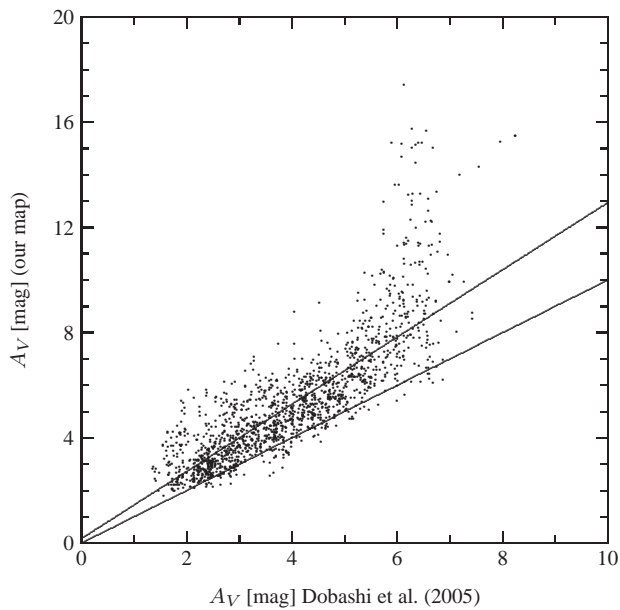

\beginpicture
\setcoordinatesystem units <7.0mm,3.5mm> point at  0 0
\setplotarea x from 0 to 10 , y from 0 to 20
\axis left label {\begin{sideways}$A_V$\,[mag] (our map)\end{sideways}}
ticks in long numbered from 0 to 20 by 4
      short unlabeled from 0 to 20 by 1 /
\axis right label {}
ticks in long unlabeled from 0 to 20 by 4
      short unlabeled from 0 to 20 by 1 /
\axis bottom label {$A_V$\,[mag] Dobashi et al. \cite{2005PASJ...57S...1D}}
ticks in long numbered from 0 to 10 by 2
      short unlabeled from 0 to 10 by 1 /
\axis top label {}
ticks in long unlabeled from 0 to 10 by 2
      short unlabeled from 0 to 10 by 1 /

\input dob_our_region2.txt
\plot 0 0 10 10 /
%{0.168927, 1.27756},
\plot 0 0.168927 10 12.944527 /
\endpicture  
\caption{\label{dobcomp} Comparison of the optical extinction values for the
region Serpens between our extinction map and the map from D05. As a solid line
we overplot a one-to-one line, as well as a linear fit. See
Table\,\ref{sigmatab} for the parameters of the fit.} 
\end{figure}

A more detailed comparison of our map with the D05 maps is shown in
Fig.\,\ref{dobcomp}, were we plot the D05 $A_V$ values against ours. The figure
contains data points from the 5\degr\,$\times$\,5\degr\ region 2 (Serpens),
whose coordinates are listed in Table\,\ref{sigmatab}. Note that we only plot a
random selection of less than 1\,\% of the data points from the entire area, to
facilitate visibility. A one-to-one line, as well as a linear fit (excluding
outliers) to the data points are overplotted. The parameters of the linear fit
are listed in Table\,\ref{sigmatab} for all regions. The individual comparison
plots for all other regions listed in Table\,\ref{sigmatab} are shown in
Appendix\,\ref{ours_vs_dob}. 

We see in Fig.\,\ref{dobcomp} that D05 systematically underestimates the
extinction by about 20\,\% compared to our values. There is on the other hand
no systematic offset, i.e. the linear fit has an offset of 0.17\,mag, well
below the one sigma variance of our $A_V$ values. At high column densities
(above 5-6\,mag $A_V$) there are very large differences between the two
methods. These are caused by two effects: i) As mentioned earlier, the optical
star counting method of D05 is not able to trace these very high extinction
regions; ii) Our better spatial resolution allows us to pick up more small
scale, high column density cores. 

In general the plots for all regions (Appendix\,\ref{ours_vs_dob}) show a
similar behaviour, i.e. a slight systematic underestimate of $A_V$ in D05
compared to our map and an offset below the variance of our data (see
Table\,\ref{sigmatab} for the linear fit parameters). The fact that the
offsets are below the variance level in our map, shows that our calibration of
the colour excess values has worked properly at all positions. The only region
tested where the offset is above the one sigma variance is Lupus. An
inspection of the actual plot (Fig.\,\ref{comp_dob_our_reg3}), however, shows
that a linear fit to the data is certainly not appropriate in this region. The
large number of small scale high $A_V$ cores in the field (see also Lombardi et
al. \cite{2008A&A...489..143L}) might cause the very different appearance of
the plot. 

With a few exceptions, the slope of the linear fits is about 1.2 (see
Table\,\ref{sigmatab}). There are two possible explanations for this: i) We
have used an incorrect extinction law; ii) There are systematic uncertainties
in the star count technique; 

There are two reasons why explanation i) is unlikely to be the dominant cause
for the discrepancy. Firstly, we do not find any residuals resembling the giant
molecular clouds when comparing the H-band extinction maps determined from
$\left< J-H \right>$ and $\left< H-K \right>$ colour excess (see
Fig.\,\ref{sysres}). Such systematic differences should occur if we had used
an incorrect extinction law. Secondly, to explain the 20\,\% difference, a
value in excess of $\beta$\,=\,2.2 has to be used in the conversion of colour
excess into extinction for all clouds. Even if such values can be found in some
clouds, generally a value of $\beta <$\,2.0 is found (e.g. Martin \& Whittet
\cite{1990ApJ...357..113M}, Drain \cite{2003ARA&A..41..241D}).

Explanation ii) seems to describe a more plausible cause for the 20\,\%
difference. The star counting technique from D05 counts the number of stars in
a certain area. With an increasing distance of the cloud an increasing fraction
of the stars are foreground. Thus, the further away the cloud is, the more the
extinction is underestimated by star counting. Our {\it median} colour excess
method on the other hand will either detect the correct colour excess of stars
behind the cloud, or not detect the cloud at all (see above). Simulations of
both extinction determination methods in Froebrich \& del Burgo
\cite{2006MNRAS.369.1901F} show this effect (their Fig.\,9), and that its
magnitude is in the order of 20\,\%. We expect that with an increasing cloud
distance the difference between our $A_V$ values and D05 increases. There is a
tentative hint of that in the regions selected in Table\,\ref{sigmatab}, but we
will analyse this in more detail in a forthcoming paper.

\subsubsection*{Detailed comparison with Schlegel et al. (1998)}

We also compare our $A_V$ values to those obtained in S98 for the same set of
regions as above. As an example we show the comparison for region2 (Serpens) in
Fig.\,\ref{schcomp}. All other individual plots are presented in
Appendix\,\ref{ours_vs_sch}. The parameters of the linear fits are also given in
Table\,\ref{sigmatab}. 

For the Serpens region there seems to be a general offset of about 1.7\,mag
$A_V$ between S98 and our map. There are, however, notable exceptions where
either the S98 or our map indicates much larger extinction values. The causes
for these are either our higher spatial resolution (our $A_V$ higher than S98)
or possibly groups of point sources not removed in the S98 maps (in particular
near the Galactic Plane) in combination with the limitations of our method to
pick up very high column density regions (S98 $A_V$ values higher than ours). In
general, we find that the S98 values overestimate the extinction by a factor of
about 1.5 to 2.0. This in in agreement with D05, who found an overestimate of
$A_V$ from S98 compared with their data by up to a factor of two. The offsets
between the two extinction values are generally higher than our one sigma
variance. The main reason for these differences are uncertainties in the dust
temperature and emissivity, as well as the much larger line of sight in the S98
data. For details we refer to the discussion in D05.

For completeness reasons we have also plotted the $A_V$ values of S98 against
D05 in Fig.\,\ref{sch_vs_dob}. The results of the linear fits are also listed
in Table\,\ref{sigmatab}.

\begin{figure}
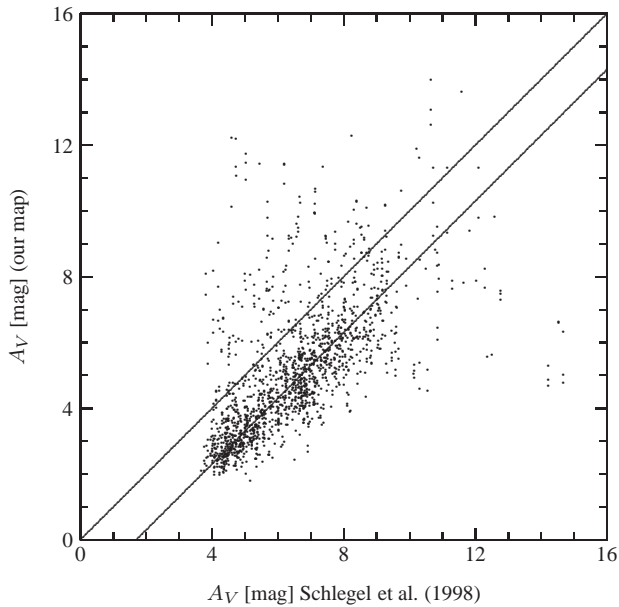

\beginpicture
\setcoordinatesystem units <4.375mm,4.375mm> point at  0 0
\setplotarea x from 0 to 16 , y from 0 to 16
\axis left label {\begin{sideways}$A_V$\,[mag] (our map)\end{sideways}}
ticks in long numbered from 0 to 16 by 4
      short unlabeled from 0 to 16 by 1 /
\axis right label {}
ticks in long unlabeled from 0 to 16 by 4
      short unlabeled from 0 to 16 by 1 /
\axis bottom label {$A_V$\,[mag] Schlegel et al. \cite{1998ApJ...500..525S}}
ticks in long numbered from 0 to 16 by 4
      short unlabeled from 0 to 16 by 1 /
\axis top label {}
ticks in long unlabeled from 0 to 16 by 4
      short unlabeled from 0 to 16 by 1 /
\input sch_our_region2.txt
\plot 0 0 16 16 /
%-1.700000 1.000000#
\plot 1.7 0 16 14.3 /
\endpicture  
\caption{\label{schcomp} Comparison of the optical extinction values for the
region Serpens between our extinction map and the map from S98. As solid line we
overplot a one-to-one line, as well as a linear fit. See Table\,\ref{sigmatab}
for the parameters of the fit.}   
\end{figure}

\section{Summary and Conclusions}
\label{summ}

We have determined a number of all sky near infrared extinction maps using the
2MASS point source catalogue. Constant spatial resolution maps (and hence
variable noise) as well as constant noise maps (and hence variable spatial
resolution) using either the nearest 25, 49, or 100 stars are presented. We
apply a {\it median} colour excess technique to accurately determine the column
density distribution of all nearby giant molecular clouds. Median colours are
converted into colour excess by means of a spline fit to extinction free
regions. 

Our {\it standard} $A_V$ map uses the nearest 49 stars around each position for
the extinction determination. The distance to the 49$^{th}$ nearest star varies
between 1\arcmin\ near the Galactic Plane to about 10\arcmin\ at the poles. The
one sigma variance in the map is 0.28\,mag $A_V$. This means that all regions
with an optical extinction of more than one magnitude (the important threshold
for self-shielding of molecular clouds - Hartmann et al.
\cite{2001ApJ...562..852H}) are detected with a signal-to-noise ratio above
3.6.

A comparison with the existing large scale extinction maps from Dobashi et al.
and Schlegel et al. has been performed. This comparison shows that our
calibration of the $A_V$ values is accurate in the entire map to within the one
sigma variance on scales of a few degrees. It further reveals systematic
differences in the extinction values compared to Dobashi et al. (our values are
$\sim$20\,\% larger) and Schlegel et al. (our values are $\sim$40\,\% smaller).
This can be attributed to the star counting technique (D05) and systematic
uncertainties in the dust temperature and emissivity (S98).

All our extinction maps have been determined in a uniform way for the entire
sky. The variety of different spatial resolutions and hence signal-to-noise
ratios provides an excellent homogeneous set of extinction values for different
needs. Our applied technique does not suffer from systematic uncertainties in
the dust temperature and emissivity, or from distance dependent offsets. The
spatial resolution is comparable or superior to other existing all sky $A_V$
maps for $|b| <$\,30\degr. Only in some regions with $|b| <$\,1\degr\ does our
map suffer from a non-detection of clouds due to the domination of foreground
stars and different methods such as the usage of longer wavelength data should
be applied (e.g. Schultheis et al. \cite{2008arXiv0811.2902S}).

\section*{acknowledgements}
\label{acks}

JR acknowledges a University of Kent scholarship. This publication makes use of
data products from the Two Micron All Sky  Survey, which is a joint project of
the University of Massachusetts and the Infrared Processing and Analysis
Center/California Institute  of Technology, funded by the National Aeronautics
and Space  Administration and the National Science Foundation.

\clearpage
\newpage
\begin{appendix}

\section{Comparison of Maps for different Regions}
\subsection{Our Map vs. Dobashi et al. (2005)}\label{ours_vs_dob}

\begin{figure}
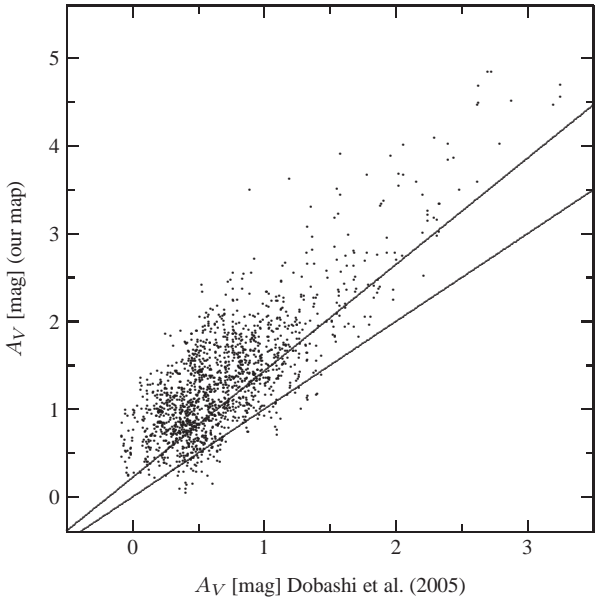

\beginpicture
\setcoordinatesystem units <17.5mm,11.67mm> point at  0 0
\setplotarea x from -0.5 to 3.5 , y from -0.4 to 5.6
\axis left label {\begin{sideways}$A_V$\,[mag] (our map)\end{sideways}}
ticks in long numbered from 0 to 5 by 1
      short unlabeled from 0 to 5 by 0.5 /
\axis right label {}
ticks in long unlabeled from 0 to 5 by 1
      short unlabeled from 0 to 5 by 0.5 /
\axis bottom label {$A_V$\,[mag] Dobashi et al. \cite{2005PASJ...57S...1D}}
ticks in long numbered from 0 to 3 by 1
      short unlabeled from 0 to 3 by 0.5 /
\axis top label {}
ticks in long unlabeled from 0 to 3 by 1
      short unlabeled from 0 to 3 by 0.5 /
\input dob_our_region1.txt
\plot -0.4 -0.4 3.5 3.5 /
%{{0.22097, 1.21353},
\plot -0.5 -0.385795 3.5 4.468325 /
\endpicture  
\caption{\label{comp_dob_our_reg1} Comparison of the optical extinction values
for region1 (Camelopardalis) in Table\,\ref{sigmatab} between our extinction map
and the map from D05. As solid line we overplot a one-to-one
line, as well as a linear fit. See Table\,\ref{sigmatab} for the parameters of
the fit.}    
\end{figure}

\begin{figure}
\beginpicture
\setcoordinatesystem units <7.0mm,3.5mm> point at  0 0
\setplotarea x from 0 to 10 , y from 0 to 20
\axis left label {\begin{sideways}$A_V$\,[mag] (our map)\end{sideways}}
ticks in long numbered from 0 to 20 by 4
      short unlabeled from 0 to 20 by 1 /
\axis right label {}
ticks in long unlabeled from 0 to 20 by 4
      short unlabeled from 0 to 20 by 1 /
\axis bottom label {$A_V$\,[mag] Dobashi et al. \cite{2005PASJ...57S...1D}}
ticks in long numbered from 0 to 10 by 2
      short unlabeled from 0 to 10 by 1 /
\axis top label {}
ticks in long unlabeled from 0 to 10 by 2
      short unlabeled from 0 to 10 by 1 /
\input dob_our_region2.txt
\plot 0 0 10 10 /
%{0.168927, 1.27756},
\plot 0 0.168927 10 12.944527 /
\endpicture  
\caption{\label{comp_dob_our_reg2} Comparison of the optical extinction values
for region2 (Serpens) in Table\,\ref{sigmatab} between our extinction map and
the map from D05. As solid line we overplot a one-to-one
line, as well as a linear fit. See Table\,\ref{sigmatab} for the parameters of
the fit.}   
\end{figure}

\begin{figure}
\beginpicture
\setcoordinatesystem units <14.0mm,7mm> point at  0 0
\setplotarea x from 0 to 5 , y from 0 to 10
\axis left label {\begin{sideways}$A_V$\,[mag] (our map)\end{sideways}}
ticks in long numbered from 0 to 10 by 2
      short unlabeled from 0 to 10 by 1 /
\axis right label {}
ticks in long unlabeled from 0 to 10 by 2
      short unlabeled from 0 to 10 by 1 /
\axis bottom label {$A_V$\,[mag] Dobashi et al. \cite{2005PASJ...57S...1D}}
ticks in long numbered from 0 to 5 by 1
      short unlabeled from 0 to 5 by 0.5 /
\axis top label {}
ticks in long unlabeled from 0 to 5 by 1
      short unlabeled from 0 to 5 by 0.5 /
\input dob_our_region3.txt
\plot 0 0 5 5 /
% {0.820032, 1.06074},
\plot 0 0.820032 5 6.1237 /
\endpicture  
\caption{\label{comp_dob_our_reg3} Comparison of the optical extinction values
for region3 (Lupus) in Table\,\ref{sigmatab} between our extinction map and the
map from D05. As solid line we overplot a one-to-one
line, as well as a linear fit. See Table\,\ref{sigmatab} for the parameters of
the fit.}   
\end{figure}

\begin{figure}
\beginpicture
\setcoordinatesystem units <17.5mm,10mm> point at  0 0
\setplotarea x from -0.5 to 3.5 , y from -0.5 to 6.5
\axis left label {\begin{sideways}$A_V$\,[mag] (our map)\end{sideways}}
ticks in long numbered from 0 to 6 by 2
      short unlabeled from 0 to 6 by 1 /
\axis right label {}
ticks in long unlabeled from 0 to 6 by 2
      short unlabeled from 0 to 6 by 1 /
\axis bottom label {$A_V$\,[mag] Dobashi et al. \cite{2005PASJ...57S...1D}}
ticks in long numbered from 0 to 3 by 1
      short unlabeled from 0 to 3 by 0.5 /
\axis top label {}
ticks in long unlabeled from 0 to 3 by 1
      short unlabeled from 0 to 3 by 0.5 /
\input dob_our_region4.txt
\plot -0.5 -0.5 3.5 3.5 /
%{0.224437,  1.40877},
\plot -0.5 -0.479948 3.5 5.155132 /
\endpicture  
\caption{\label{comp_dob_our_reg4} Comparison of the optical extinction values
for region4 (Vela) in Table\,\ref{sigmatab} between our extinction map and the
map from D05. As solid line we overplot a one-to-one
line, as well as a linear fit. See Table\,\ref{sigmatab} for the parameters of
the fit.}   
\end{figure}

\begin{figure}
\beginpicture
\setcoordinatesystem units <14.0mm,7mm> point at  0 0
\setplotarea x from 0 to 5 , y from -0.5 to 9.5
\axis left label {\begin{sideways}$A_V$\,[mag] (our map)\end{sideways}}
ticks in long numbered from 0 to 9 by 2
      short unlabeled from 0 to 9 by 1 /
\axis right label {}
ticks in long unlabeled from 0 to 9 by 2
      short unlabeled from 0 to 9 by 1 /
\axis bottom label {$A_V$\,[mag] Dobashi et al. \cite{2005PASJ...57S...1D}}
ticks in long numbered from 0 to 5 by 1
      short unlabeled from 0 to 5 by 0.5 /
\axis top label {}
ticks in long unlabeled from 0 to 5 by 1
      short unlabeled from 0 to 5 by 0.5 /
\input dob_our_region5.txt
\plot 0 0 5 5 /
% {0.193034, 1.12891},
\plot 0 0.193034 5 5.83755 /
\endpicture  
\caption{\label{comp_dob_our_reg5} Comparison of the optical extinction values
for region5 (Taurus) in Table\,\ref{sigmatab} between our extinction map and the
map from D05. As solid line we overplot a one-to-one
line, as well as a linear fit. See Table\,\ref{sigmatab} for the parameters of
the fit.}   
\end{figure}

\begin{figure}
\beginpicture
\setcoordinatesystem units <10.77mm,6.36mm> point at  0 0
\setplotarea x from 0 to 6.5 , y from -0.5 to 10.5
\axis left label {\begin{sideways}$A_V$\,[mag] (our map)\end{sideways}}
ticks in long numbered from 0 to 10 by 2
      short unlabeled from 0 to 10 by 1 /
\axis right label {}
ticks in long unlabeled from 0 to 10 by 2
      short unlabeled from 0 to 10 by 1 /
\axis bottom label {$A_V$\,[mag] Dobashi et al. \cite{2005PASJ...57S...1D}}
ticks in long numbered from 0 to 6 by 2
      short unlabeled from 0 to 6 by 1 /
\axis top label {}
ticks in long unlabeled from 0 to 6 by 2
      short unlabeled from 0 to 6 by 1 /
\input dob_our_region6.txt
\plot 0 0 6.5 6.5 /
%{0.193361, 1.34346},
\plot 0 0.193361 6.5 8.925851 /
\endpicture  
\caption{\label{comp_dob_our_reg6} Comparison of the optical extinction values
for region6 (North America Nebula) in Table\,\ref{sigmatab} between our
extinction map and the map from D05. As solid line we overplot a one-to-one
line, as well as a linear fit. See Table\,\ref{sigmatab} for the parameters of
the fit.}   
\end{figure}

\begin{figure}
\beginpicture
\setcoordinatesystem units <17.5mm,11.67mm> point at  0 0
\setplotarea x from -0.5 to 3.5 , y from -0.62 to 5.38
\axis left label {\begin{sideways}$A_V$\,[mag] (our map)\end{sideways}}
ticks in long numbered from 0 to 5 by 1
      short unlabeled from 0 to 5 by 0.5 /
\axis right label {}
ticks in long unlabeled from 0 to 5 by 1
      short unlabeled from 0 to 5 by 0.5 /
\axis bottom label {$A_V$\,[mag] Dobashi et al. \cite{2005PASJ...57S...1D}}
ticks in long numbered from 0 to 3 by 1
      short unlabeled from 0 to 3 by 0.5 /
\axis top label {}
ticks in long unlabeled from 0 to 3 by 1
      short unlabeled from 0 to 3 by 0.5 /
\input dob_our_region7.txt
\plot -0.5 -0.5 3.5 3.5 /
%{-0.0145357, 1.22787},
\plot -0.5 -0.6284707 3.5 4.2830093 /
\endpicture  
\caption{\label{comp_dob_our_reg7} Comparison of the optical extinction values
for region7 (Monoceros) in Table\,\ref{sigmatab} between our extinction map and
the map from D05. As solid line we overplot a one-to-one
line, as well as a linear fit. See Table\,\ref{sigmatab} for the parameters of
the fit.}   
\end{figure}

\begin{figure}
\beginpicture
\setcoordinatesystem units <17.5mm,11.67mm> point at  0 0
\setplotarea x from -0.5 to 3.5 , y from -0.5 to 5.5
\axis left label {\begin{sideways}$A_V$\,[mag] (our map)\end{sideways}}
ticks in long numbered from 0 to 5 by 1
      short unlabeled from 0 to 5 by 0.5 /
\axis right label {}
ticks in long unlabeled from 0 to 5 by 1
      short unlabeled from 0 to 5 by 0.5 /
\axis bottom label {$A_V$\,[mag] Dobashi et al. \cite{2005PASJ...57S...1D}}
ticks in long numbered from 0 to 3 by 1
      short unlabeled from 0 to 3 by 0.5 /
\axis top label {}
ticks in long unlabeled from 0 to 3 by 1
      short unlabeled from 0 to 3 by 0.5 /
\input dob_our_region8.txt
\plot -0.5 -0.5 3.5 3.5 /
% {0.110873, 1.22741}}
\plot -0.5 -0.502832 3.5 4.406808 /
\endpicture  
\caption{\label{comp_dob_our_reg8} Comparison of the optical extinction values
for region8 (Auriga) in Table\,\ref{sigmatab} between our extinction map and the
map from D05. As solid line we overplot a one-to-one
line, as well as a linear fit. See Table\,\ref{sigmatab} for the parameters of
the fit.}   
\end{figure}

\clearpage
\newpage

\subsection{Our Map vs. Schlegel et al. (1998)}\label{ours_vs_sch}

\begin{figure}
\beginpicture
\setcoordinatesystem units <8.75mm,11.67mm> point at  0 0
\setplotarea x from 0 to 8 , y from -0.52 to 5.48
\axis left label {\begin{sideways}$A_V$\,[mag] (our map)\end{sideways}}
ticks in long numbered from 0 to 5 by 1
      short unlabeled from 0 to 5 by 0.5 /
\axis right label {}
ticks in long unlabeled from 0 to 5 by 1
      short unlabeled from 0 to 5 by 0.5 /
\axis bottom label {$A_V$\,[mag] Schlegel et al. \cite{1998ApJ...500..525S}}
ticks in long numbered from 0 to 8 by 2
      short unlabeled from 0 to 8 by 1 /
\axis top label {}
ticks in long unlabeled from 0 to 8 by 2
      short unlabeled from 0 to 8 by 1 /
\input sch_our_region1.txt
\plot 0 0 5.5 5.5 /
%-0.518412 0.520406
\plot 0 -0.518412 8 3.644848 /
\endpicture  
\caption{\label{comp_sch_our_reg1} Comparison of the optical extinction values
for region1 (Camelopardalis) in Table\,\ref{sigmatab} between our extinction map
and the map from S98. As solid line we overplot a one-to-one
line, as well as a linear fit. See Table\,\ref{sigmatab} for the parameters of
the fit.}   
\end{figure}

\begin{figure}
\beginpicture
\setcoordinatesystem units <4.375mm,4.375mm> point at  0 0
\setplotarea x from 0 to 16 , y from 0 to 16
\axis left label {\begin{sideways}$A_V$\,[mag] (our map)\end{sideways}}
ticks in long numbered from 0 to 16 by 4
      short unlabeled from 0 to 16 by 1 /
\axis right label {}
ticks in long unlabeled from 0 to 16 by 4
      short unlabeled from 0 to 16 by 1 /
\axis bottom label {$A_V$\,[mag] Schlegel et al. \cite{1998ApJ...500..525S}}
ticks in long numbered from 0 to 16 by 4
      short unlabeled from 0 to 16 by 1 /
\axis top label {}
ticks in long unlabeled from 0 to 16 by 4
      short unlabeled from 0 to 16 by 1 /
\input sch_our_region2.txt
\plot 0 0 16 16 /
%-1.700000 1.000000#
\plot 1.7 0 16 14.3 /
\endpicture  
\caption{\label{comp_sch_our_reg2} Comparison of the optical extinction values
for region2 (Serpens) in Table\,\ref{sigmatab} between our extinction map and
the map from S98. As solid line we overplot a one-to-one
line, as well as a linear fit. See Table\,\ref{sigmatab} for the parameters of
the fit.}   
\end{figure}

\begin{figure}
\beginpicture
\setcoordinatesystem units <4.67mm,7mm> point at  0 0
\setplotarea x from 0 to 15 , y from -0.5 to 9.5
\axis left label {\begin{sideways}$A_V$\,[mag] (our map)\end{sideways}}
ticks in long numbered from 0 to 9 by 2
      short unlabeled from 0 to 9 by 1 /
\axis right label {}
ticks in long unlabeled from 0 to 9 by 2
      short unlabeled from 0 to 9 by 1 /
\axis bottom label {$A_V$\,[mag] Schlegel et al. \cite{1998ApJ...500..525S}}
ticks in long numbered from 0 to 15 by 2
      short unlabeled from 0 to 15 by 1 /
\axis top label {}
ticks in long unlabeled from 0 to 15 by 2
      short unlabeled from 0 to 15 by 1 /
\input sch_our_region3.txt
\plot 0 0 9.5 9.5 /
%-0.339223 0.725608
\plot 0 -0.339223 13.56 9.5 /
\endpicture  
\caption{\label{comp_sch_our_reg3} Comparison of the optical extinction values
for region3 (Lupus) in Table\,\ref{sigmatab} between our extinction map and the
map from S98. As solid line we overplot a one-to-one
line, as well as a linear fit. See Table\,\ref{sigmatab} for the parameters of
the fit.}   
\end{figure}

\begin{figure}
\beginpicture
\setcoordinatesystem units <7mm,10mm> point at  0 0
\setplotarea x from 0 to 10 , y from -0.5 to 6.5
\axis left label {\begin{sideways}$A_V$\,[mag] (our map)\end{sideways}}
ticks in long numbered from 0 to 6 by 2
      short unlabeled from 0 to 6 by 1 /
\axis right label {}
ticks in long unlabeled from 0 to 6 by 2
      short unlabeled from 0 to 6 by 1 /
\axis bottom label {$A_V$\,[mag] Schlegel et al. \cite{1998ApJ...500..525S}}
ticks in long numbered from 0 to 10 by 2
      short unlabeled from 0 to 10 by 1 /
\axis top label {}
ticks in long unlabeled from 0 to 10 by 2
      short unlabeled from 0 to 10 by 1 /
\input sch_our_region4.txt
\plot 0 0 6.5 6.5 /
%-0.856302 0.700092
\plot 0.5089 -0.5 10 6.14462 /
\endpicture  
\caption{\label{comp_sch_our_reg4} Comparison of the optical extinction values
for region4 (Vela) in Table\,\ref{sigmatab} between our extinction map and the
map from S98. As solid line we overplot a one-to-one
line, as well as a linear fit. See Table\,\ref{sigmatab} for the parameters of
the fit.}   
\end{figure}

\begin{figure}
\beginpicture
\setcoordinatesystem units <11.67mm,10mm> point at  0 0
\setplotarea x from 0 to 6 , y from -0.5 to 6.5
\axis left label {\begin{sideways}$A_V$\,[mag] (our map)\end{sideways}}
ticks in long numbered from 0 to 6 by 2
      short unlabeled from 0 to 6 by 1 /
\axis right label {}
ticks in long unlabeled from 0 to 6 by 2
      short unlabeled from 0 to 6 by 1 /
\axis bottom label {$A_V$\,[mag] Schlegel et al. \cite{1998ApJ...500..525S}}
ticks in long numbered from 0 to 6 by 1
      short unlabeled from 0 to 6 by 0.5 /
\axis top label {}
ticks in long unlabeled from 0 to 6 by 1
      short unlabeled from 0 to 6 by 0.5 /
\input sch_our_region5.txt
\plot 0 0 6 6 /
%-0.309924 0.621611
\plot 0 -0.309924 6 3.419766 /
\endpicture  
\caption{\label{comp_sch_our_reg5} Comparison of the optical extinction values
for region5 (Taurus) in Table\,\ref{sigmatab} between our extinction map and the
map from S98. As solid line we overplot a one-to-one
line, as well as a linear fit. See Table\,\ref{sigmatab} for the parameters of
the fit.}   
\end{figure}

\begin{figure}
\beginpicture
\setcoordinatesystem units <4.67mm,6.36mm> point at  0 0
\setplotarea x from 0 to 15 , y from -0.5 to 10.5
\axis left label {\begin{sideways}$A_V$\,[mag] (our map)\end{sideways}}
ticks in long numbered from 0 to 10 by 2
      short unlabeled from 0 to 10 by 1 /
\axis right label {}
ticks in long unlabeled from 0 to 10 by 2
      short unlabeled from 0 to 10 by 1 /
\axis bottom label {$A_V$\,[mag] Schlegel et al. \cite{1998ApJ...500..525S}}
ticks in long numbered from 0 to 15 by 2
      short unlabeled from 0 to 15 by 1 /
\axis top label {}
ticks in long unlabeled from 0 to 15 by 2
      short unlabeled from 0 to 15 by 1 /
\input sch_our_region6.txt
\plot 0 0 10.5 10.5 /
%-0.738508 0.734131
\plot 0.324887617 -0.5 15 10.272 /
\endpicture  
\caption{\label{comp_sch_our_reg6} Comparison of the optical extinction values
for region6 (North America Nebula) in Table\,\ref{sigmatab} between our
extinction map and the map from S98. As solid line we overplot a one-to-one
line, as well as a linear fit. See Table\,\ref{sigmatab} for the parameters of
the fit.}   
\end{figure}

\begin{figure}
\beginpicture
\setcoordinatesystem units <7mm,11.67mm> point at  0 0
\setplotarea x from 0 to 10 , y from -0.5 to 5.5
\axis left label {\begin{sideways}$A_V$\,[mag] (our map)\end{sideways}}
ticks in long numbered from 0 to 5 by 1
      short unlabeled from 0 to 5 by 0.5 /
\axis right label {}
ticks in long unlabeled from 0 to 5 by 1
      short unlabeled from 0 to 5 by 0.5 /
\axis bottom label {$A_V$\,[mag] Schlegel et al. \cite{1998ApJ...500..525S}}
ticks in long numbered from 0 to 10 by 2
      short unlabeled from 0 to 10 by 1 /
\axis top label {}
ticks in long unlabeled from 0 to 10 by 2
      short unlabeled from 0 to 10 by 1 /
\input sch_our_region7.txt
\plot 0 0 5.5 5.5 /
%-0.484028 0.577421
\plot 0 -0.484028 10 5.29021 /
\endpicture  
\caption{\label{comp_sch_our_reg7} Comparison of the optical extinction values
for region7 (Monoceros) in Table\,\ref{sigmatab} between our extinction map and
the map from S98. As solid line we overplot a one-to-one
line, as well as a linear fit. See Table\,\ref{sigmatab} for the parameters of
the fit.}   
\end{figure}

\begin{figure}
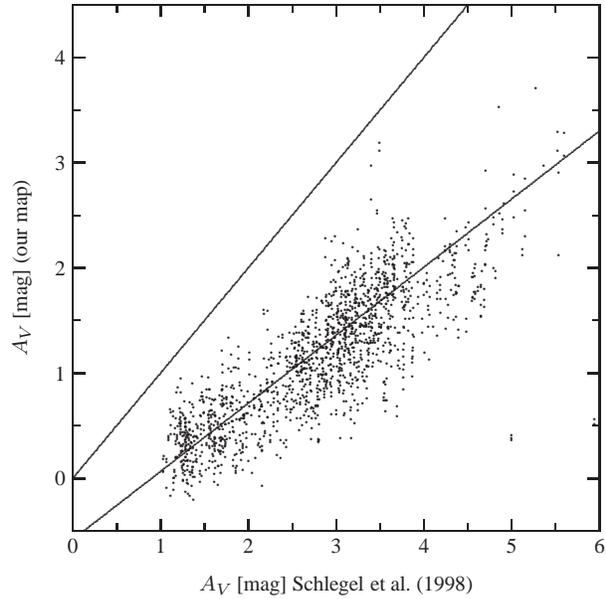

\beginpicture
\setcoordinatesystem units <11.67mm,14mm> point at  0 0
\setplotarea x from 0 to 6 , y from -0.5 to 4.5
\axis left label {\begin{sideways}$A_V$\,[mag] (our map)\end{sideways}}
ticks in long numbered from 0 to 4 by 1
      short unlabeled from 0 to 4 by 0.5 /
\axis right label {}
ticks in long unlabeled from 0 to 4 by 1
      short unlabeled from 0 to 4 by 0.5 /
\axis bottom label {$A_V$\,[mag] Schlegel et al. \cite{1998ApJ...500..525S}}
ticks in long numbered from 0 to 6 by 1
      short unlabeled from 0 to 6 by 0.5 /
\axis top label {}
ticks in long unlabeled from 0 to 6 by 1
      short unlabeled from 0 to 6 by 0.5 /
\input sch_our_region8.txt
\plot 0 0 4.5 4.5 /
%-0.581162 0.647421
\plot 0.125440495 -0.5 6 3.301 /
\endpicture  
\caption{\label{comp_sch_our_reg8} Comparison of the optical extinction values
for region8 (Auriga) in Table\,\ref{sigmatab} between our extinction map and the
map from S98. As solid line we overplot a one-to-one
line, as well as a linear fit. See Table\,\ref{sigmatab} for the parameters of
the fit.}   
\end{figure}

\clearpage
\newpage

\subsection{Dobashi et al. (2005) vs. Schlegel et al. (1998)}\label{sch_vs_dob}

\begin{figure}
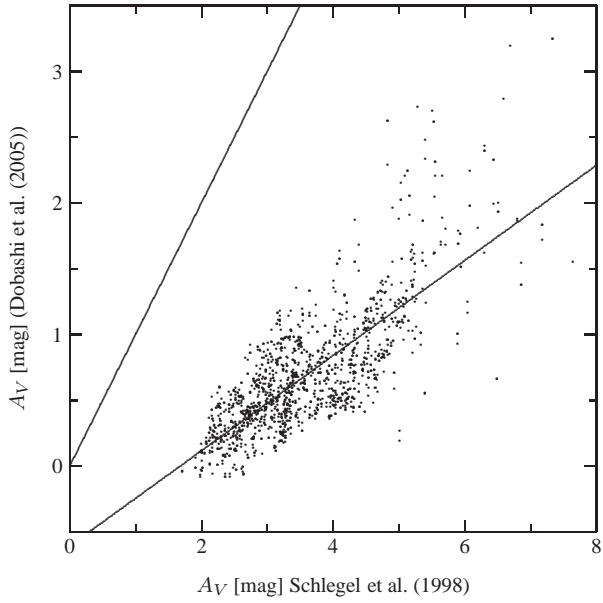

\beginpicture
\setcoordinatesystem units <8.75mm,17.5mm> point at  0 0
\setplotarea x from 0 to 8 , y from -0.5 to 3.5
\axis left label {\begin{sideways}$A_V$\,[mag] (Dobashi et al. (2005))\end{sideways}}
ticks in long numbered from 0 to 3 by 1
      short unlabeled from 0 to 3 by 0.5 /
\axis right label {}
ticks in long unlabeled from 0 to 3 by 1
      short unlabeled from 0 to 3 by 0.5 /
\axis bottom label {$A_V$\,[mag] Schlegel et al. \cite{1998ApJ...500..525S}}
ticks in long numbered from 0 to 8 by 2
      short unlabeled from 0 to 8 by 1 /
\axis top label {}
ticks in long unlabeled from 0 to 8 by 2
      short unlabeled from 0 to 8 by 1 /
\input dob_sch_region1.txt
\plot 0 0 3.5 3.5 /
%-0.606836 0.361866
\plot 0.295136874 -0.5 8 2.286568 /
\endpicture  
\caption{\label{comp_dob_sch_reg1} Comparison of the optical extinction values
for region1 (Camelopardalis) in Table\,\ref{sigmatab} between the map of D05
and the map from S98. As solid line we overplot a one-to-one
line, as well as a linear fit. See Table\,\ref{sigmatab} for the parameters of
the fit.}   
\end{figure}

\begin{figure}
\beginpicture
\setcoordinatesystem units <5mm,8.75mm> point at  0 0
\setplotarea x from 0 to 14 , y from 0 to 8
\axis left label {\begin{sideways}$A_V$\,[mag] (Dobashi et al. (2005))\end{sideways}}
ticks in long numbered from 0 to 8 by 2
      short unlabeled from 0 to 8 by 1 /
\axis right label {}
ticks in long unlabeled from 0 to 8 by 2
      short unlabeled from 0 to 8 by 1 /
\axis bottom label {$A_V$\,[mag] Schlegel et al. \cite{1998ApJ...500..525S}}
ticks in long numbered from 0 to 14 by 4
      short unlabeled from 0 to 14 by 1 /
\axis top label {}
ticks in long unlabeled from 0 to 14 by 4
      short unlabeled from 0 to 14 by 1 /
\input dob_sch_region2.txt
\plot 0 0 8 8 /
%-1.400000 0.800000#
\plot 1.75 0 11.75 8  /
\endpicture  
\caption{\label{comp_dob_sch_reg2} Comparison of the optical extinction values
for region2 (Serpens) in Table\,\ref{sigmatab} between the map of D05 and the
map from S98. As solid line we overplot a one-to-one line,
as well as a linear fit. See Table\,\ref{sigmatab} for the parameters of the
fit.} 
\end{figure}

\begin{figure}
\beginpicture
\setcoordinatesystem units <7mm,14mm> point at  0 0
\setplotarea x from 0 to 10 , y from 0 to 5
\axis left label {\begin{sideways}$A_V$\,[mag] (Dobashi et al. (2005))\end{sideways}}
ticks in long numbered from 0 to 5 by 1
      short unlabeled from 0 to 5 by 0.5 /
\axis right label {}
ticks in long unlabeled from 0 to 5 by 1
      short unlabeled from 0 to 5 by 0.5 /
\axis bottom label {$A_V$\,[mag] Schlegel et al. \cite{1998ApJ...500..525S}}
ticks in long numbered from 0 to 10 by 2
      short unlabeled from 0 to 10 by 1 /
\axis top label {}
ticks in long unlabeled from 0 to 10 by 2
      short unlabeled from 0 to 10 by 1 /
\input dob_sch_region3.txt
\plot 0 0 5 5 /
%-0.800000 0.600000#
\plot 1.33 0 9.67 5 /
\endpicture  
\caption{\label{comp_dob_sch_reg3} Comparison of the optical extinction values
for region3 (Lupus) in Table\,\ref{sigmatab} between the map of D05 and the map
from S98. As solid line we overplot a one-to-one line, as
well as a linear fit. See Table\,\ref{sigmatab} for the parameters of the fit.}   
\end{figure}

\begin{figure}
\beginpicture
\setcoordinatesystem units <7mm,17.5mm> point at  0 0
\setplotarea x from 0 to 10 , y from -0.5 to 3.5
\axis left label {\begin{sideways}$A_V$\,[mag] (Dobashi et al. (2005))\end{sideways}}
ticks in long numbered from 0 to 3 by 1
      short unlabeled from 0 to 3 by 0.5 /
\axis right label {}
ticks in long unlabeled from 0 to 3 by 1
      short unlabeled from 0 to 3 by 0.5 /
\axis bottom label {$A_V$\,[mag] Schlegel et al. \cite{1998ApJ...500..525S}}
ticks in long numbered from 0 to 10 by 2
      short unlabeled from 0 to 10 by 1 /
\axis top label {}
ticks in long unlabeled from 0 to 10 by 2
      short unlabeled from 0 to 10 by 1 /
\input dob_sch_region4.txt
\plot 0 0 3.5 3.5 /
%-0.481181 0.372409
\plot 0 -0.481181 10 3.2429 /
\endpicture  
\caption{\label{comp_dob_sch_reg4} Comparison of the optical extinction values
for region4 (Vela) in Table\,\ref{sigmatab} between the map of D05 and the map
from S98. As solid line we overplot a one-to-one line, as
well as a linear fit. See Table\,\ref{sigmatab} for the parameters of the fit.}   
\end{figure}

\begin{figure}
\beginpicture
\setcoordinatesystem units <11.67mm,14mm> point at  0 0
\setplotarea x from 0 to 6 , y from -0.5 to 4.5
\axis left label {\begin{sideways}$A_V$\,[mag] (Dobashi et al. (2005))\end{sideways}}
ticks in long numbered from 0 to 4 by 1
      short unlabeled from 0 to 4 by 0.5 /
\axis right label {}
ticks in long unlabeled from 0 to 4 by 1
      short unlabeled from 0 to 4 by 0.5 /
\axis bottom label {$A_V$\,[mag] Schlegel et al. \cite{1998ApJ...500..525S}}
ticks in long numbered from 0 to 6 by 1
      short unlabeled from 0 to 6 by 0.5 /
\axis top label {}
ticks in long unlabeled from 0 to 6 by 1
      short unlabeled from 0 to 6 by 0.5 /
\input dob_sch_region5.txt
\plot 0 0 4.5 4.5 /
%-0.537058 0.560896
\plot 0.065965883 -0.5 6 2.828376 /
\endpicture  
\caption{\label{comp_dob_sch_reg5} Comparison of the optical extinction values
for region5 (Taurus) in Table\,\ref{sigmatab} between the map of D05 and the
map from S98. As solid line we overplot a one-to-one line,
as well as a linear fit. See Table\,\ref{sigmatab} for the parameters of the
fit.} 
\end{figure}

\begin{figure}
\beginpicture
\setcoordinatesystem units <5.83mm,10mm> point at  0 0
\setplotarea x from 0 to 12 , y from -0.5 to 6.5
\axis left label {\begin{sideways}$A_V$\,[mag] (Dobashi et al. (2005))\end{sideways}}
ticks in long numbered from 0 to 6 by 2
      short unlabeled from 0 to 6 by 1 /
\axis right label {}
ticks in long unlabeled from 0 to 6 by 2
      short unlabeled from 0 to 6 by 1 /
\axis bottom label {$A_V$\,[mag] Schlegel et al. \cite{1998ApJ...500..525S}}
ticks in long numbered from 0 to 12 by 2
      short unlabeled from 0 to 12 by 1 /
\axis top label {}
ticks in long unlabeled from 0 to 12 by 2
      short unlabeled from 0 to 12 by 1 /
\input dob_sch_region6.txt
\plot 0 0 6.5 6.5 /
%-0.418972 0.461069
\plot 0 -0.418972 12 5.1131 /
\endpicture  
\caption{\label{comp_dob_sch_reg6} Comparison of the optical extinction values
for region6 (North America Nebula) in Table\,\ref{sigmatab} between our
extinction map and the map from S98. As solid line we overplot a one-to-one
line, as well as a linear fit. See Table\,\ref{sigmatab} for the parameters of
the fit.}   
\end{figure}

\begin{figure}
\beginpicture
\setcoordinatesystem units <8.75mm,17.5mm> point at  0 0
\setplotarea x from 0 to 8 , y from -0.5 to 3.5
\axis left label {\begin{sideways}$A_V$\,[mag] (Dobashi et al. (2005))\end{sideways}}
ticks in long numbered from 0 to 3 by 1
      short unlabeled from 0 to 3 by 0.5 /
\axis right label {}
ticks in long unlabeled from 0 to 3 by 1
      short unlabeled from 0 to 3 by 0.5 /
\axis bottom label {$A_V$\,[mag] Schlegel et al. \cite{1998ApJ...500..525S}}
ticks in long numbered from 0 to 8 by 2
      short unlabeled from 0 to 8 by 1 /
\axis top label {}
ticks in long unlabeled from 0 to 8 by 2
      short unlabeled from 0 to 8 by 1 /
\input dob_sch_region7.txt
\plot 0 0 3.5 3.5 /
%-0.700000 0.550000#
\plot 0.363636364 -0.5 7.636363636 3.5 /
\endpicture  
\caption{\label{comp_dob_sch_reg7} Comparison of the optical extinction values
for region7 (Monoceros) in Table\,\ref{sigmatab} between the map of D05 and the
map from S98. As solid line we overplot a one-to-one line,
as well as a linear fit. See Table\,\ref{sigmatab} for the parameters of the
fit.}    
\end{figure}

\begin{figure}
\beginpicture
\setcoordinatesystem units <11.67mm,17.5mm> point at  0 0
\setplotarea x from 0 to 6 , y from -0.5 to 3.5
\axis left label {\begin{sideways}$A_V$\,[mag] (Dobashi et al. (2005))\end{sideways}}
ticks in long numbered from 0 to 3 by 1
      short unlabeled from 0 to 3 by 0.5 /
\axis right label {}
ticks in long unlabeled from 0 to 3 by 1
      short unlabeled from 0 to 3 by 0.5 /
\axis bottom label {$A_V$\,[mag] Schlegel et al. \cite{1998ApJ...500..525S}}
ticks in long numbered from 0 to 6 by 1
      short unlabeled from 0 to 6 by 0.5 /
\axis top label {}
ticks in long unlabeled from 0 to 6 by 1
      short unlabeled from 0 to 6 by 0.5 /
\input dob_sch_region8.txt
\plot 0 0 3.5 3.5 /
%-0.627058 0.462164
\plot 0.274796607 -0.5 6 2.145 /
\endpicture  
\caption{\label{comp_dob_sch_reg8} Comparison of the optical extinction values
for region8 (Auriga) in Table\,\ref{sigmatab} between the map of D05 and the
map from S98. As solid line we overplot a one-to-one line,
as well as a linear fit. See Table\,\ref{sigmatab} for the parameters of the
fit.}    
\end{figure}

\section{High resolution $A_V$ maps}
\label{highres}

\clearpage
\newpage
\begin{figure*}
\beginpicture
\setcoordinatesystem units <-5mm,5mm> point at  0 0
\setplotarea x from 31 to -1 , y from -1 to 40
\put {\includegraphics[width=42.5cm]{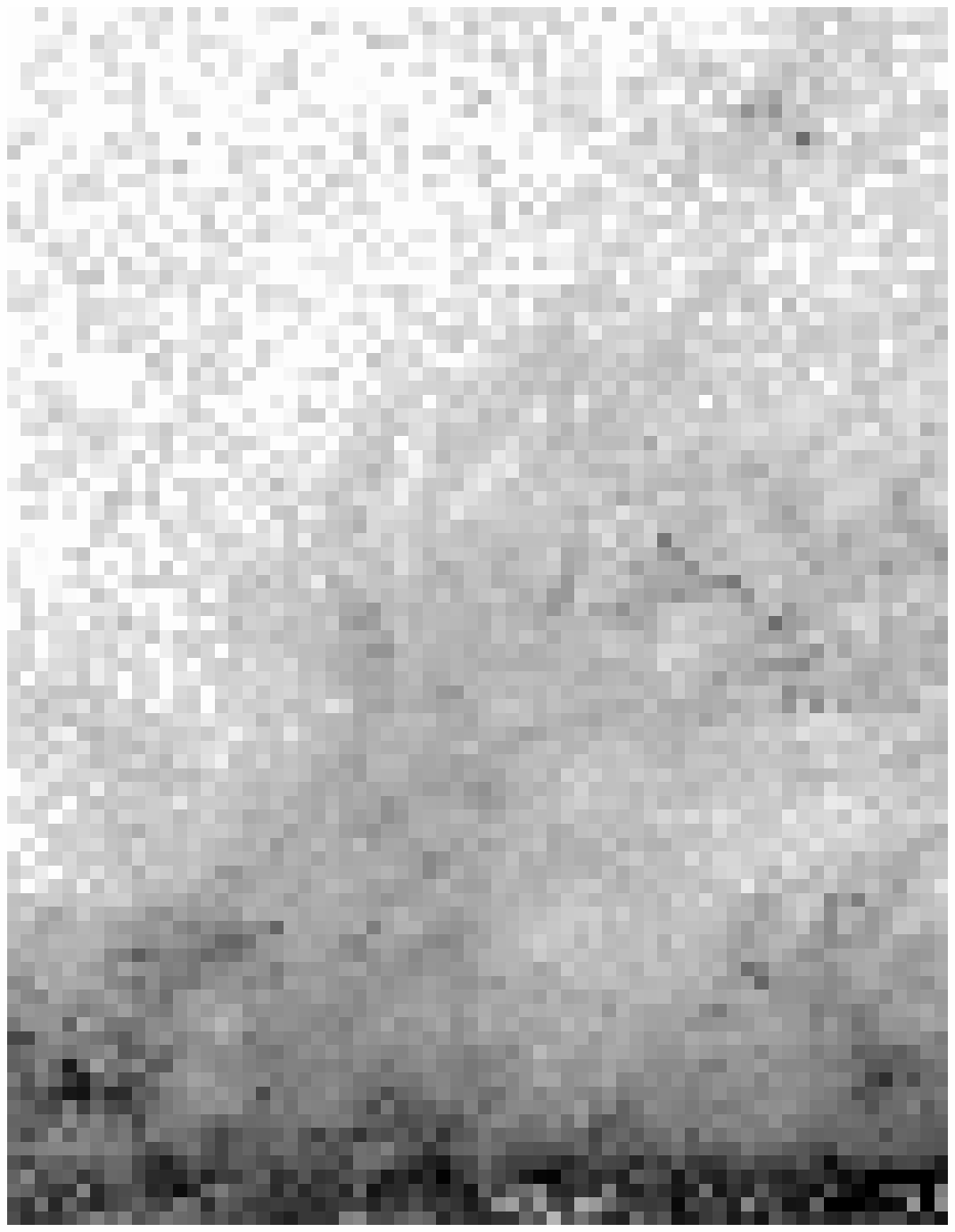}} at 15 19.5
\axis left label {\begin{sideways}b [\degr]\end{sideways}}
ticks in long numbered from 0 to 40 by 5
      short unlabeled from 0 to 40 by 1 /
\axis right label {}
ticks in long unlabeled from 0 to 40 by 5
      short unlabeled from 0 to 40 by 1 /
\axis bottom label {l [\degr]}
ticks in long numbered from 0 to 30 by 5
      short unlabeled from 0 to 30 by 1 /
\axis top label {}
ticks in long unlabeled from 0 to 30 by 5
      short unlabeled from 0 to 30 by 1 /
\endpicture  
\caption{\label{detail1} Grey scale representation of a detail of our
extinction map based on the nearest 49 stars. Extinction values are square root
scaled from 0\,mag (white) to 15\,mag (black) of optical extinction. }    
\end{figure*}

\clearpage
\newpage
\begin{figure*}
\beginpicture
\setcoordinatesystem units <-5mm,5mm> point at  0 0
\setplotarea x from 61 to 29 , y from -1 to 40
\put {\includegraphics[width=42.5cm]{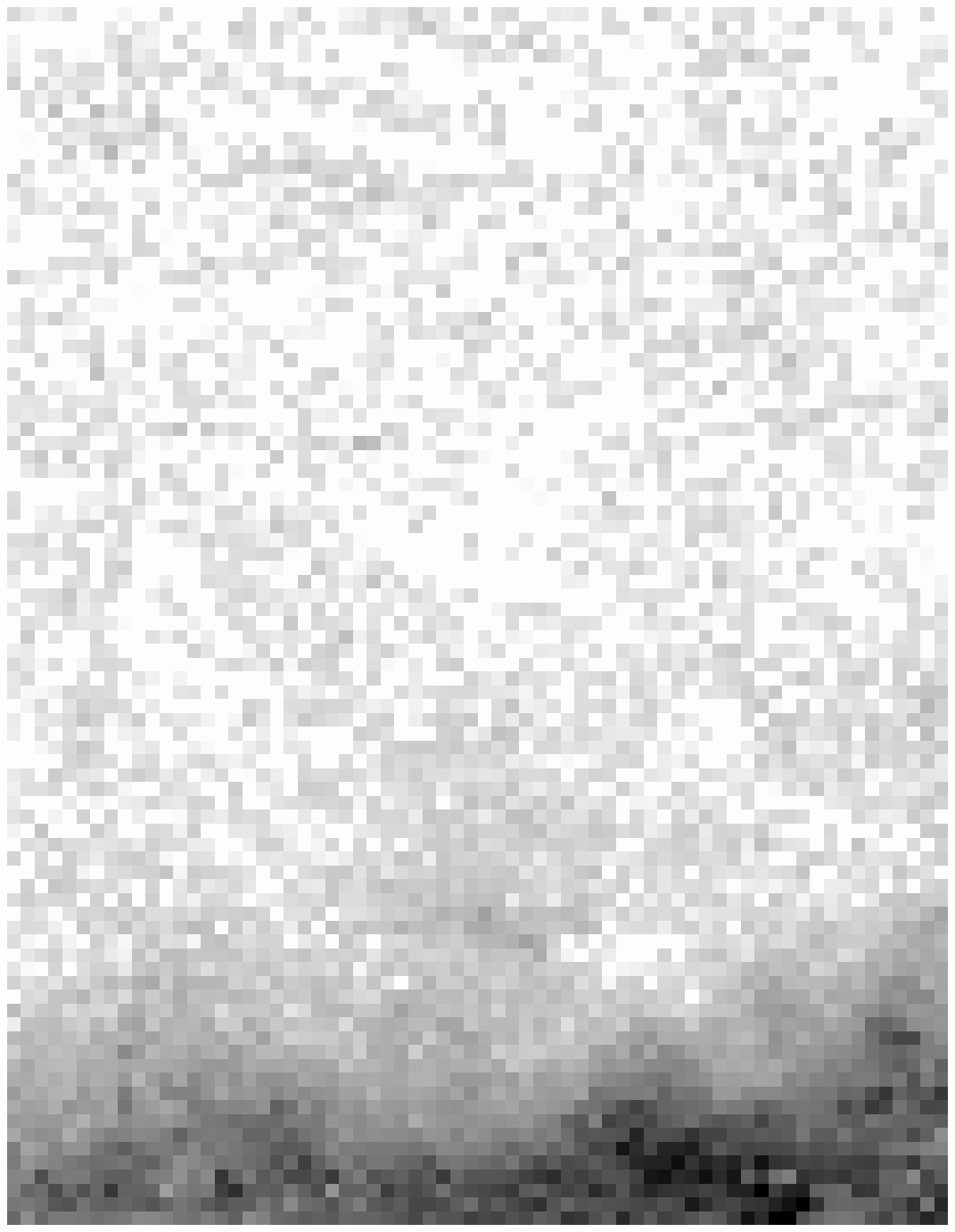}} at 45 19.5
\axis left label {\begin{sideways}b [\degr]\end{sideways}}
ticks in long numbered from 0 to 40 by 5
      short unlabeled from 0 to 40 by 1 /
\axis right label {}
ticks in long unlabeled from 0 to 40 by 5
      short unlabeled from 0 to 40 by 1 /
\axis bottom label {l [\degr]}
ticks in long numbered from 30 to 60 by 5
      short unlabeled from 30 to 60 by 1 /
\axis top label {}
ticks in long unlabeled from 30 to 60 by 5
      short unlabeled from 30 to 60 by 1 /
\endpicture  
\caption{\label{detail2} Grey scale representation of a detail of our
extinction map based on the nearest 49 stars. Extinction values are square root
scaled from 0\,mag (white) to 15\,mag (black) of optical extinction. }    
\end{figure*}

\clearpage
\newpage
\begin{figure*}
\beginpicture
\setcoordinatesystem units <-5mm,5mm> point at  0 0
\setplotarea x from 91 to 59 , y from -1 to 40
\put {\includegraphics[width=42.5cm]{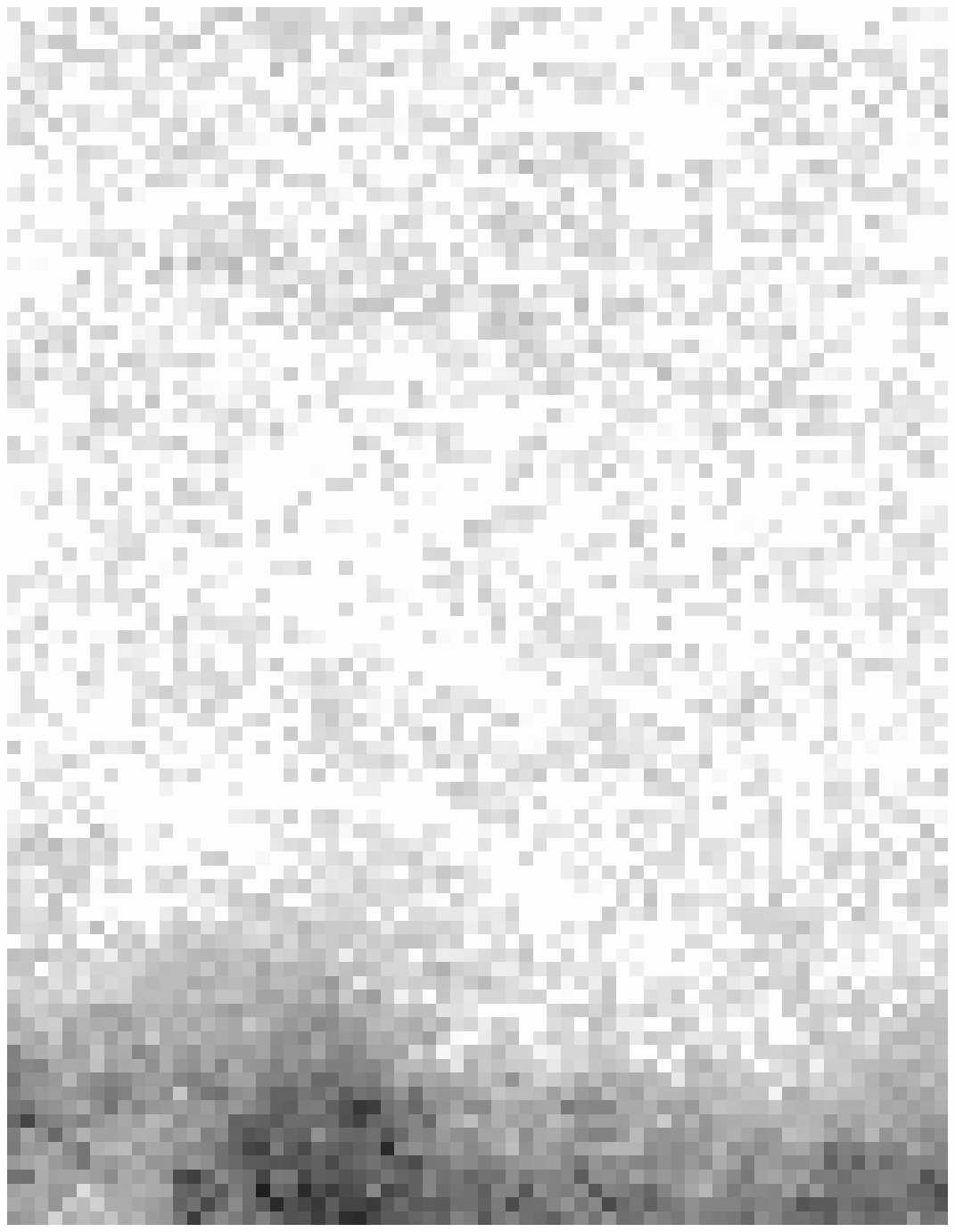}} at 75 19.5
\axis left label {\begin{sideways}b [\degr]\end{sideways}}
ticks in long numbered from 0 to 40 by 5
      short unlabeled from 0 to 40 by 1 /
\axis right label {}
ticks in long unlabeled from 0 to 40 by 5
      short unlabeled from 0 to 40 by 1 /
\axis bottom label {l [\degr]}
ticks in long numbered from 60 to 90 by 5
      short unlabeled from 60 to 90 by 1 /
\axis top label {}
ticks in long unlabeled from 60 to 90 by 5
      short unlabeled from 60 to 90 by 1 /
\endpicture  
\caption{\label{detail3} Grey scale representation of a detail of our
extinction map based on the nearest 49 stars. Extinction values are square root
scaled from 0\,mag (white) to 15\,mag (black) of optical extinction. }    
\end{figure*}

\clearpage
\newpage
\begin{figure*}
\beginpicture
\setcoordinatesystem units <-5mm,5mm> point at  0 0
\setplotarea x from 121 to 89 , y from -1 to 40
\put {\includegraphics[width=42.5cm]{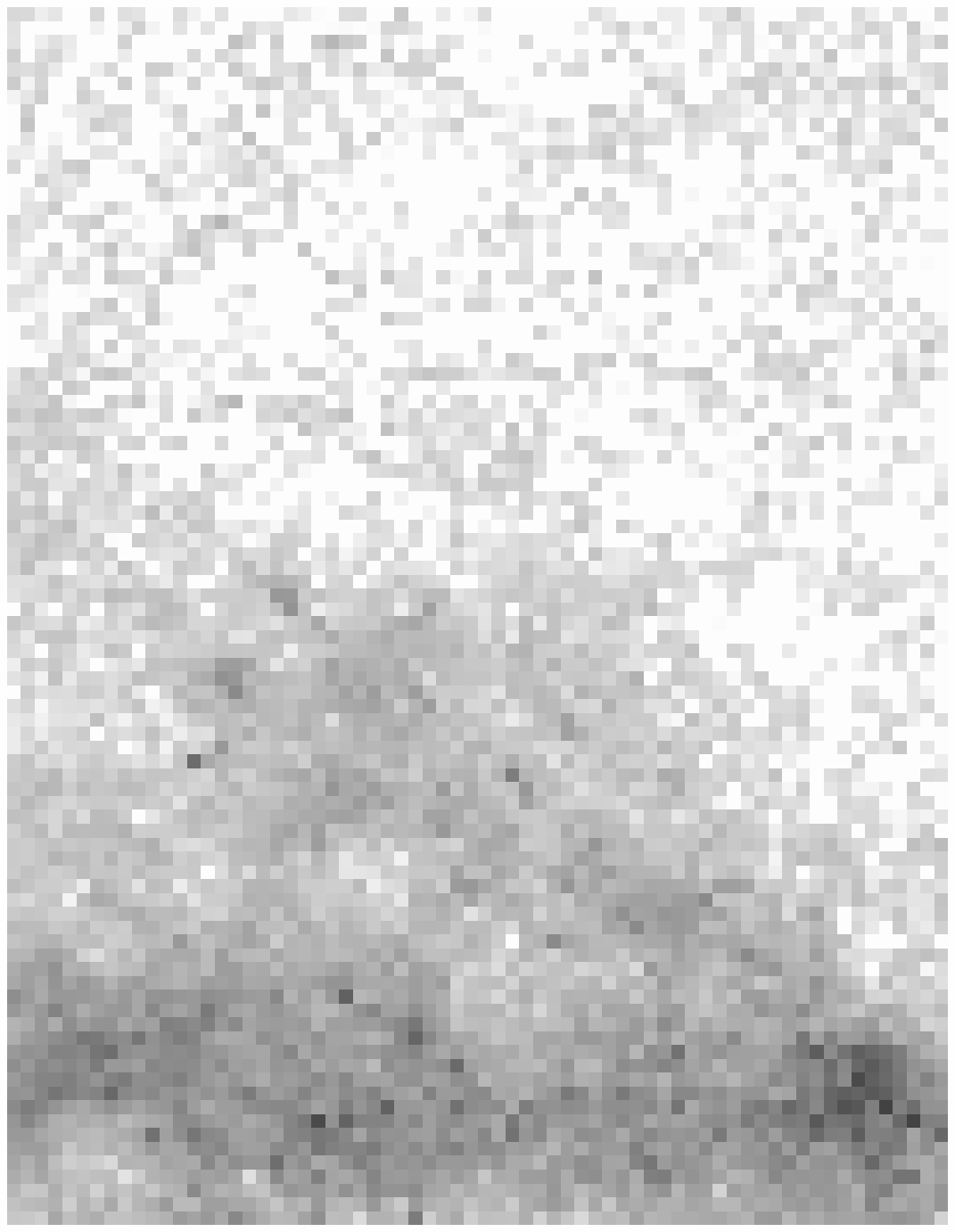}} at 105 19.5
\axis left label {\begin{sideways}b [\degr]\end{sideways}}
ticks in long numbered from 0 to 40 by 5
      short unlabeled from 0 to 40 by 1 /
\axis right label {}
ticks in long unlabeled from 0 to 40 by 5
      short unlabeled from 0 to 40 by 1 /
\axis bottom label {l [\degr]}
ticks in long numbered from 90 to 120 by 5
      short unlabeled from 90 to 120 by 1 /
\axis top label {}
ticks in long unlabeled from 90 to 120 by 5
      short unlabeled from 90 to 120 by 1 /
\endpicture  
\caption{\label{detail4} Grey scale representation of a detail of our
extinction map based on the nearest 49 stars. Extinction values are square root
scaled from 0\,mag (white) to 15\,mag (black) of optical extinction. }    
\end{figure*}

\clearpage
\newpage
\begin{figure*}
\beginpicture
\setcoordinatesystem units <-5mm,5mm> point at  0 0
\setplotarea x from 151 to 119 , y from -1 to 40
\put {\includegraphics[width=42.5cm]{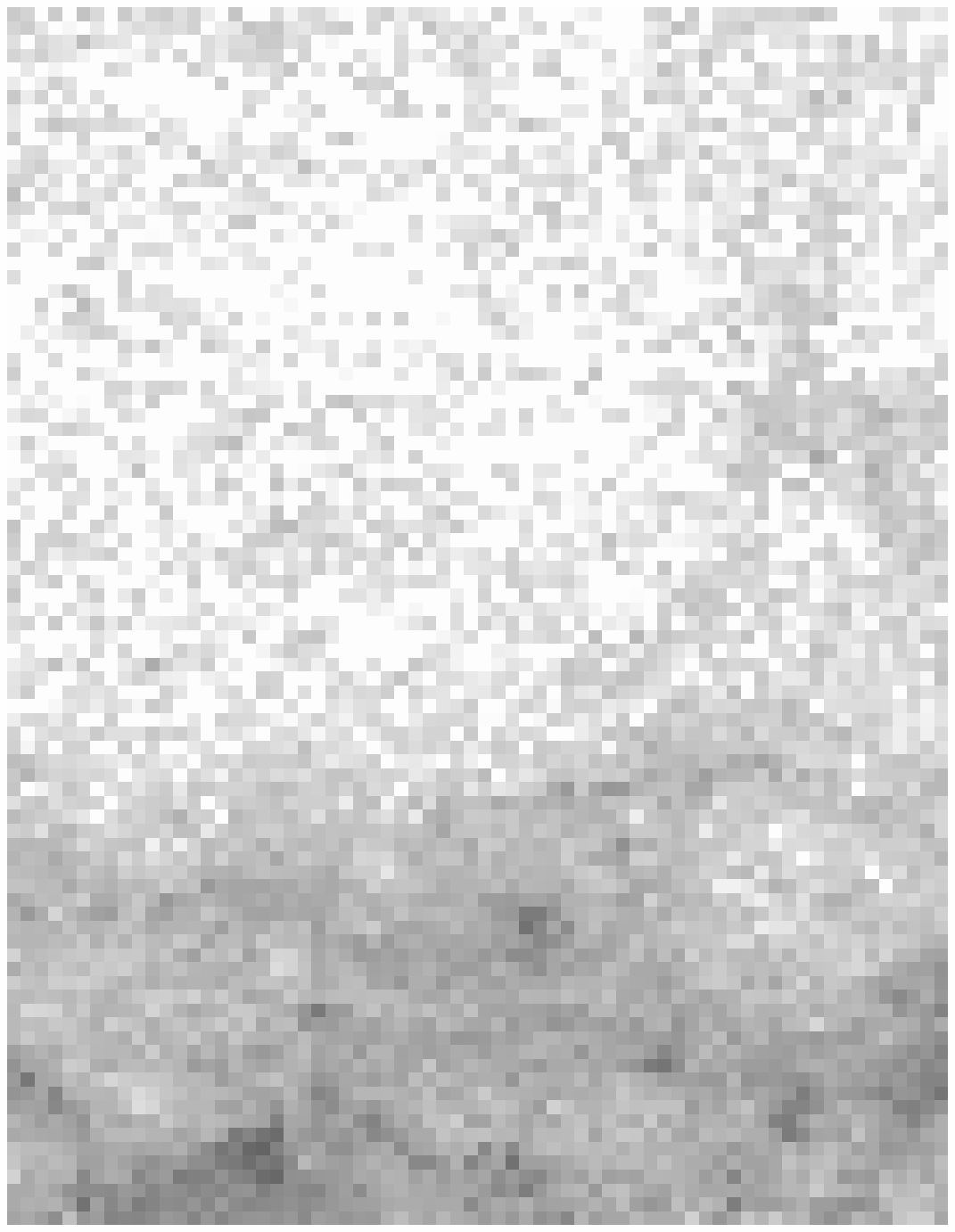}} at 135 19.5
\axis left label {\begin{sideways}b [\degr]\end{sideways}}
ticks in long numbered from 0 to 40 by 5
      short unlabeled from 0 to 40 by 1 /
\axis right label {}
ticks in long unlabeled from 0 to 40 by 5
      short unlabeled from 0 to 40 by 1 /
\axis bottom label {l [\degr]}
ticks in long numbered from 120 to 150 by 5
      short unlabeled from 120 to 150 by 1 /
\axis top label {}
ticks in long unlabeled from 120 to 150 by 5
      short unlabeled from 120 to 150 by 1 /
\endpicture  
\caption{\label{detail5} Grey scale representation of a detail of our
extinction map based on the nearest 49 stars. Extinction values are square root
scaled from 0\,mag (white) to 15\,mag (black) of optical extinction. }    
\end{figure*}

\clearpage
\newpage
\begin{figure*}
\beginpicture
\setcoordinatesystem units <-5mm,5mm> point at  0 0
\setplotarea x from 181 to 149 , y from -1 to 40
\put {\includegraphics[width=42.5cm]{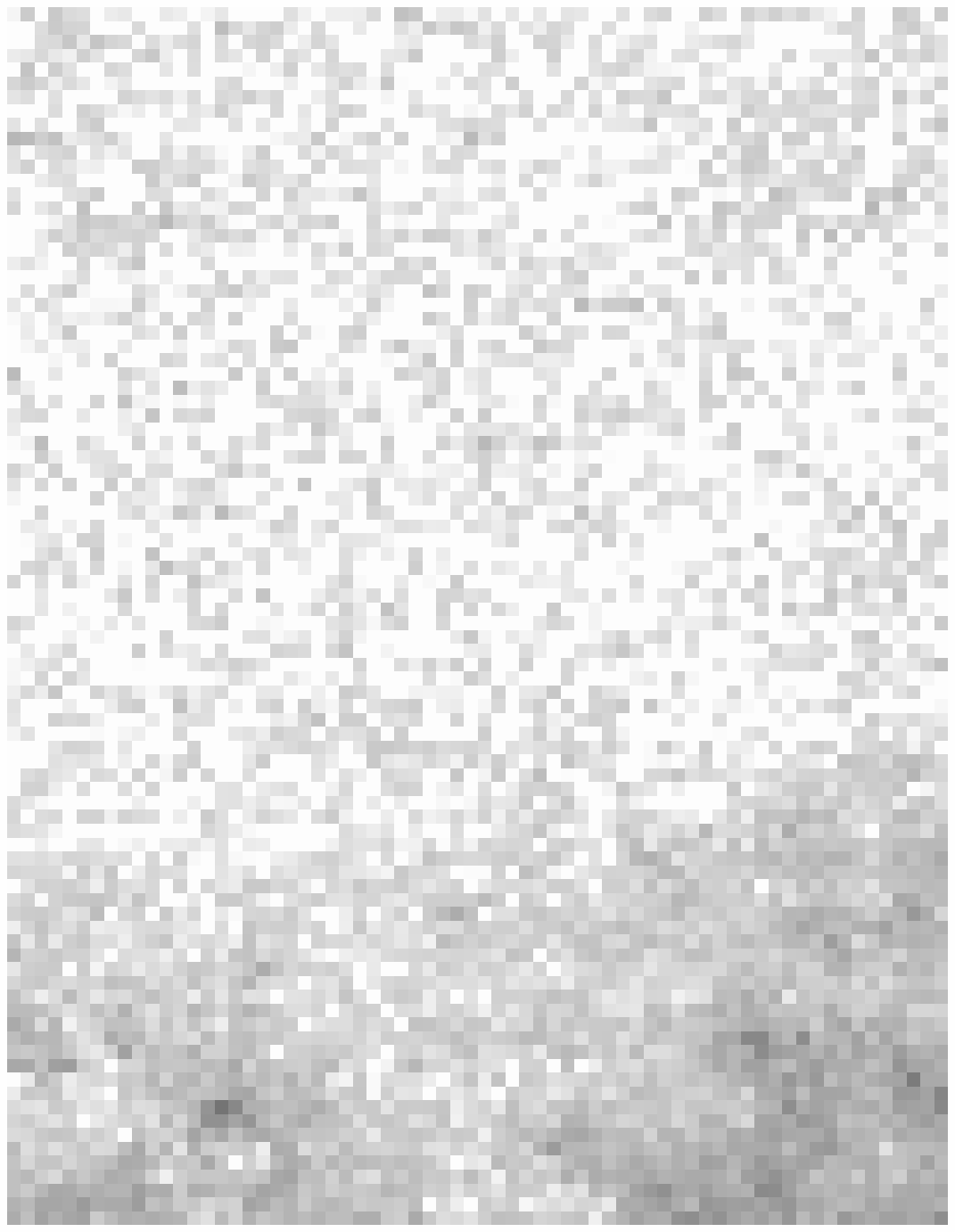}} at 165 19.5
\axis left label {\begin{sideways}b [\degr]\end{sideways}}
ticks in long numbered from 0 to 40 by 5
      short unlabeled from 0 to 40 by 1 /
\axis right label {}
ticks in long unlabeled from 0 to 40 by 5
      short unlabeled from 0 to 40 by 1 /
\axis bottom label {l [\degr]}
ticks in long numbered from 150 to 180 by 5
      short unlabeled from 150 to 180 by 1 /
\axis top label {}
ticks in long unlabeled from 150 to 180 by 5
      short unlabeled from 150 to 180 by 1 /
\endpicture  
\caption{\label{detail6} Grey scale representation of a detail of our
extinction map based on the nearest 49 stars. Extinction values are square root
scaled from 0\,mag (white) to 15\,mag (black) of optical extinction. }    
\end{figure*}

\clearpage
\newpage
\begin{figure*}
\beginpicture
\setcoordinatesystem units <-5mm,5mm> point at  0 0
\setplotarea x from 211 to 179 , y from -1 to 40
\put {\includegraphics[width=42.5cm]{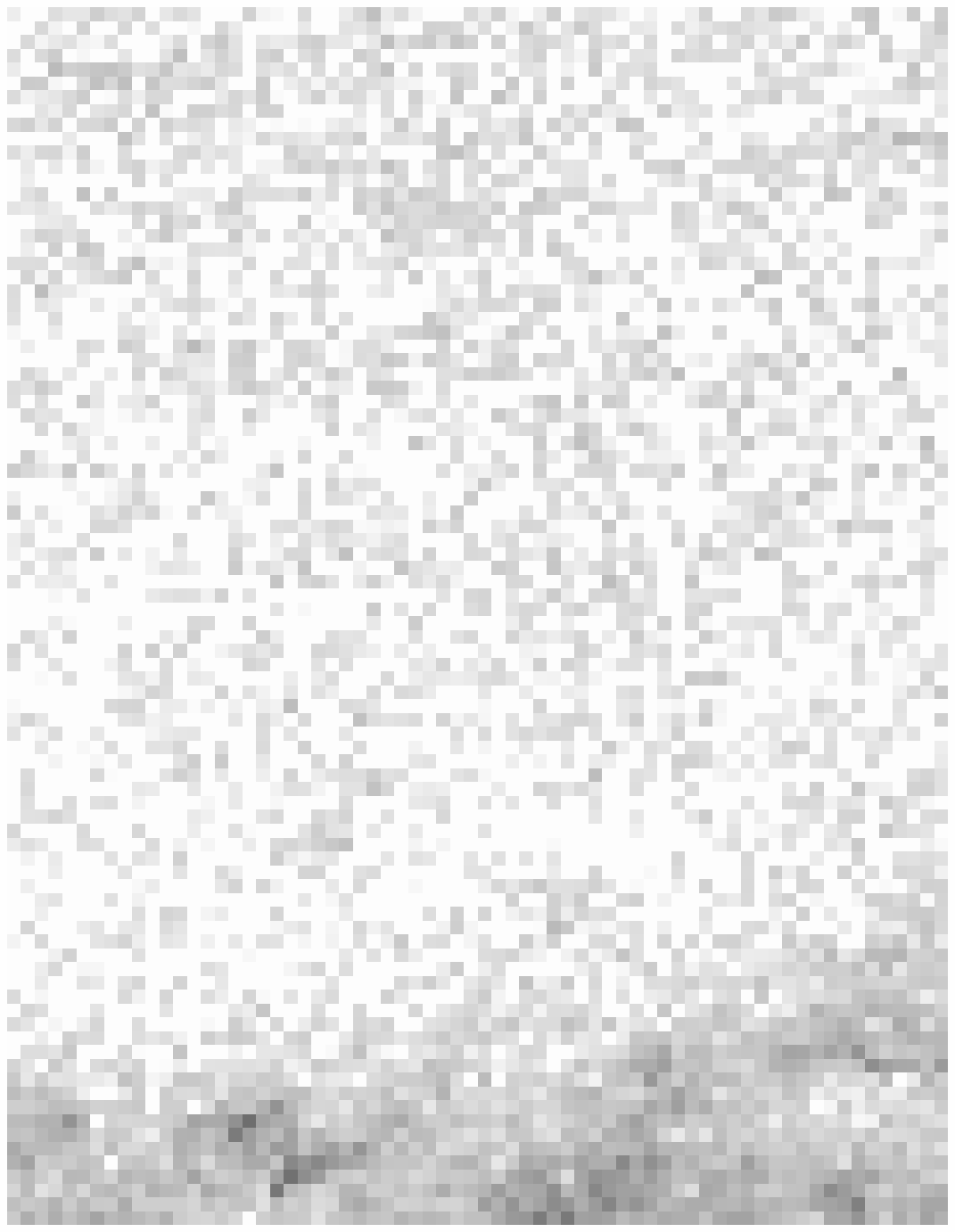}} at 195 19.5
\axis left label {\begin{sideways}b [\degr]\end{sideways}}
ticks in long numbered from 0 to 40 by 5
      short unlabeled from 0 to 40 by 1 /
\axis right label {}
ticks in long unlabeled from 0 to 40 by 5
      short unlabeled from 0 to 40 by 1 /
\axis bottom label {l [\degr]}
ticks in long numbered from 180 to 210 by 5
      short unlabeled from 180 to 210 by 1 /
\axis top label {}
ticks in long unlabeled from 180 to 210 by 5
      short unlabeled from 180 to 210 by 1 /
\endpicture  
\caption{\label{detail7} Grey scale representation of a detail of our
extinction map based on the nearest 49 stars. Extinction values are square root
scaled from 0\,mag (white) to 15\,mag (black) of optical extinction. }    
\end{figure*}

\clearpage
\newpage
\begin{figure*}
\beginpicture
\setcoordinatesystem units <-5mm,5mm> point at  0 0
\setplotarea x from 241 to 209 , y from -1 to 40
\put {\includegraphics[width=42.5cm]{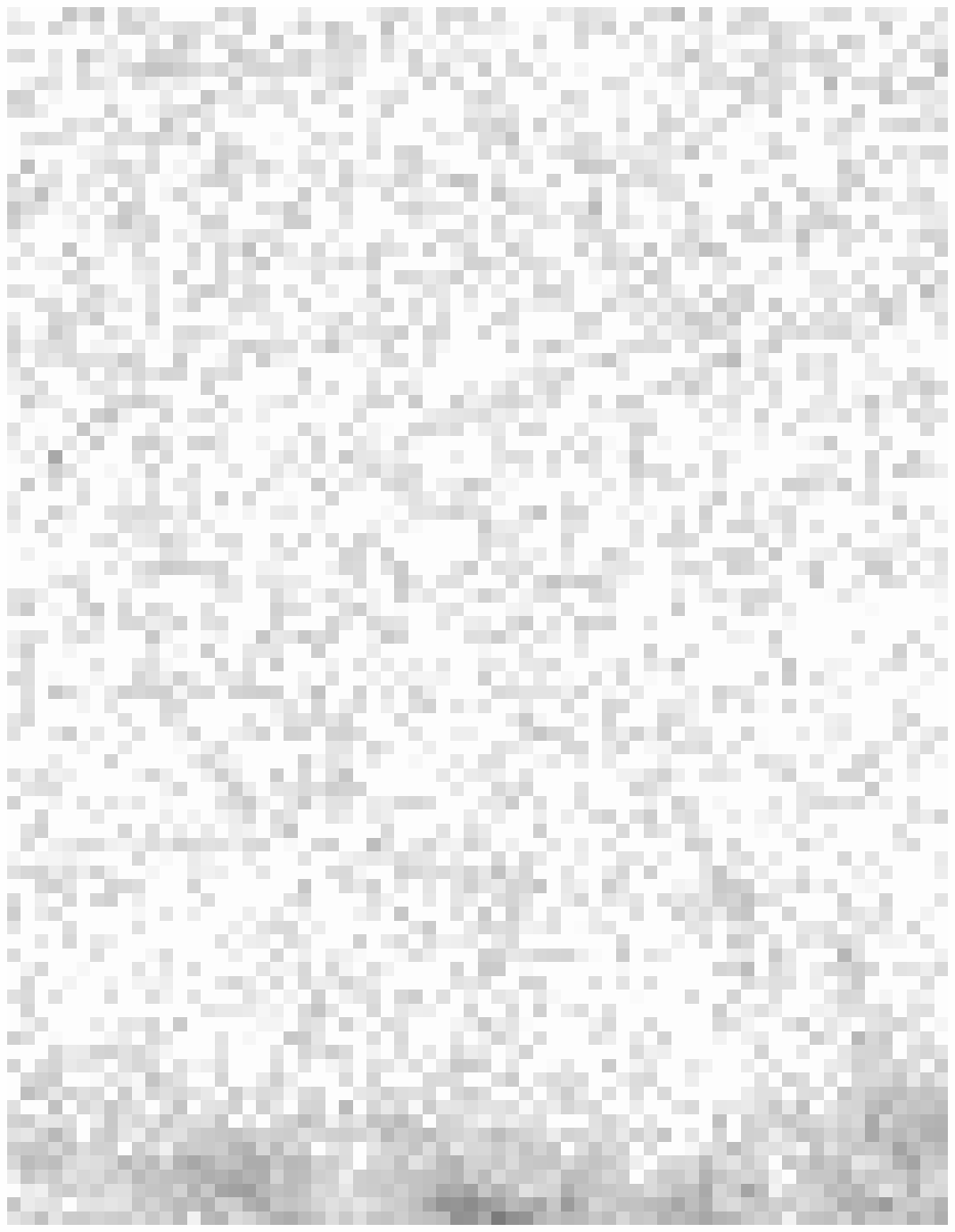}} at 225 19.5
\axis left label {\begin{sideways}b [\degr]\end{sideways}}
ticks in long numbered from 0 to 40 by 5
      short unlabeled from 0 to 40 by 1 /
\axis right label {}
ticks in long unlabeled from 0 to 40 by 5
      short unlabeled from 0 to 40 by 1 /
\axis bottom label {l [\degr]}
ticks in long numbered from 210 to 240 by 5
      short unlabeled from 210 to 240 by 1 /
\axis top label {}
ticks in long unlabeled from 210 to 240 by 5
      short unlabeled from 210 to 240 by 1 /
\endpicture  
\caption{\label{detail8} Grey scale representation of a detail of our
extinction map based on the nearest 49 stars. Extinction values are square root
scaled from 0\,mag (white) to 15\,mag (black) of optical extinction. }    
\end{figure*}

\clearpage
\newpage
\begin{figure*}
\beginpicture
\setcoordinatesystem units <-5mm,5mm> point at  0 0
\setplotarea x from 271 to 239 , y from -1 to 40
\put {\includegraphics[width=42.5cm]{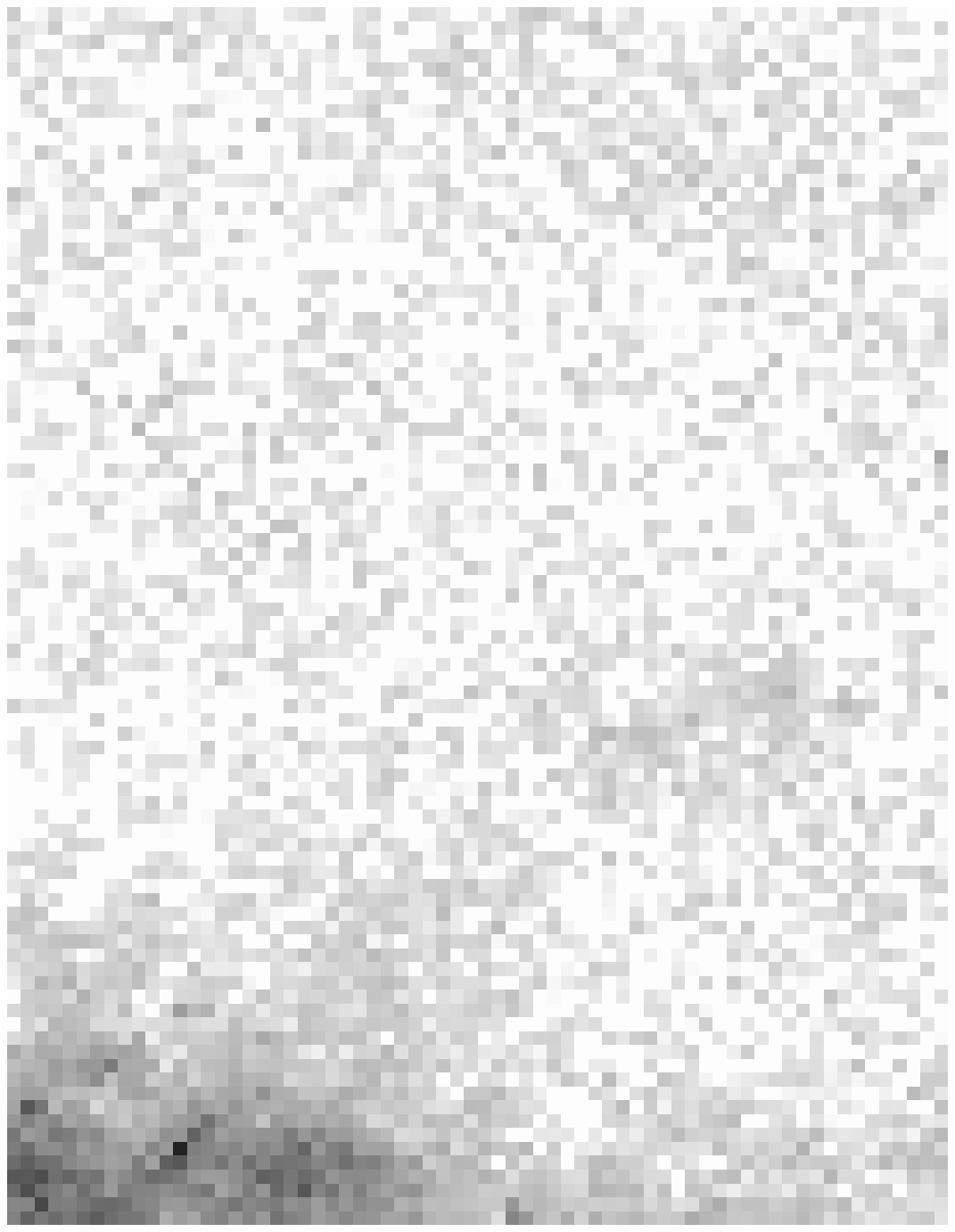}} at 255 19.5
\axis left label {\begin{sideways}b [\degr]\end{sideways}}
ticks in long numbered from 0 to 40 by 5
      short unlabeled from 0 to 40 by 1 /
\axis right label {}
ticks in long unlabeled from 0 to 40 by 5
      short unlabeled from 0 to 40 by 1 /
\axis bottom label {l [\degr]}
ticks in long numbered from 240 to 270 by 5
      short unlabeled from 240 to 270 by 1 /
\axis top label {}
ticks in long unlabeled from 240 to 270 by 5
      short unlabeled from 240 to 270 by 1 /
\endpicture  
\caption{\label{detail9} Grey scale representation of a detail of our
extinction map based on the nearest 49 stars. Extinction values are square root
scaled from 0\,mag (white) to 15\,mag (black) of optical extinction. }    
\end{figure*}

\clearpage
\newpage
\begin{figure*}
\beginpicture
\setcoordinatesystem units <-5mm,5mm> point at  0 0
\setplotarea x from 301 to 269 , y from -1 to 40
\put {\includegraphics[width=42.5cm]{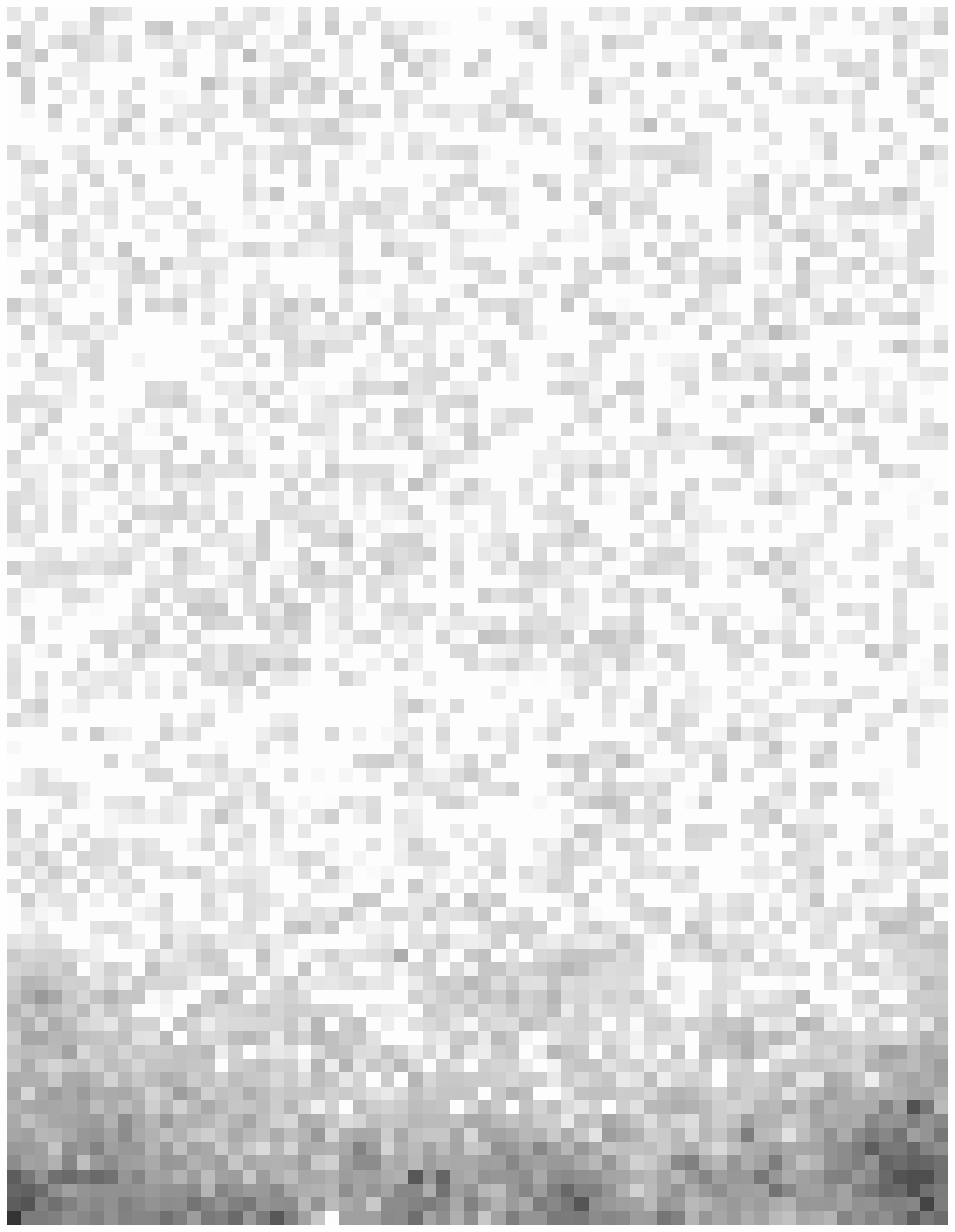}} at 285 19.5
\axis left label {\begin{sideways}b [\degr]\end{sideways}}
ticks in long numbered from 0 to 40 by 5
      short unlabeled from 0 to 40 by 1 /
\axis right label {}
ticks in long unlabeled from 0 to 40 by 5
      short unlabeled from 0 to 40 by 1 /
\axis bottom label {l [\degr]}
ticks in long numbered from 270 to 300 by 5
      short unlabeled from 270 to 300 by 1 /
\axis top label {}
ticks in long unlabeled from 270 to 300 by 5
      short unlabeled from 270 to 300 by 1 /
\endpicture  
\caption{\label{detail10} Grey scale representation of a detail of our
extinction map based on the nearest 49 stars. Extinction values are square root
scaled from 0\,mag (white) to 15\,mag (black) of optical extinction. }    
\end{figure*}

\clearpage
\newpage
\begin{figure*}
\beginpicture
\setcoordinatesystem units <-5mm,5mm> point at  0 0
\setplotarea x from 331 to 299 , y from -1 to 40
\put {\includegraphics[width=42.5cm]{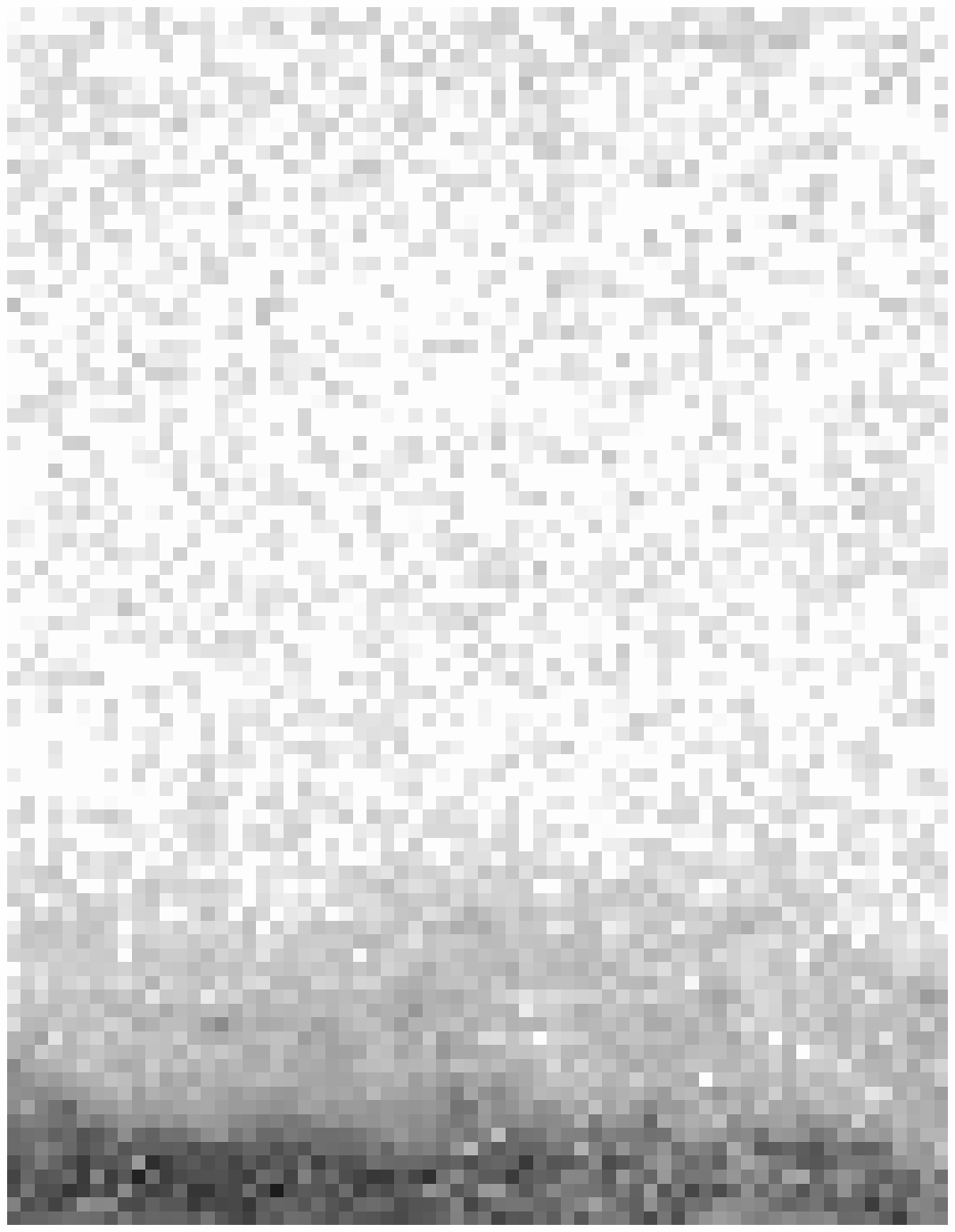}} at 315 19.5
\axis left label {\begin{sideways}b [\degr]\end{sideways}}
ticks in long numbered from 0 to 40 by 5
      short unlabeled from 0 to 40 by 1 /
\axis right label {}
ticks in long unlabeled from 0 to 40 by 5
      short unlabeled from 0 to 40 by 1 /
\axis bottom label {l [\degr]}
ticks in long numbered from 300 to 330 by 5
      short unlabeled from 300 to 330 by 1 /
\axis top label {}
ticks in long unlabeled from 300 to 330 by 5
      short unlabeled from 300 to 330 by 1 /
\endpicture  
\caption{\label{detail11} Grey scale representation of a detail of our
extinction map based on the nearest 49 stars. Extinction values are square root
scaled from 0\,mag (white) to 15\,mag (black) of optical extinction. }    
\end{figure*}

\clearpage
\newpage
\begin{figure*}
\beginpicture
\setcoordinatesystem units <-5mm,5mm> point at  0 0
\setplotarea x from 361 to 329 , y from -1 to 40
\put {\includegraphics[width=42.5cm]{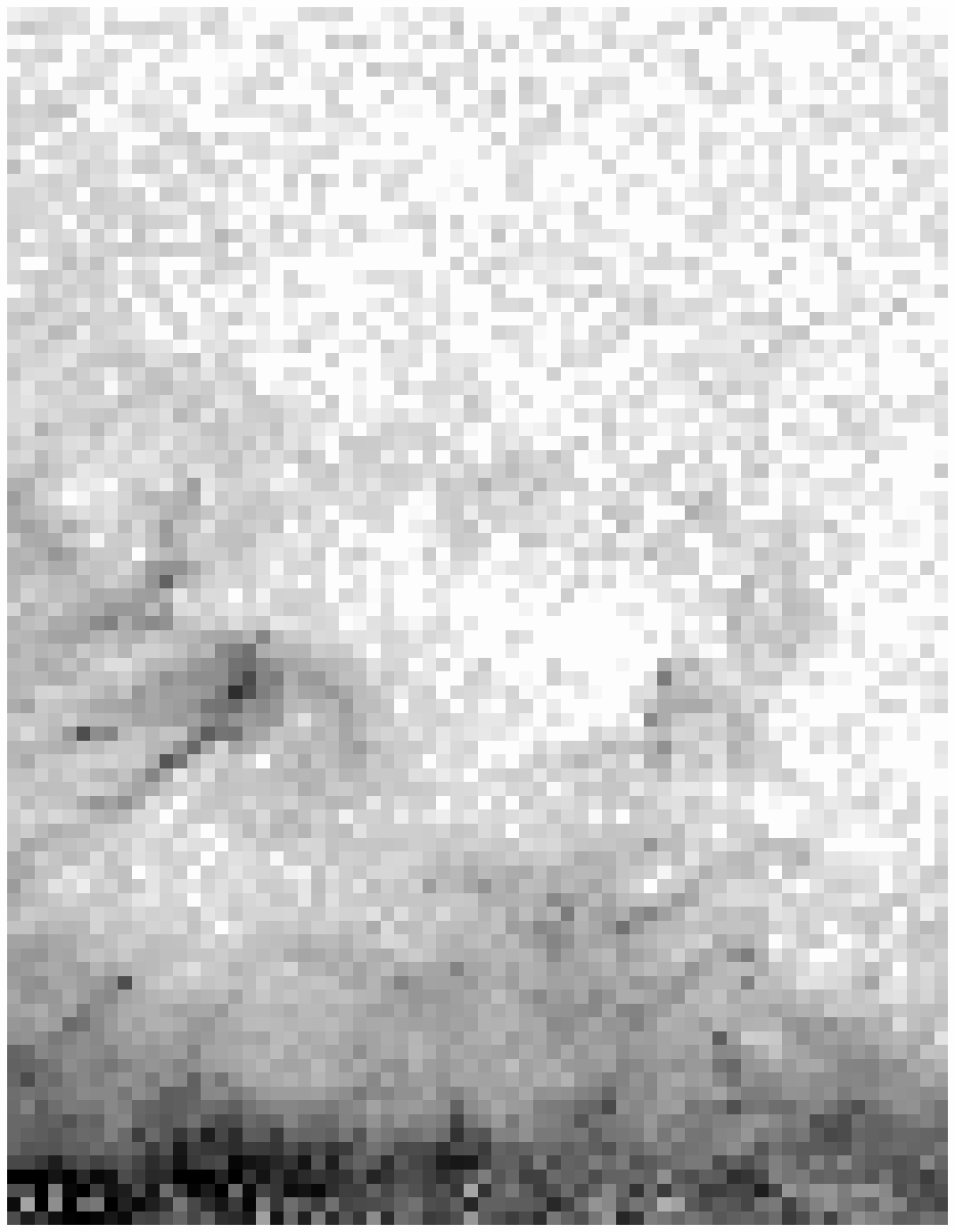}} at 345 19.5
\axis left label {\begin{sideways}b [\degr]\end{sideways}}
ticks in long numbered from 0 to 40 by 5
      short unlabeled from 0 to 40 by 1 /
\axis right label {}
ticks in long unlabeled from 0 to 40 by 5
      short unlabeled from 0 to 40 by 1 /
\axis bottom label {l [\degr]}
ticks in long numbered from 330 to 360 by 5
      short unlabeled from 330 to 360 by 1 /
\axis top label {}
ticks in long unlabeled from 330 to 360 by 5
      short unlabeled from 330 to 360 by 1 /
\endpicture  
\caption{\label{detail12} Grey scale representation of a detail of our
extinction map based on the nearest 49 stars. Extinction values are square root
scaled from 0\,mag (white) to 15\,mag (black) of optical extinction. }    
\end{figure*}

\clearpage
\newpage
\begin{figure*}
\beginpicture
\setcoordinatesystem units <-5mm,5mm> point at  0 0
\setplotarea x from 31 to -1 , y from -40 to 1
\put {\includegraphics[width=42.5cm]{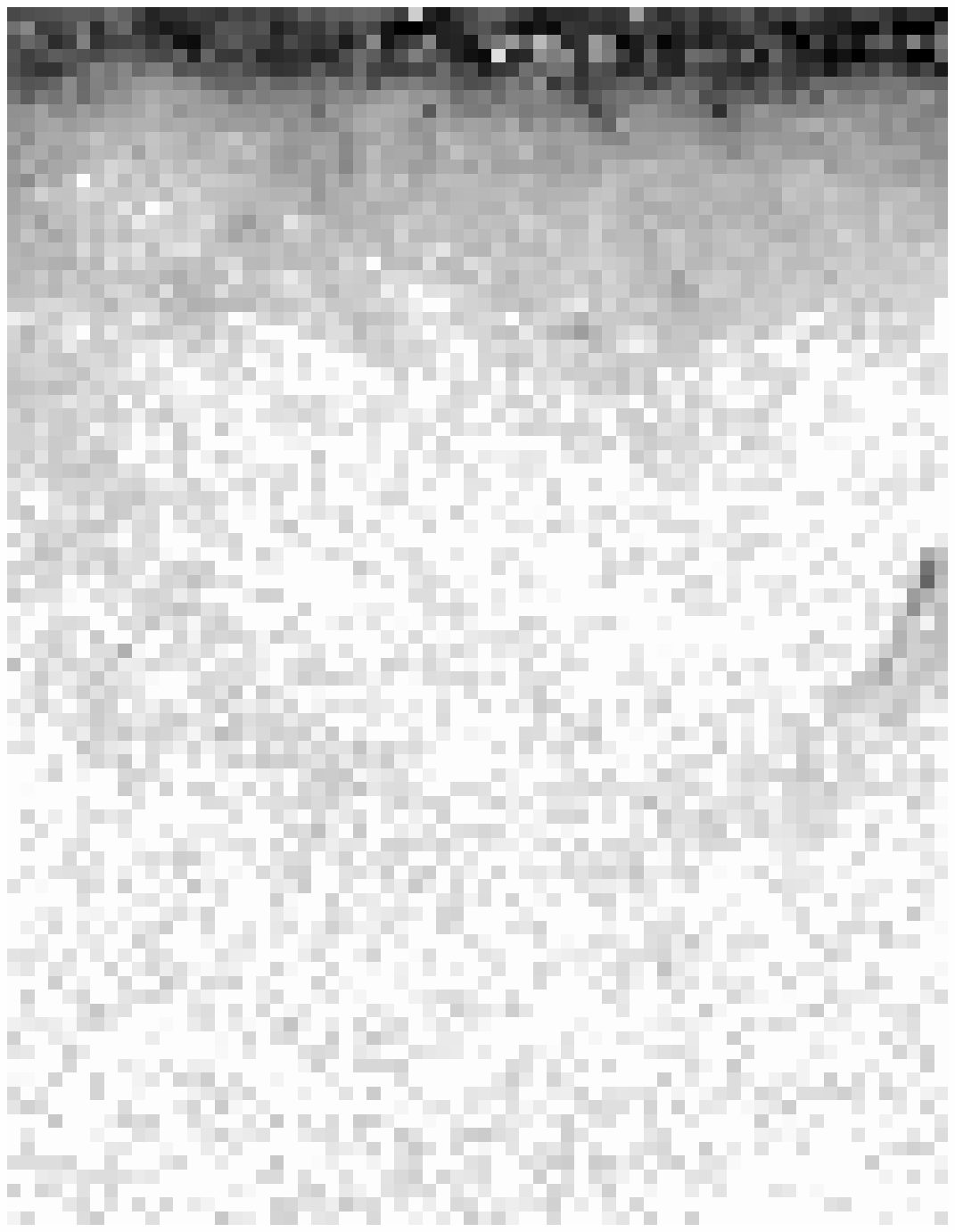}} at 15 -19.5
\axis left label {\begin{sideways}b [\degr]\end{sideways}}
ticks in long numbered from -40 to 0 by 5
      short unlabeled from -40 to 0 by 1 /
\axis right label {}
ticks in long unlabeled from -40 to 0 by 5
      short unlabeled from -40 to 0 by 1 /
\axis bottom label {l [\degr]}
ticks in long numbered from 0 to 30 by 5
      short unlabeled from 0 to 30 by 1 /
\axis top label {}
ticks in long unlabeled from 0 to 30 by 5
      short unlabeled from 0 to 30 by 1 /
\endpicture  
\caption{\label{detail13} Grey scale representation of a detail of our
extinction map based on the nearest 49 stars. Extinction values are square root
scaled from 0\,mag (white) to 15\,mag (black) of optical extinction. }    
\end{figure*}

\clearpage
\newpage
\begin{figure*}
\beginpicture
\setcoordinatesystem units <-5mm,5mm> point at  0 0
\setplotarea x from 61 to 29 , y from -40 to 1
\put {\includegraphics[width=42.5cm]{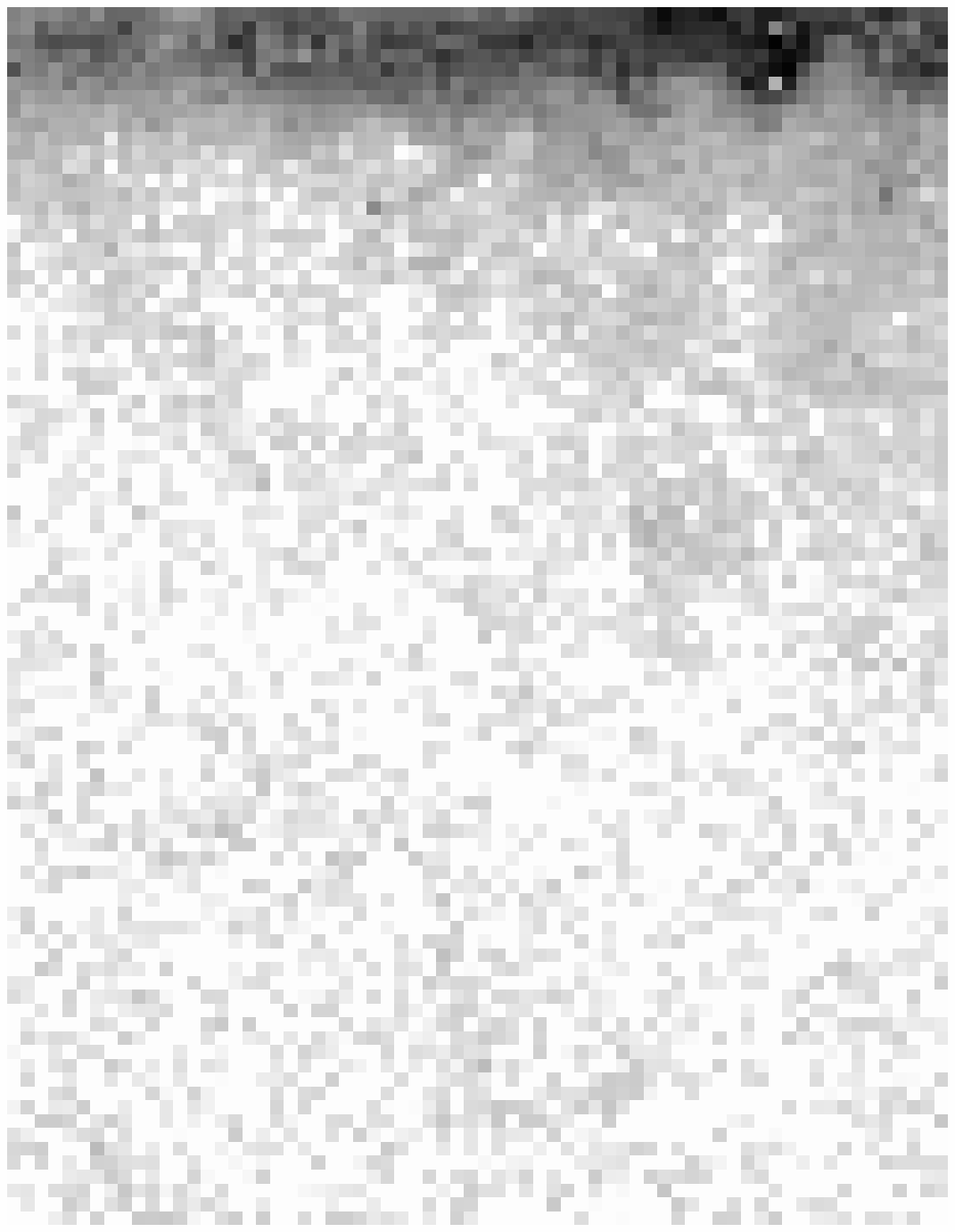}} at 45 -19.5
\axis left label {\begin{sideways}b [\degr]\end{sideways}}
ticks in long numbered from -40 to 0 by 5
      short unlabeled from -40 to 0 by 1 /
\axis right label {}
ticks in long unlabeled from -40 to 0 by 5
      short unlabeled from -40 to 0 by 1 /
\axis bottom label {l [\degr]}
ticks in long numbered from 30 to 60 by 5
      short unlabeled from 30 to 60 by 1 /
\axis top label {}
ticks in long unlabeled from 30 to 60 by 5
      short unlabeled from 30 to 60 by 1 /
\endpicture  
\caption{\label{detail14} Grey scale representation of a detail of our
extinction map based on the nearest 49 stars. Extinction values are square root
scaled from 0\,mag (white) to 15\,mag (black) of optical extinction. }    
\end{figure*}

\clearpage
\newpage
\begin{figure*}
\beginpicture
\setcoordinatesystem units <-5mm,5mm> point at  0 0
\setplotarea x from 91 to 59 , y from -40 to 1
\put {\includegraphics[width=42.5cm]{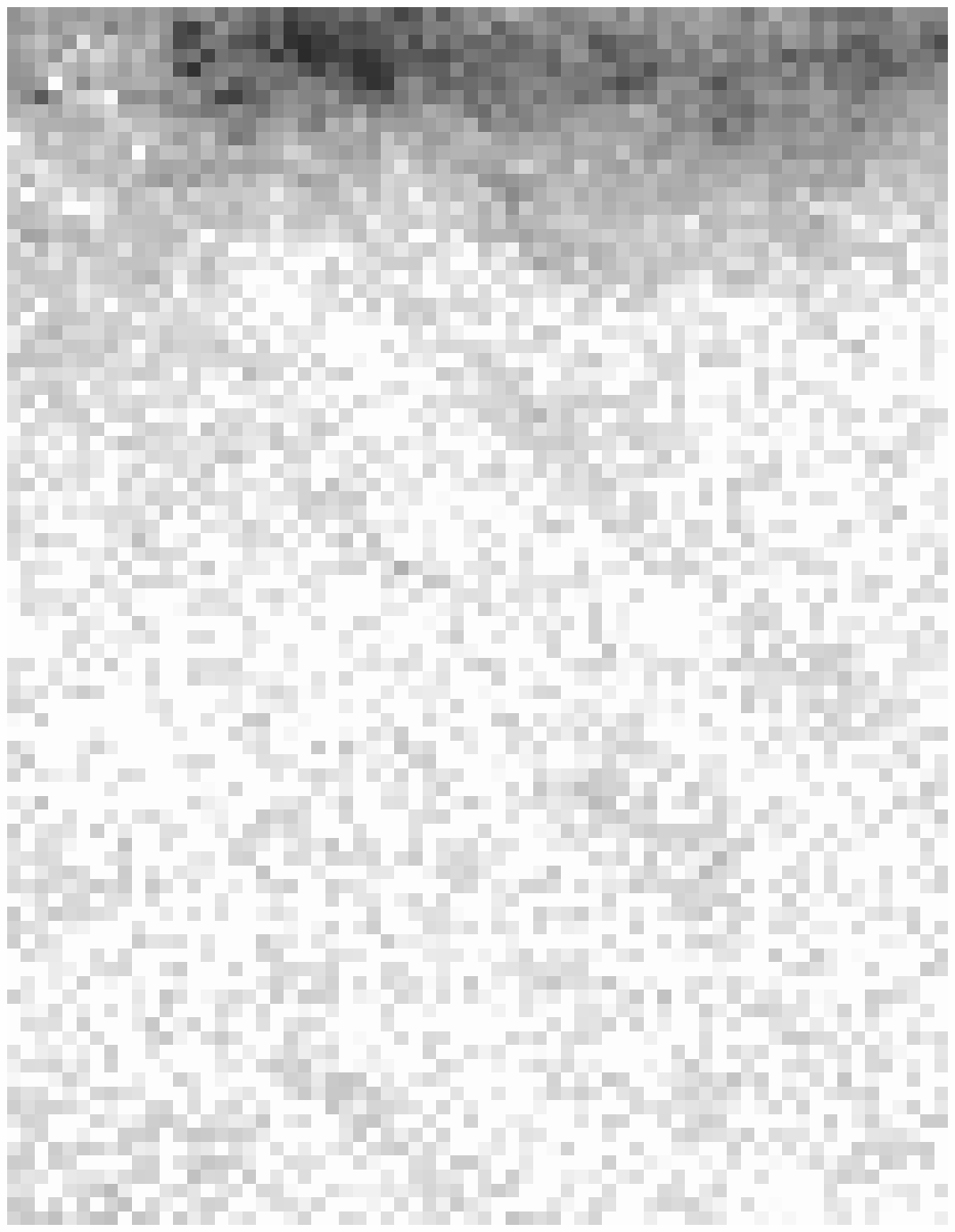}} at 75 -19.5
\axis left label {\begin{sideways}b [\degr]\end{sideways}}
ticks in long numbered from -40 to 0 by 5
      short unlabeled from -40 to 0 by 1 /
\axis right label {}
ticks in long unlabeled from -40 to 0 by 5
      short unlabeled from -40 to 0 by 1 /
\axis bottom label {l [\degr]}
ticks in long numbered from 60 to 90 by 5
      short unlabeled from 60 to 90 by 1 /
\axis top label {}
ticks in long unlabeled from 60 to 90 by 5
      short unlabeled from 60 to 90 by 1 /
\endpicture  
\caption{\label{detail15} Grey scale representation of a detail of our
extinction map based on the nearest 49 stars. Extinction values are square root
scaled from 0\,mag (white) to 15\,mag (black) of optical extinction. }    
\end{figure*}

\clearpage
\newpage
\begin{figure*}
\beginpicture
\setcoordinatesystem units <-5mm,5mm> point at  0 0
\setplotarea x from 121 to 89 , y from -40 to 1
\put {\includegraphics[width=42.5cm]{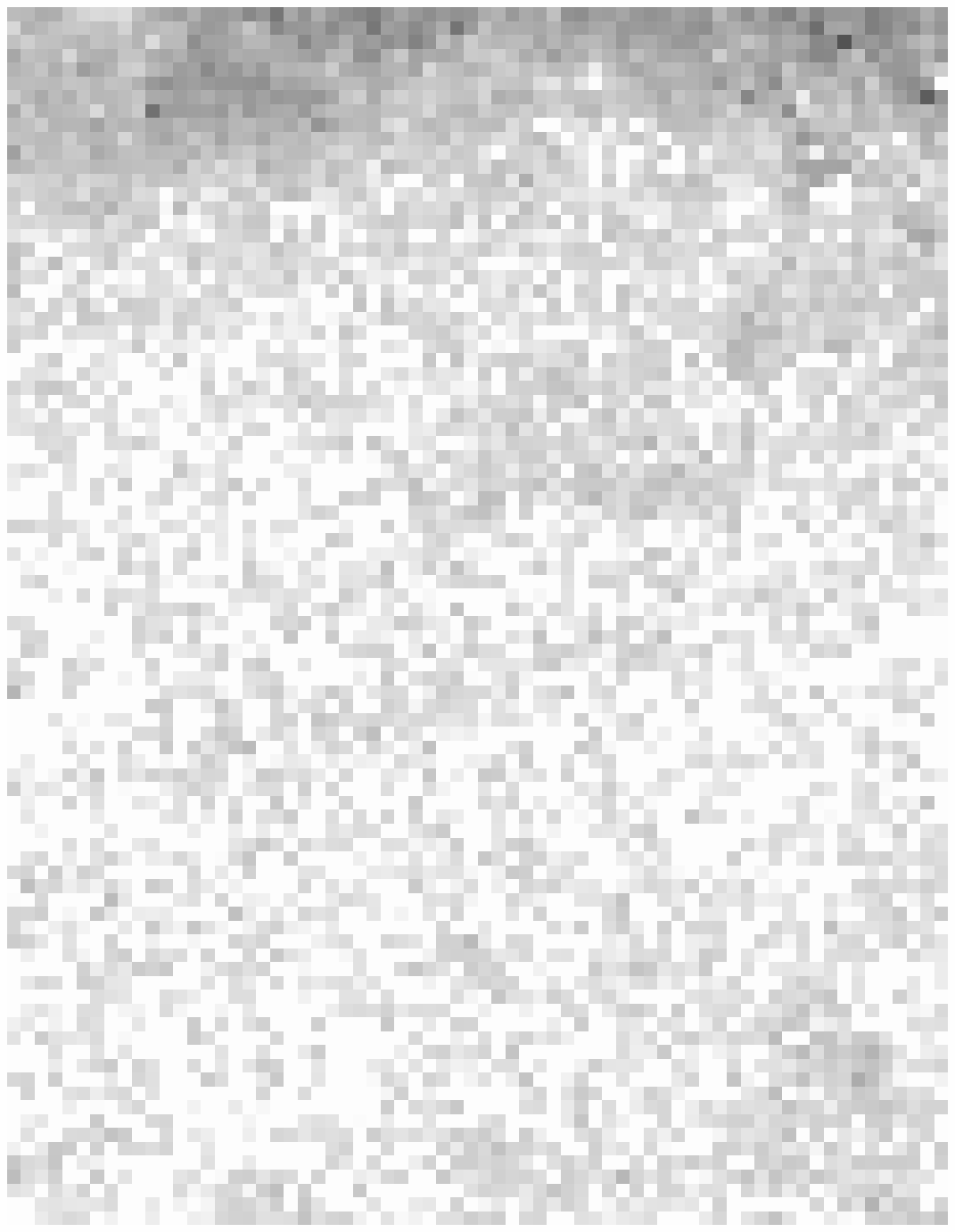}} at 105 -19.5
\axis left label {\begin{sideways}b [\degr]\end{sideways}}
ticks in long numbered from -40 to 0 by 5
      short unlabeled from -40 to 0 by 1 /
\axis right label {}
ticks in long unlabeled from -40 to 0 by 5
      short unlabeled from -40 to 0 by 1 /
\axis bottom label {l [\degr]}
ticks in long numbered from 90 to 120 by 5
      short unlabeled from 90 to 120 by 1 /
\axis top label {}
ticks in long unlabeled from 90 to 120 by 5
      short unlabeled from 90 to 120 by 1 /
\endpicture  
\caption{\label{detail16} Grey scale representation of a detail of our
extinction map based on the nearest 49 stars. Extinction values are square root
scaled from 0\,mag (white) to 15\,mag (black) of optical extinction. }    
\end{figure*}

\clearpage
\newpage
\begin{figure*}
\beginpicture
\setcoordinatesystem units <-5mm,5mm> point at  0 0
\setplotarea x from 151 to 119 , y from -40 to 1
\put {\includegraphics[width=42.5cm]{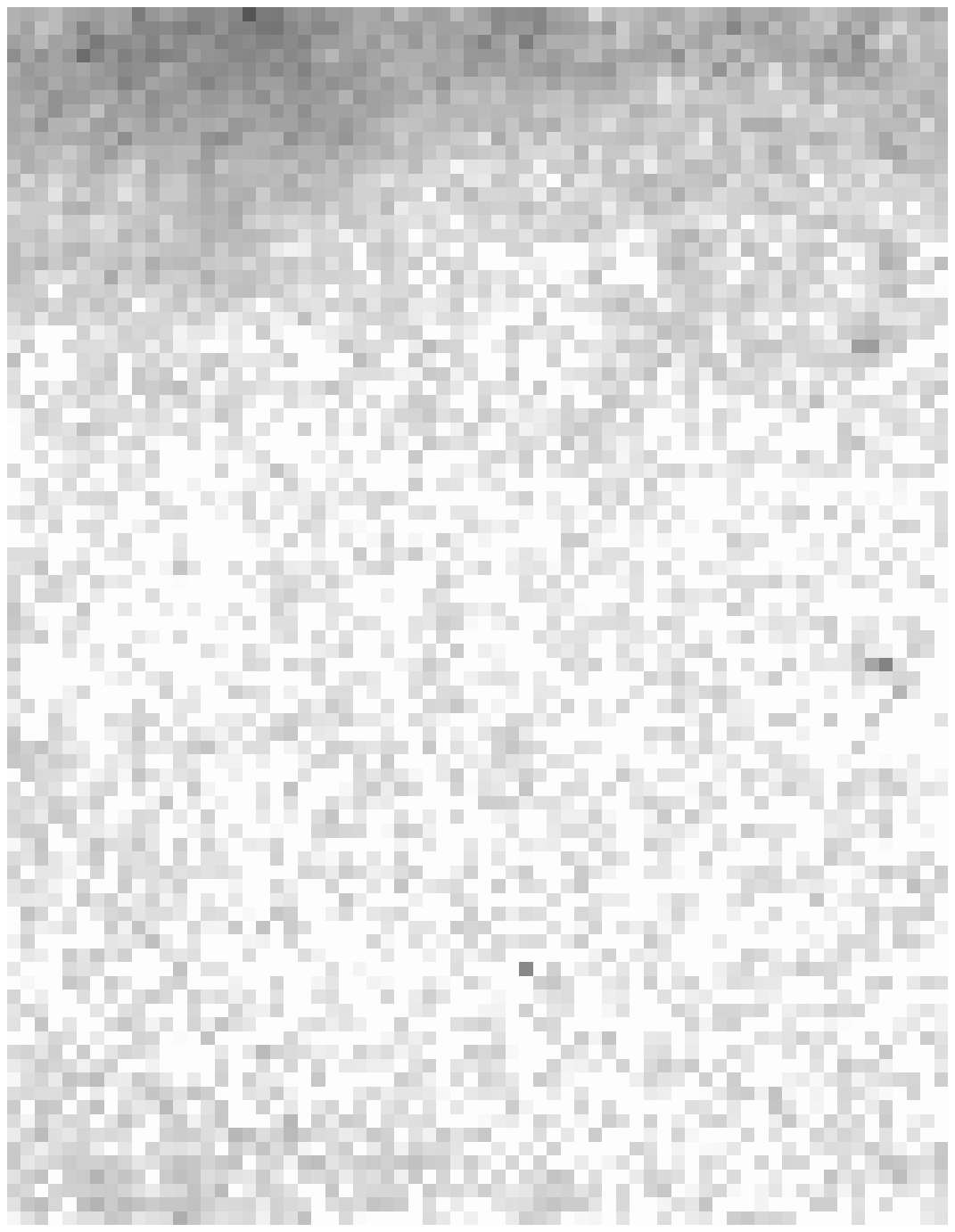}} at 135 -19.5
\axis left label {\begin{sideways}b [\degr]\end{sideways}}
ticks in long numbered from -40 to 0 by 5
      short unlabeled from -40 to 0 by 1 /
\axis right label {}
ticks in long unlabeled from -40 to 0 by 5
      short unlabeled from -40 to 0 by 1 /
\axis bottom label {l [\degr]}
ticks in long numbered from 120 to 150 by 5
      short unlabeled from 120 to 150 by 1 /
\axis top label {}
ticks in long unlabeled from 120 to 150 by 5
      short unlabeled from 120 to 150 by 1 /
\endpicture  
\caption{\label{detail17} Grey scale representation of a detail of our
extinction map based on the nearest 49 stars. Extinction values are square root
scaled from 0\,mag (white) to 15\,mag (black) of optical extinction. }    
\end{figure*}

\clearpage
\newpage
\begin{figure*}
\beginpicture
\setcoordinatesystem units <-5mm,5mm> point at  0 0
\setplotarea x from 181 to 149 , y from -40 to 1
\put {\includegraphics[width=42.5cm]{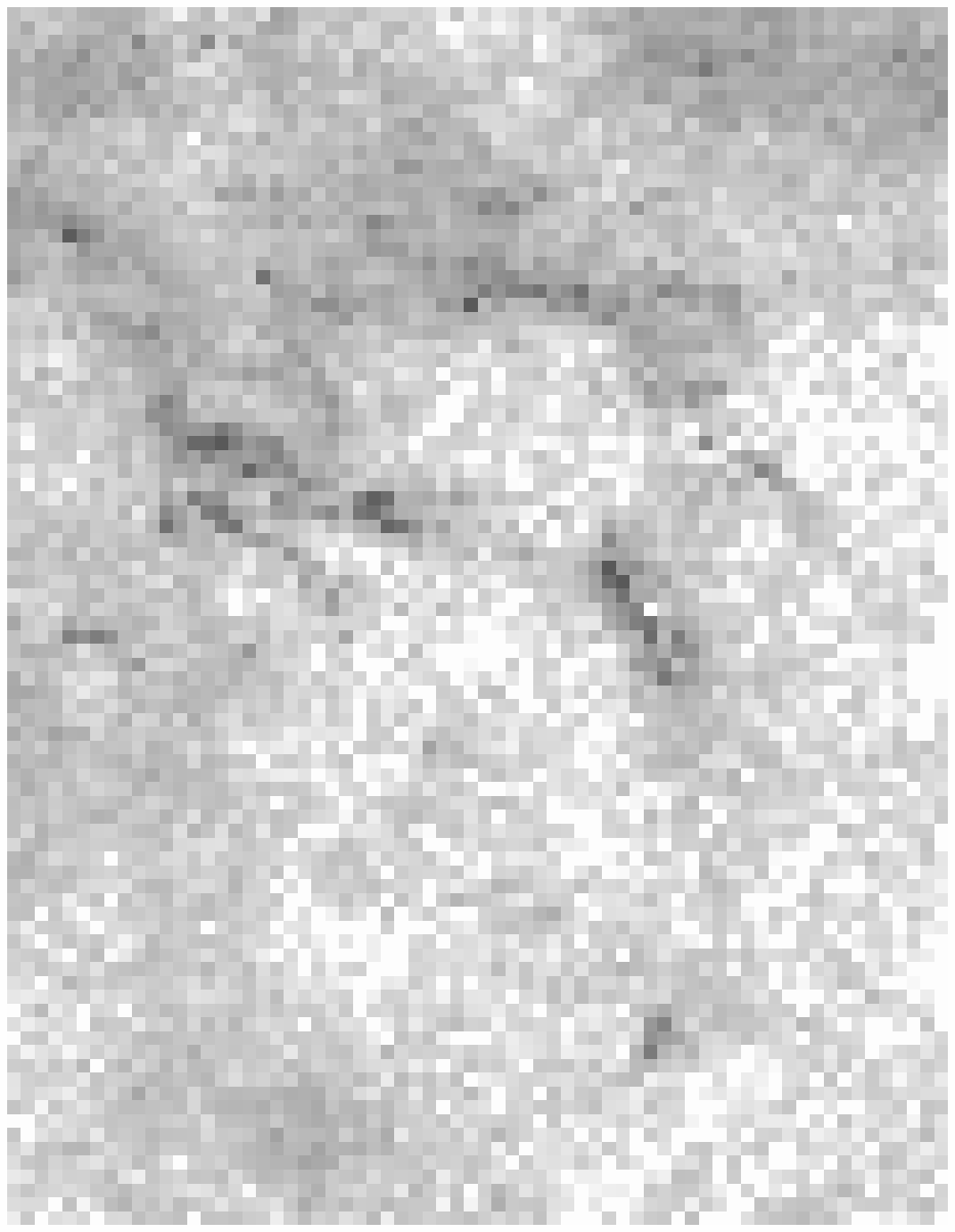}} at 165 -19.5
\axis left label {\begin{sideways}b [\degr]\end{sideways}}
ticks in long numbered from -40 to 0 by 5
      short unlabeled from -40 to 0 by 1 /
\axis right label {}
ticks in long unlabeled from -40 to 0 by 5
      short unlabeled from -40 to 0 by 1 /
\axis bottom label {l [\degr]}
ticks in long numbered from 150 to 180 by 5
      short unlabeled from 150 to 180 by 1 /
\axis top label {}
ticks in long unlabeled from 150 to 180 by 5
      short unlabeled from 150 to 180 by 1 /
\endpicture  
\caption{\label{detail18} Grey scale representation of a detail of our
extinction map based on the nearest 49 stars. Extinction values are square root
scaled from 0\,mag (white) to 15\,mag (black) of optical extinction. }    
\end{figure*}

\clearpage
\newpage
\begin{figure*}
\beginpicture
\setcoordinatesystem units <-5mm,5mm> point at  0 0
\setplotarea x from 211 to 179 , y from -40 to 1
\put {\includegraphics[width=42.5cm]{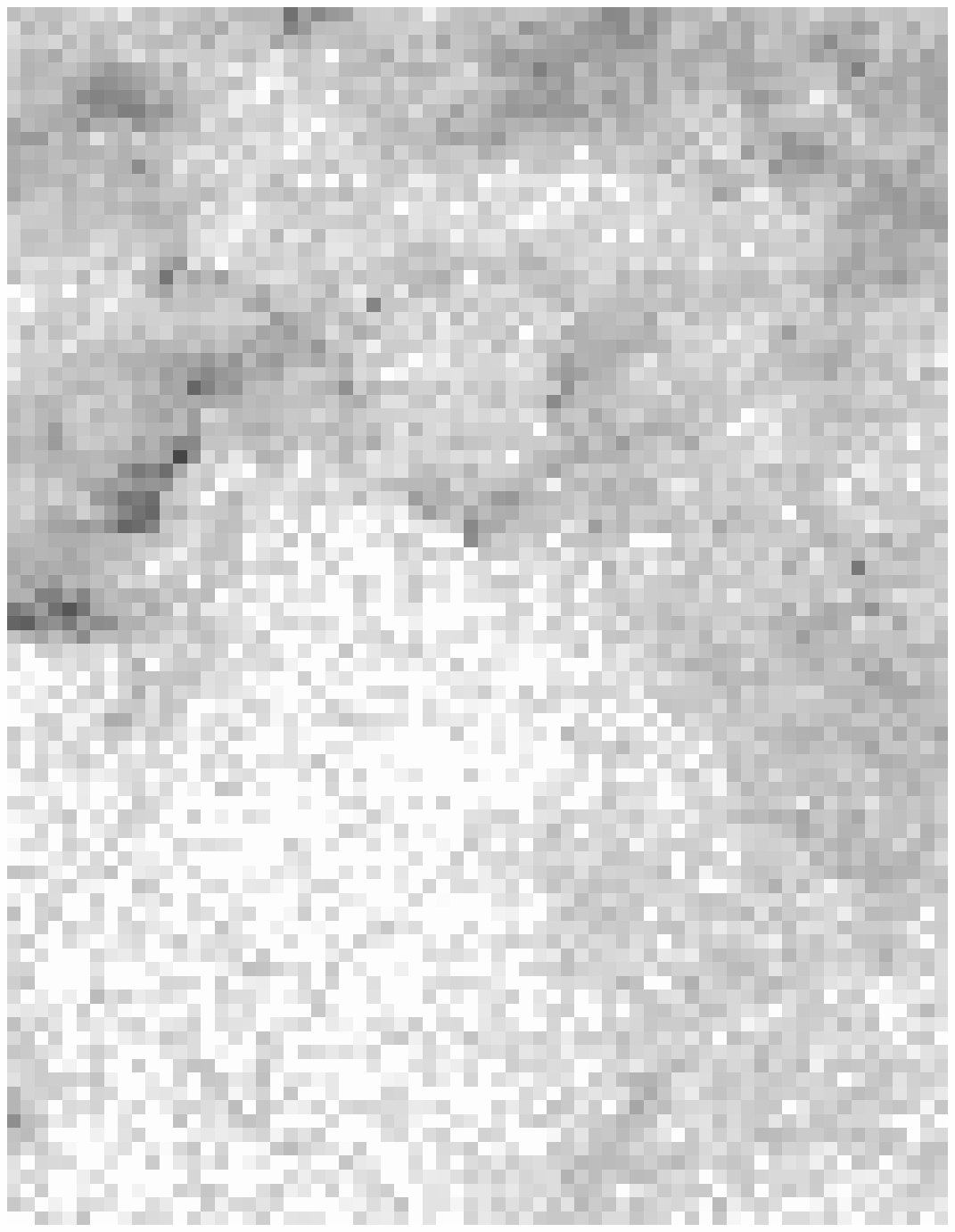}} at 195 -19.5
\axis left label {\begin{sideways}b [\degr]\end{sideways}}
ticks in long numbered from -40 to 0 by 5
      short unlabeled from -40 to 0 by 1 /
\axis right label {}
ticks in long unlabeled from -40 to 0 by 5
      short unlabeled from -40 to 0 by 1 /
\axis bottom label {l [\degr]}
ticks in long numbered from 180 to 210 by 5
      short unlabeled from 180 to 210 by 1 /
\axis top label {}
ticks in long unlabeled from 180 to 210 by 5
      short unlabeled from 180 to 210 by 1 /
\endpicture  
\caption{\label{detail19} Grey scale representation of a detail of our
extinction map based on the nearest 49 stars. Extinction values are square root
scaled from 0\,mag (white) to 15\,mag (black) of optical extinction. }    
\end{figure*}

\clearpage
\newpage
\begin{figure*}
\beginpicture
\setcoordinatesystem units <-5mm,5mm> point at  0 0
\setplotarea x from 241 to 209 , y from -40 to 1
\put {\includegraphics[width=42.5cm]{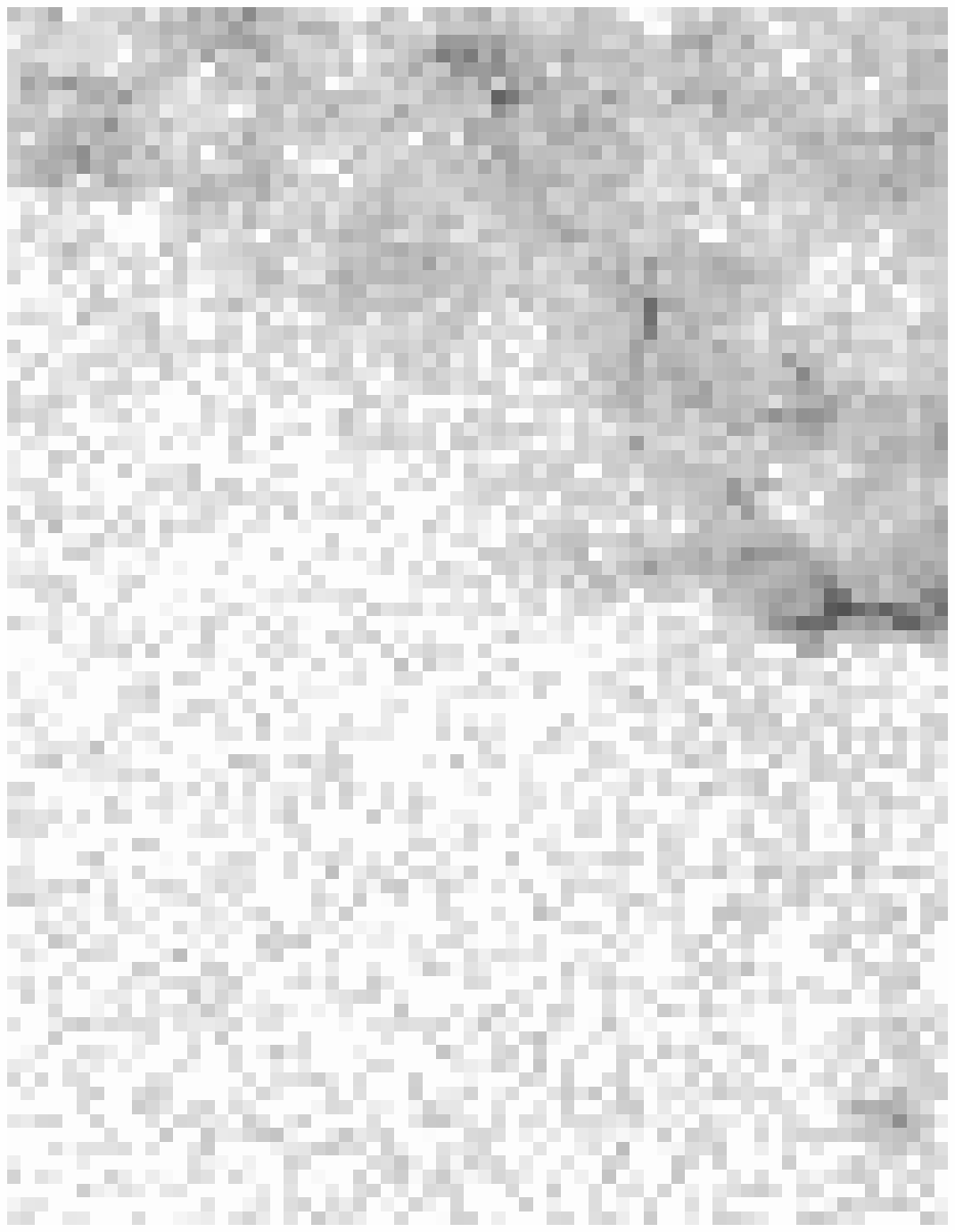}} at 225 -19.5
\axis left label {\begin{sideways}b [\degr]\end{sideways}}
ticks in long numbered from -40 to 0 by 5
      short unlabeled from -40 to 0 by 1 /
\axis right label {}
ticks in long unlabeled from -40 to 0 by 5
      short unlabeled from -40 to 0 by 1 /
\axis bottom label {l [\degr]}
ticks in long numbered from 210 to 240 by 5
      short unlabeled from 210 to 240 by 1 /
\axis top label {}
ticks in long unlabeled from 210 to 240 by 5
      short unlabeled from 210 to 240 by 1 /
\endpicture  
\caption{\label{detail20} Grey scale representation of a detail of our
extinction map based on the nearest 49 stars. Extinction values are square root
scaled from 0\,mag (white) to 15\,mag (black) of optical extinction. }    
\end{figure*}

\clearpage
\newpage
\begin{figure*}
\beginpicture
\setcoordinatesystem units <-5mm,5mm> point at  0 0
\setplotarea x from 271 to 239 , y from -40 to 1
\put {\includegraphics[width=42.5cm]{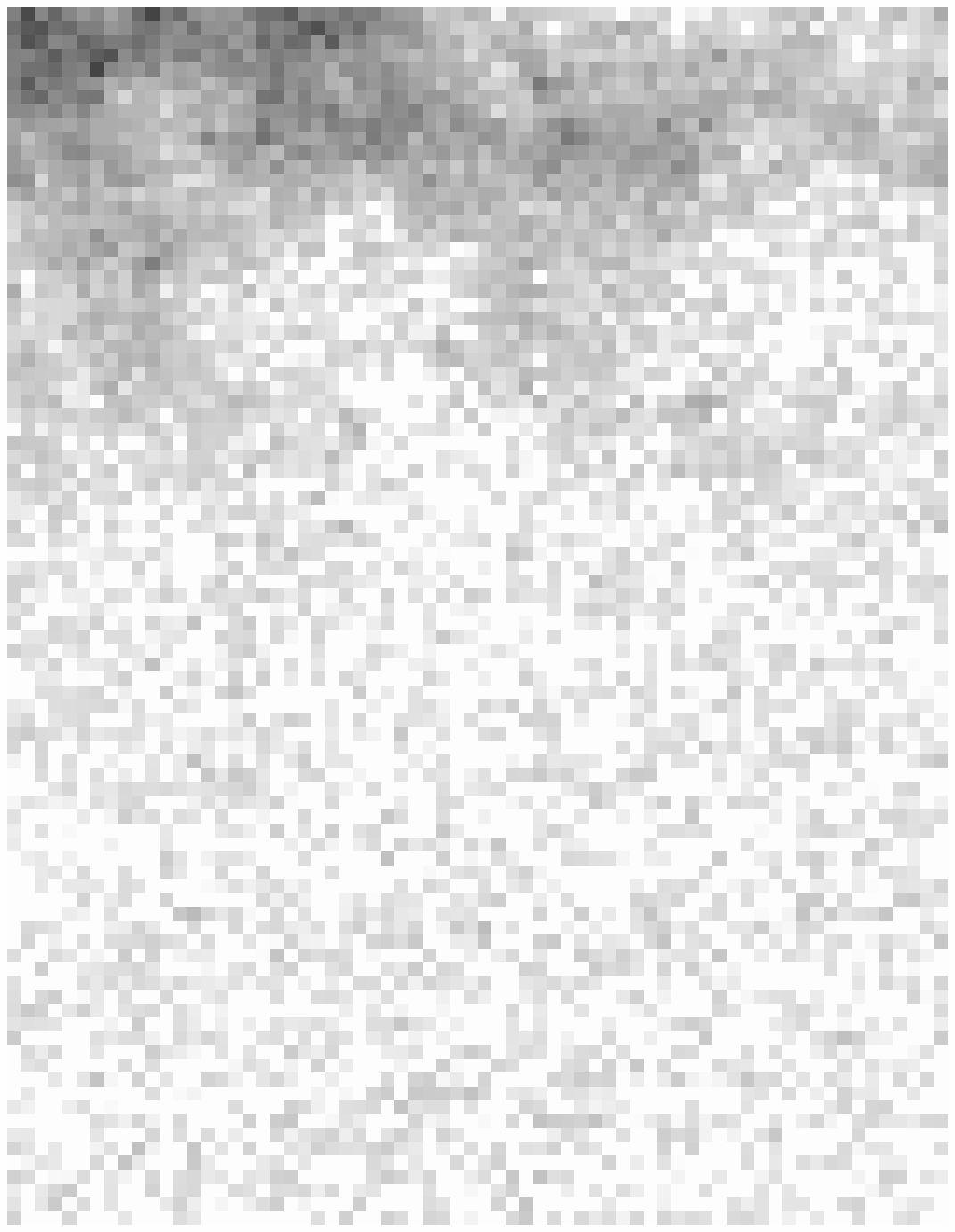}} at 255 -19.5
\axis left label {\begin{sideways}b [\degr]\end{sideways}}
ticks in long numbered from -40 to 0 by 5
      short unlabeled from -40 to 0 by 1 /
\axis right label {}
ticks in long unlabeled from -40 to 0 by 5
      short unlabeled from -40 to 0 by 1 /
\axis bottom label {l [\degr]}
ticks in long numbered from 240 to 270 by 5
      short unlabeled from 240 to 270 by 1 /
\axis top label {}
ticks in long unlabeled from 240 to 270 by 5
      short unlabeled from 240 to 270 by 1 /
\endpicture  
\caption{\label{detail21} Grey scale representation of a detail of our
extinction map based on the nearest 49 stars. Extinction values are square root
scaled from 0\,mag (white) to 15\,mag (black) of optical extinction. }    
\end{figure*}

\clearpage
\newpage
\begin{figure*}
\beginpicture
\setcoordinatesystem units <-5mm,5mm> point at  0 0
\setplotarea x from 301 to 269 , y from -40 to 1
\put {\includegraphics[width=42.5cm]{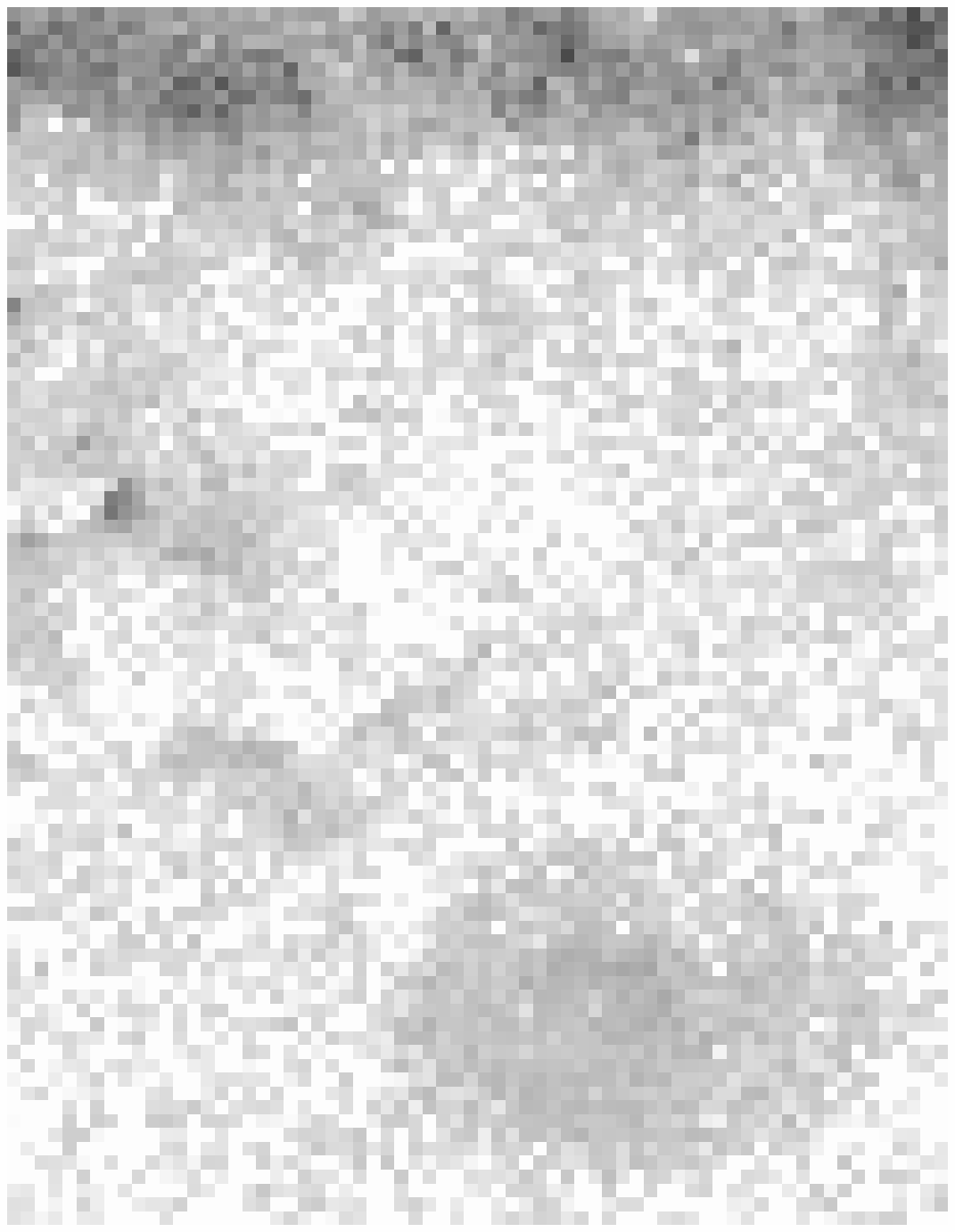}} at 285 -19.5
\axis left label {\begin{sideways}b [\degr]\end{sideways}}
ticks in long numbered from -40 to 0 by 5
      short unlabeled from -40 to 0 by 1 /
\axis right label {}
ticks in long unlabeled from -40 to 0 by 5
      short unlabeled from -40 to 0 by 1 /
\axis bottom label {l [\degr]}
ticks in long numbered from 270 to 300 by 5
      short unlabeled from 270 to 300 by 1 /
\axis top label {}
ticks in long unlabeled from 270 to 300 by 5
      short unlabeled from 270 to 300 by 1 /
\endpicture  
\caption{\label{detail22} Grey scale representation of a detail of our
extinction map based on the nearest 49 stars. Extinction values are square root
scaled from 0\,mag (white) to 15\,mag (black) of optical extinction. }    
\end{figure*}

\clearpage
\newpage
\begin{figure*}
\beginpicture
\setcoordinatesystem units <-5mm,5mm> point at  0 0
\setplotarea x from 331 to 299 , y from -40 to 1
\put {\includegraphics[width=42.5cm]{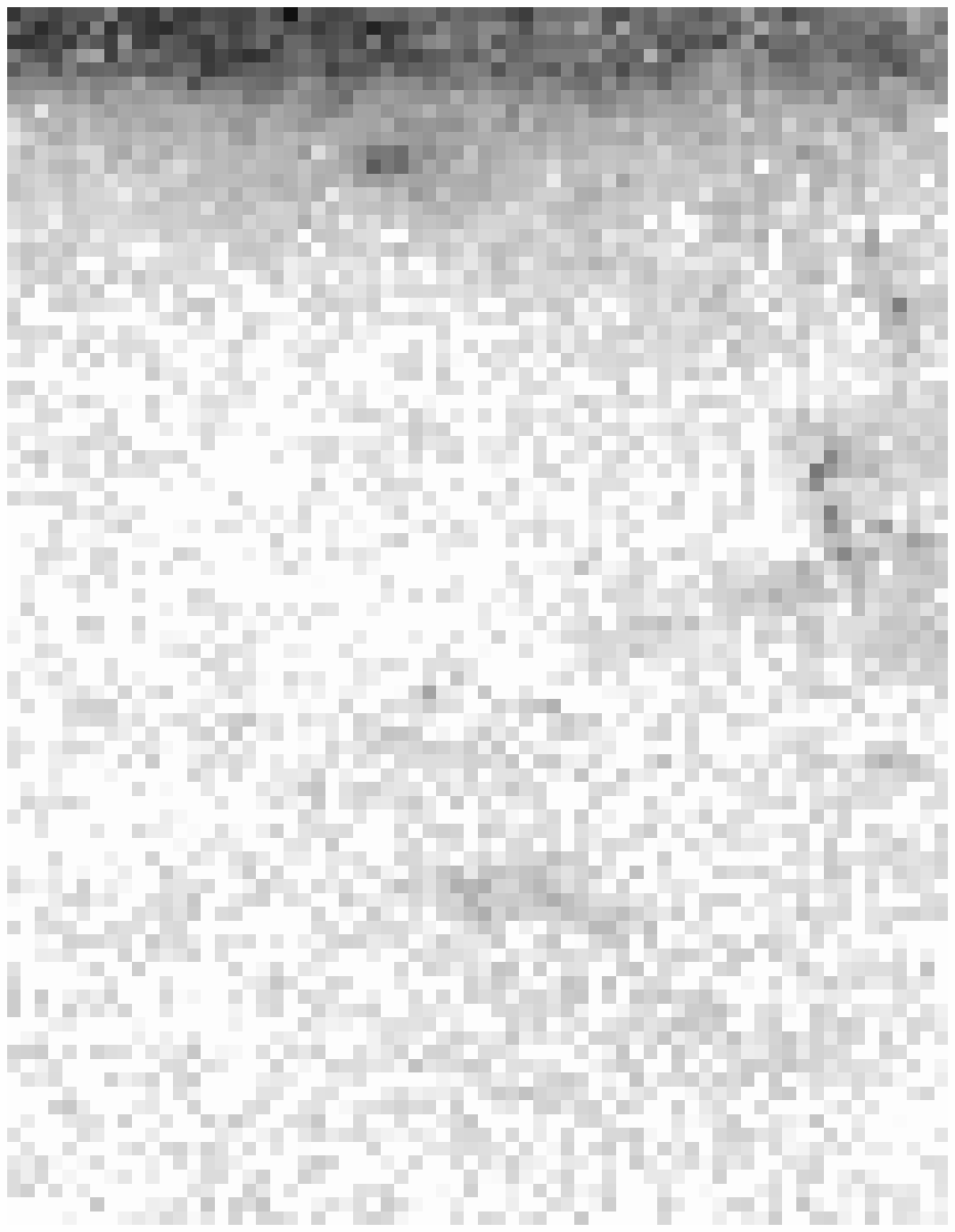}} at 315 -19.5
\axis left label {\begin{sideways}b [\degr]\end{sideways}}
ticks in long numbered from -40 to 0 by 5
      short unlabeled from -40 to 0 by 1 /
\axis right label {}
ticks in long unlabeled from -40 to 0 by 5
      short unlabeled from -40 to 0 by 1 /
\axis bottom label {l [\degr]}
ticks in long numbered from 300 to 330 by 5
      short unlabeled from 300 to 330 by 1 /
\axis top label {}
ticks in long unlabeled from 300 to 330 by 5
      short unlabeled from 300 to 330 by 1 /
\endpicture  
\caption{\label{detail23} Grey scale representation of a detail of our
extinction map based on the nearest 49 stars. Extinction values are square root
scaled from 0\,mag (white) to 15\,mag (black) of optical extinction. }    
\end{figure*}

\clearpage
\newpage
\begin{figure*}
\beginpicture
\setcoordinatesystem units <-5mm,5mm> point at  0 0
\setplotarea x from 361 to 329 , y from -40 to 1
\put {\includegraphics[width=42.5cm]{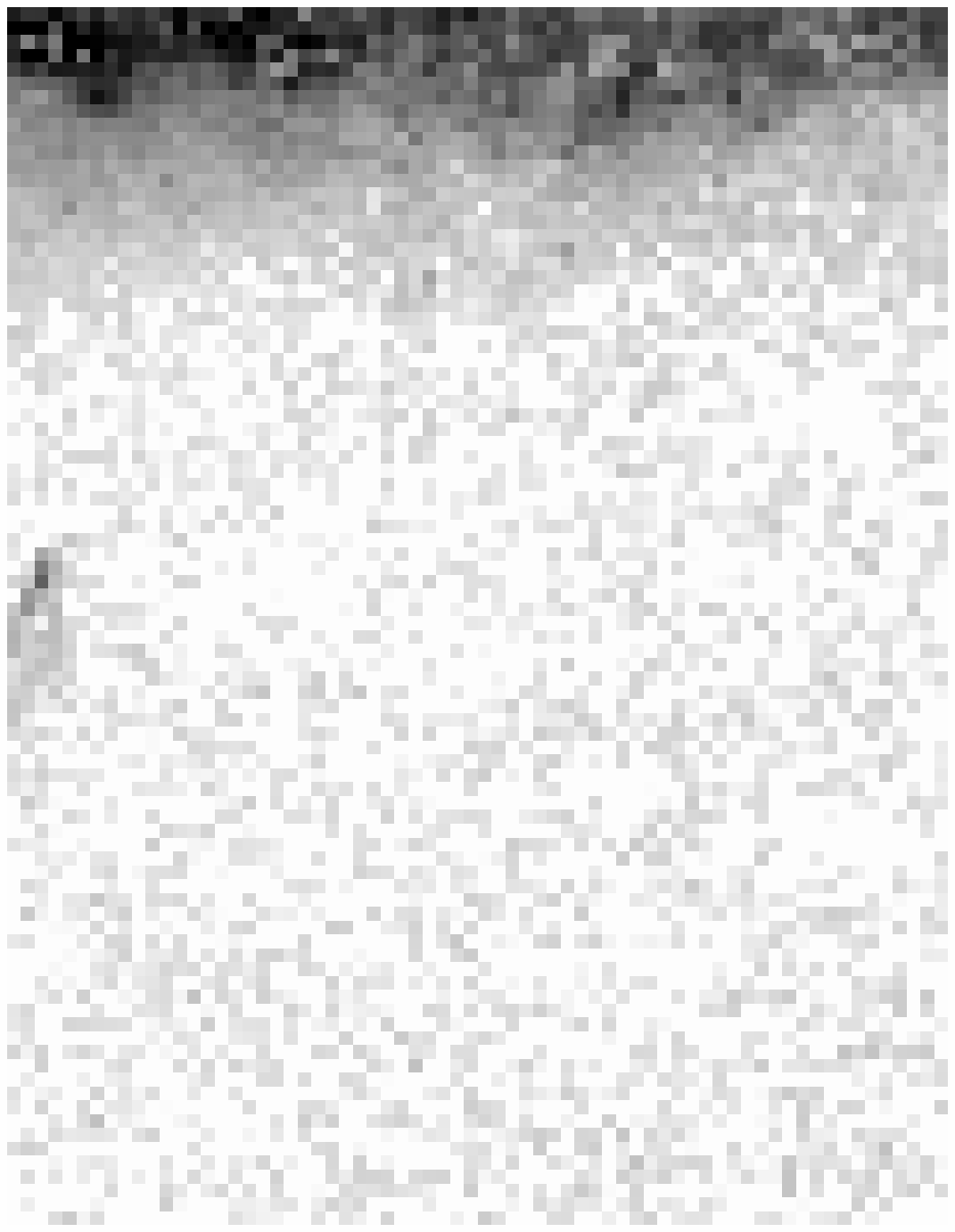}} at 345 -19.5
\axis left label {\begin{sideways}b [\degr]\end{sideways}}
ticks in long numbered from -40 to 0 by 5
      short unlabeled from -40 to 0 by 1 /
\axis right label {}
ticks in long unlabeled from -40 to 0 by 5
      short unlabeled from -40 to 0 by 1 /
\axis bottom label {l [\degr]}
ticks in long numbered from 330 to 360 by 5
      short unlabeled from 330 to 360 by 1 /
\axis top label {}
ticks in long unlabeled from 330 to 360 by 5
      short unlabeled from 330 to 360 by 1 /
\endpicture  
\caption{\label{detail24} Grey scale representation of a detail of our
extinction map based on the nearest 49 stars. Extinction values are square root
scaled from 0\,mag (white) to 15\,mag (black) of optical extinction. }    
\end{figure*}

\end{appendix}


\begin{thebibliography}{}

\bibitem[2005]{2005PASJ...57S...1D}
Dobashi, K., Uehara, H., Kandori, R., Sakurai, T., Kaiden, M., Umemoto, T., Sato, F., 2005, \pasj, 57, 1
%Atlas and Catalog of Dark Clouds Based on Digitized Sky Survey I

\bibitem[2003]{2003ARA&A..41..241D}
Draine, B.T., 2003, \araa, 41, 241-289
%Interstellar Dust Grains

\bibitem[2006]{2006MNRAS.369.1901F}
Froebrich, D. \& del Burgo, C. 2006, MNRAS, 369, 1901
%Extinction techniques and impact on dust property determination

\bibitem[2007]{2007MNRAS.378.1447F}
Froebrich, D., Murphy, G.C., Smith, M.D., Walsh, J., Del Burgo, C., 2007, \mnras, 378, 1447-1460
%A large-scale extinction map of the Galactic Anticentre from 2MASS

\bibitem[2005]{2005A&A...432L..67F}
Froebrich, D., Ray, T.P., Murphy, G.C., Scholz, A., 2005, \aap, 432, L67-L70
%A Galactic Plane relative extinction map from 2MASS

\bibitem[2008]{2008arXiv0806.3441G}
Goodman, A.A., Pineda, J.E., Schnee, S.L., 2008, 0806.3441
%The ''True'' Column Density Distribution in Star-Forming Molecular Clouds

\bibitem[2001]{2001ApJ...562..852H}
Hartmann, L., Ballesteros-Paredes, J. \& Bergin, E.A. 2001, ApJ, 562, 852
%Rapid Formation of Molecular Clouds and Stars in the Solar Neighborhood

\bibitem[1962]{1962..............K}
Kenney, J.F. Keeping, E.S., 1962, in Mathematics of Statistics, Pt. 1, 3rd ed. Princeton, NJ: Van Nostrand, p. 211
%Dunno the title

\bibitem[2003]{2003ARA&A..41...57L}
Lada, C.J., Lada, E.A., 2003, \araa, 41, 57
%Embedded Clusters in Molecular Clouds

\bibitem[1994]{1994ApJ...429..694L}
Lada, C.J., Lada, E.A., Clemens, D.P., Bally, J., 1994, \apj, 429, 694-709
%Dust extinction and molecular gas in the dark cloud IC 5146

\bibitem[2005]{2005ApJ...623..897L}
Larson, K.A., Whittet, D.C.B. 2005, ApJ, 623, 897
%Reddening and the Extinction Law at High Galactic Latitude

\bibitem[1994]{1994AAS...184.3501L}
Lasker, B.~M., 1994, in Bulletin of the American Astronomical Society, 26, 914
%Digitization and Distribution of the Large Photographic Surveys

\bibitem[2005]{2005A&A...438..169L}
Lombardi, M., 2005, \aap, 438, 169-185
%Optimal column density measurements from multiband near-infrared observations

\bibitem[2001]{2001A&A...377.1023L}
Lombardi, M., Alves, J., 2001, \aap, 377, 1023-1034
%Mapping the interstellar dust with near-infrared observations: An optimized multi-band technique

\bibitem[2006]{2006A&A...454..781L}
Lombardi, M., Alves, J., Lada, C.J., 2006, \aap, 454, 781-796
%2MASS wide field extinction maps. I. The Pipe nebula

\bibitem[2008]{2008A&A...489..143L}
Lombardi, M., Lada, C.J., Alves, J., 2008, \aap, 489, 143-156
%2MASS wide field extinction maps. II. The Ophiuchus and the Lupus cloud complexes

\bibitem[2006]{2006A&A...453..635M}
Marshall, D.J., Robin, A.C., Reyl\'e, C., Schultheis, M., Picaud, S. 2006, A\&A, 453, 635
%Modelling the Galactic interstellar extinction distribution in three dimensions

\bibitem[1990]{1990ApJ...357..113M}
Martin, P.G., Whittet, D.C.B., 1990, \apj, 357, 113-124
%Interstellar extinction and polarization in the infrared

\bibitem[1990]{1990ARA&A..28...37M}
Mathis, J.S., 1990, \araa, 28, 37-70
%Interstellar dust and extinction

\bibitem[1993]{1993A&A...280..617O}
Ossenkopf, V., 1993, \aap, 280, 617-646
%Dust coagulation in dense molecular clouds: The formation of fluffy aggregates

\bibitem[1998]{1998PhRvE..58.4501P}
Passot, T., V\'{a}zquez-Semadeni, E., 1998, PhRvE, 58, 4501
%Density probability distribution in one-dimensional polytropic gas dynamics

\bibitem[1993]{1993A&A...279..577P}
Preibisch, T., Ossenkopf, V., Yorke, H.W., Henning, T., 1993, \aap, 279, 577-588
%The influence of ice-coated grains on protostellar spectra

\bibitem[1998]{1998ApJ...500..525S}
Schlegel, D.J., Finkbeiner, D.P., Davis, M., 1998, \apj, 500, 525
%Maps of Dust Infrared Emission for Use in Estimation of Reddening and Cosmic Microwave Background Radiation Foregrounds

\bibitem[2008]{2008arXiv0811.2902S}
Schultheis, M., Sellgren, K., Ramirez, S., Stolovy, S., Ganesh, S., Glass, I. S., Girardi, L. 2008, \aa, in press
%Interstellar Extinction and Long-Period Variables in the Galactic Center

\bibitem[2006]{2006AJ....131.1163S}
Skrutskie, M.F., Cutri, R.M., Stiening, R., Weinberg, M.D., Schneider, S., Carpenter, J.M., Beichman, C., and et al.,  2006, \aj, 131, 1163-1183
%The Two Micron All Sky Survey (2MASS)

\bibitem[2001]{2001ApJ...557..727V}
V\'{a}zquez-Semadeni, E., Garc\'{i}a, N., 2001, \apj, 557, 727
%The Probability Distribution Function of Column Density in Molecular Clouds

\bibitem[1923]{1923AN....219..109W}
Wolf, M., 1923, Astronomische Nachrichten, 219, 109
%\"Uber den dunklen Nebel NGC 6960

\end{thebibliography}
\end{document}